\documentclass[review]{elsarticle}
\usepackage{amssymb}
\usepackage{lineno,hyperref}
\modulolinenumbers[5]

\journal{Physics of the Dark Universe}
\newcommand{\beq}{\begin{equation}}
\newcommand{\eeq}{\end{equation}}
\newcommand{\bea}{\begin{eqnarray}}
\newcommand{\eea}{\end{eqnarray}}
\newcommand{\eqref}[1]{(\ref{#1})}
\newcommand{\calC}{\mathcal{C}}
\newcommand{\calS}{\mathcal{S}}
\newcommand{\calN}{\mathcal{N}}

\begin{document}
\begin{frontmatter}
\title{The Evens and Odds of CMB Anomalies}

\author[mymainaddressAG,mysecondaryaddressAG]{A.Gruppuso\corref{mycorrespondingauthor}}
\cortext[mycorrespondingauthor]{Corresponding author}
\ead{gruppuso@iasfbo.inaf.it}
\author[mymainaddressNK]{N.~Kitazawa}
\ead{noriaki.kitazawa@tmu.ac.jp}
\author[mymainaddressML]{M.~Lattanzi}
\ead{lattanzi@fe.infn.it}
\author[mymainaddressNM,mymainaddressAG]{N.~Mandolesi}
\ead{mandolesi@iasfbo.inaf.it}
\author[mymainaddressNM,mymainaddressML]{P.~Natoli}
\ead{paolo.natoli@unife.it}
\author[mymainaddressAS]{A.~Sagnotti}
\ead{sagnotti@sns.it}
\address[mymainaddressAG]{INAF-OAS Bologna,
Osservatorio di Astrofisica e Scienza dello Spazio di Bologna,
Istituto Nazionale di Astrofisica,
via Gobetti 101, I-40129 Bologna, Italy}
%\footnote{On January 1 2018, INAF-IASF Bologna merged with Osservatorio Astronomico di Bologna to form a new institute called INAF-OAS Bologna.}}
%\context[mynewinstitutename]{new affiliation: INAF-OAS Bologna, Osservatorio di Astrofisica e Scienza dello Spazio di Bologna, Istituto Nazionale di Astrofisica,
%via Gobetti 101, I-40129 Bologna, Italy}
\address[mysecondaryaddressAG]{INFN, Sezione di Bologna,
Via Irnerio 46, I-40126 Bologna, Italy}
\address[mymainaddressNK]{Department of Physics, Tokyo Metropolitan University,
Hachioji, Tokyo 192-0397, Japan}
\address[mymainaddressML]{INFN -- Sezione di Ferrara, Via Saragat 1, I-44100 Ferrara, Italy}
\address[mymainaddressNM]{Dipartimento di Fisica e Scienze della Terra, Universit\`a degli Studi di Ferrara, Via Saragat 1, I-44100 Ferrara, Italy,
INAF-IASF Bologna}
\address[mymainaddressAS]{Scuola Normale Superiore and INFN,
Piazza dei Cavalieri 7\
I-56126 Pisa, Italy}

\begin{abstract}
\noindent
The lack of power of large--angle CMB anisotropies is known to increase its statistical significance at higher Galactic latitudes, where a string--inspired pre--inflationary scale $\Delta$ can also be detected. { Considering the Planck 2015 data, and relying largely on a Bayesian approach, we show that the effect is mostly driven by the \emph{even}--$\ell$ harmonic multipoles with $\ell \lesssim 20$, which appear sizably suppressed in a way that is robust with respect to Galactic masking, along with the corresponding detections of $\Delta$.} On the other hand, the first \emph{odd}--$\ell$ multipoles are only suppressed at high Galactic latitudes. We investigate this behavior in different sky masks, constraining $\Delta$ through even and odd multipoles, and we elaborate on possible implications. {We include low--$\ell$ polarization data which, despite being noise--limited, help in attaining confidence levels of about 3 $\sigma$ in the detection of $\Delta$. }We also show by direct forecasts that a future all--sky $E$--mode cosmic--variance--limited polarization survey may push the constraining power for $\Delta$ beyond 5 $\sigma$.
\end{abstract}
\begin{keyword}
CMB \sep Low-$\ell$ anomaly \sep Inflation \sep Supergravity \sep String Theory

\end{keyword}

\end{frontmatter}

%\linenumbers
%\pacs{}
%
%
%\maketitle
%
%
\section{Introduction}
\label{intro}
Cosmic Microwave Background (CMB) observations have been instrumental in defining the $\Lambda$CDM concordance model, and have constrained its parameters to the percent level or better~\cite{Ade:2015xua}.
Anomalies in CMB temperature maps, however, have long surfaced, especially at large angular scales, with a typical 2 to 3 $\sigma$ significance. Still, these anomalies are not a priori irrelevant in view of their potential physical implications.
The lack of angular correlation in the CMB two--point function was originally noted in COBE data~\cite{Hinshaw:1996ut}, and was later confirmed in all WMAP~\cite{Copi:2006tu,Copi:2008hw,Sarkar:2010yj,Gruppuso:2013dba} and {\sc Planck} releases~\cite{Ade:2015hxq,Schwarz:2015cma,Copi:2013cya}. The low--variance anomaly~\cite{Monteserin:2007fv,Cruz:2010ud,Gruppuso:2013xba} is a related observation, suggesting that the low--$\ell$ CMB anisotropy contains less power, with respect to smaller angular scales, than $\Lambda$CDM would prefer. As pointed out by Copi \emph{et al.}~\cite{Copi:2016hhq}, if this behavior is not a statistical fluke only a physical mechanism impacting the CMB up to last scattering could explain it, while the integrated Sachs-Wolfe effect \cite{ISW} would not be able to screen it.

In \cite{Gruppuso:2015zia,Gruppuso:2015xqa} we have searched, in {\sc Planck} temperature (and polarization) data, for signatures of modified primordial power spectra of the type
\beq
P(k) \ \sim \ \frac{k^3}{\left[k^2 \ + \ \Delta^2 \right]^{2\,-\,\frac{n_s}{2}}} \ . \label{cutoff}
\eeq
Here $\Delta$ controls the transition from a large--scale depression to the usual Chibisov--Mukhanov tilt~\cite{cm} $P(k) \sim k^{\,n_s\,-\,1}$.  Starting from the slow--roll Coul\-omb--barrier Mukhanov--Sasaki potential~\cite{inflation} $\alpha/\eta^2$, where $\eta$ denotes conformal time and $\alpha=(n_s-3)(n_s-5)/4$, the power spectrum of eq.~\eqref{cutoff} can be readily obtained via a vertical shift into $\alpha/\eta^2 - \Delta^2$, which also models the sign change accompanying transitions from fast--roll to slow--roll. A power cut is indeed the main signature of an inflaton decelerating to slow-roll. It is also accompanied, in general, by a peak and some nearby oscillations, whose positions and sizes are however model dependent.

String Theory \cite{strings} and Supergravity \cite{supergravity} may provide some clues on how slow--roll started. In orientifold models \cite{orientifolds} with ``brane supersymmetry breaking'' \cite{bsb}, vacuum effects force a scalar to climb up a steep exponential potential as it emerges from the initial singularity, before attaining slow--roll in the ensuing descent \cite{climb}.  Up to the local features that we have just mentioned, this dynamics gives rise to primordial power spectra exhibiting the cut of eq.~\eqref{cutoff}, which could account for the observed low--$\ell$ depression of the CMB temperature angular power spectrum. This CMB anomaly might thus reflect an early epoch of deceleration, in principle accessible in scenarios with relatively few $e$--folds~\cite{depression}.

The string--inspired cosmology of~\cite{climb}~\footnote{A number of scenarios leading to infrared cuts were explored over the years, see \cite{CMB_cut} for a list of references.} (see \cite{erice15} for reviews)  motivated  our work in \cite{Gruppuso:2015zia,Gruppuso:2015xqa}, where we showed that at higher Galactic latitudes the detection level of $\Delta$ in {\sc Planck} data improves, reaching up a $99.4\%$ C.L.\ in a blind $30^\circ$ extension of the standard mask (corresponding to an available sky fraction $f_\mathrm{sky}=39\%$), where one finds
\beq
\Delta \ = \ \left(0.35 \ \pm \ 0.11 \right) \ \times \ 10^{-3} \ {\rm Mpc}^{-1} \qquad \mathrm{(68\% \,\, C.L.)\ .} \label{delta}
\eeq
The lack--of--power anomaly was known to grow when the Galactic mask is widened \cite{Gruppuso:2013xba,Gruppuso:2013dba}, and $\Lambda$CDM extended with eq.~\eqref{cutoff} nicely captures this feature. Sampling for $\Delta$ while altering the Galactic mask does not affect significantly the standard cosmological parameters, while $\Delta$ increases from $\left(0.17 \ \pm 0.09 \right) \ \times 10^{-3} \ {\rm Mpc}^{-1}$ $\mathrm{(68\% \,\, C.L.)}$ to the result in eq.~\eqref{delta} in going from $f_\mathrm{sky}=94\%$ to $f_\mathrm{sky}=39\%$.  {Still, the dependence on the sky fraction of $\Delta$ -- a parameter that is potentially of cosmological origin -- appears intriguing. This very fact has led us to reconsider the issue from a different perspective, which is the purpose of the present work.}

We thus continue our investigation of latitude effects, relying on the 2015 low~-~$\ell$ {\sc Planck} likelihood~\cite{Aghanim:2015xee}, which was already used in~\cite{Gruppuso:2015xqa} and is publicly available from the Planck Legacy Archive~\footnote{\url{http://www.cosmos.esa.int/web/planck/pla}}.
Our aim was initially a critical assessment of the current overall setup, in order to try and highlight local features that might or might not support further excisions near the Galactic plane, or perhaps local improvements of the component separation algorithms, for the sake of present and future observations.  However, the scope of this inquiry has broadened somewhat along the way.

In detail, we focus on $\Delta$, both in view of its potential theoretical significance and, more phenomenologically, as a proxy for two well--known CMB anomalies, the lack of power and the even--odd asymmetry, and of their dependence on the sky fraction. To this end, we consider the same set of masks used in~\cite{Gruppuso:2015xqa}, which correspond to a \emph{sequence} of blind extensions of the standard~\texttt{Commander}~\cite{Eriksen:2004ss,Eriksen:2007mx,Adam:2015tpy} mask by \emph{six--degree} steps~\footnote{\texttt{Commander} is one of four component separation algorithms used to analyze {\sc Planck} data. Although we focus on \texttt{Commander}, which is employed in the standard {\sc Planck} likelihood, for the sake of comparison, and in order to verify that our results are not unique to \texttt{Commander}, we have also analyzed a subset of cases with the \texttt{Smica} CMB temperature map~\cite{Adam:2015tpy}, with consistent findings.}, {together with some additional ``complementary'' masks that we created, }in combination with (anti)symmetrized temperature maps that disentangle the contributions of low--$\ell$ even and odd multipoles.

We believe that our main result, the different behaviour of even and odd multipoles, which was detected via different estimators, provides new insights into the low--$\ell$ tension between $\Lambda$CDM and the CMB. Moreover, we also confirm that this tension, which is largely driven by the even multipoles, increases at higher Galactic latitudes as described in \cite{Gruppuso:2013xba,Gruppuso:2013dba}. Initially we dwelled at length on a possible origin of the phenomenon from spurious contaminations around the Galactic plane, but the very behavior of even multipoles, which appear to conform to an isotropic pattern, strengthens somewhat the case for a primordial interpretation of $\Delta$.

In \cite{Gruppuso:2015xqa} we also attempted to place joint constraints on $\Delta$ and on a second parameter, $\gamma$, which was originally introduced in~\cite{depression}. A positive value of $\gamma$ would model a peak located at the transition to an almost scale--invariant primordial power spectrum, albeit without nearby oscillations. We did find a scant evidence for a positive $\gamma$, but adding a second free parameter, as expected, weakened somewhat the detection level of $\Delta$. Here we are thus restricting our attention to $\Delta$, which suffices to model the large--scale depression that a decelerating inflaton would introduce in primordial power spectra.

The future of CMB experimental observations lies in providing high--quality cosmic--variance--limited polarization data. The main drive for this effort is at present the quest for $B$--modes. One of the purposes of this work is to stress that even high--quality $E$--mode polarization data, if fully exploited, can have far reaching implications. As we shall see, they can shed light on CMB anomalies, and thus potentially on the onset of inflation, via better determinations of the parameter $\Delta$ that is the main focus of this work.

The paper is organized as follows. In Section~\ref{dataandsims} we describe the adopted estimators, the tools used in the analysis, the datasets considered and the corresponding simulations. In Section \ref{results} we describe our results for the cosmological parameters, paying special attention to $\Delta$ and to its connection with the even--odd asymmetry. We also report on our analysis for the three estimators of Section~\ref{dataandsims} and explore the constraining power of future all--sky polarization surveys. Finally, Section \ref{sec:final_remarks} collects some general considerations on this work and on its relation to future measurements, whose highlights are briefly summarized in Section~\ref{sec:conclusions}.
\begin{table}
\centering
\caption{\small The masks that we have used, with the notation that identifies them. Here $f_\mathrm{sky}$ denotes the sky fraction available for the analysis of the even+odd $TT$ contributions, while $f_\mathrm{sky}^\pm$ denote the corresponding fractions available after parity (anti)symmetrization of the maps. The sky fractions of the last two columns refer to the same low--$\ell$ polarization mask, which relies on both {\sc Planck} and WMAP9 data, as described in Section~\ref{dataandsims}.}
\vskip 5pt
\label{tab:masks}
\begin{tabular}{ccccccc}
\hline
Case & Label & $f_\mathrm{sky}$ &  $f_\mathrm{sky}^\pm$ & $f_\mathrm{sky}^{\rm Pol}$ & $f_\mathrm{sky}^{\rm Pol, \pm}$\\
\hline

$a$ & Standard   & 93.6  & 90.7 & 73.9 & 65.3\\
$b$ & \phantom{mmmmm} ${\rm Ext}_{06}$ \phantom{mmmmm}  &   83.8 & 83.0 & 73.9 & 65.3\\
$c$ & ${\rm Ext}_{12}$   &   70.8 & 70.6 & 73.9 & 65.3\\
$d$ & ${\rm Ext}_{18}$   &   59.1 & 59.1 & 73.9 & 65.3\\
$e$ & ${\rm Ext}_{24}$   &  48.7  & 48.7 & 73.9 & 65.3\\
$f$ & ${\rm Ext}_{30}$   & 39.4  & 39.4 & 73.9 & 65.3\\
$g$ & ${\rm Ext}_{36}$   &  30.9 & 30.9 & 73.9 & 65.3\\
$h$ & ${\rm Compl}_{18}$ &  34.5 & 31.6 & 73.9 & 65.3\\
$i$ & ${\rm Compl}_{24}$ &  44.9 & 42.0 & 73.9 & 65.3\\
$j$ & ${\rm Compl}_{30}$ & 54.2  & 51.3 & 73.9 & 65.3\\
$k$ & ${\rm Compl}_{36}$ & 62.7 & 59.8 & 73.9 & 65.3\\
\hline
\end{tabular}
\end{table}

\section{Dataset and simulations}
\label{dataandsims}

Our analysis rests largely on the latest public {\it Planck} satellite CMB temperature data (2015 release) for $\ell < 30$ and on a sequence of blind extensions of the standard {\sc Planck} mask, together with some complements, whose features are summarized in table~\ref{tab:masks}. It is worth stressing here that the {\sc Planck} temperature measurements have already reached the cosmic variance limit, surely in the low--multipole range $\ell < 30$ that is central to this work.

The first row refers to the standard $f_\mathrm{sky}=93.6\%$ mask provided by {\sc Planck}, rows $b-g$ refer to masks obtained via blind extensions of the standard one by $6^\circ$ steps, as in \cite{Gruppuso:2015xqa}, and finally rows $h-k$ refer to masks obtained adjoining to the standard one in $a$ the complements of those in $d-g$. Notice that the (anti)symmetrization of maps reduces the sky fraction at the expense of regions close to the Galactic plane.
Notice also that the polarization maps used
%in rows $b-k$
were obtained combining WMAP9 and {\sc Planck} data, as in~\cite{Lattanzi:2016dzq}, thus improving the signal--to--noise ratio while also extending the corresponding sky fractions. In the following, we shall refer to this combination as {\sc Planck}LFI+WMAP, and to the corresponding mask as the ``union'' mask~\footnote{We do not widen polarization masks, since current polarization data are already largely noise limited.}.

\begin{figure}[ht]
\centering
\begin{tabular}{c}
\includegraphics[width=90mm]{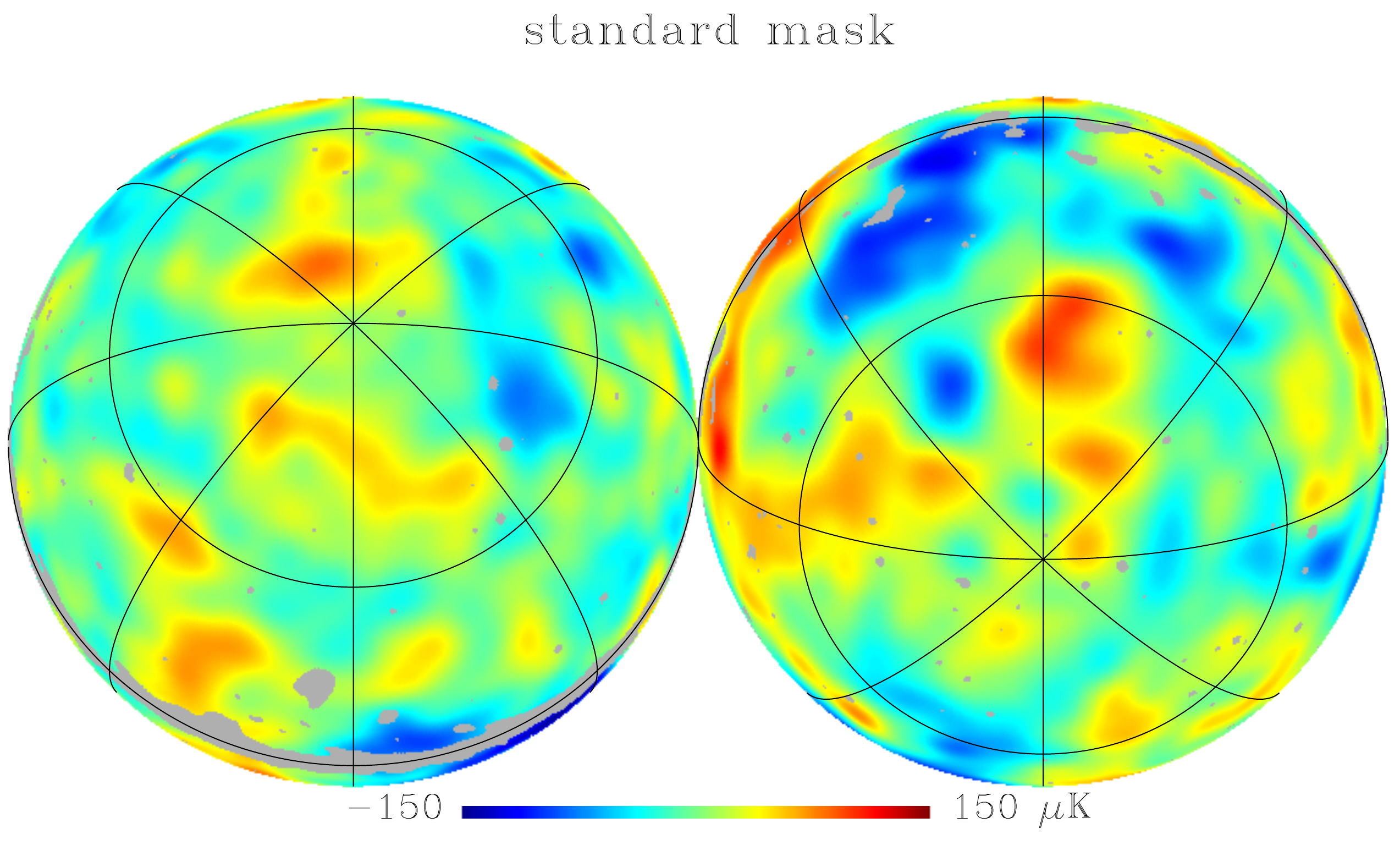}  \\
\includegraphics[width=90mm]{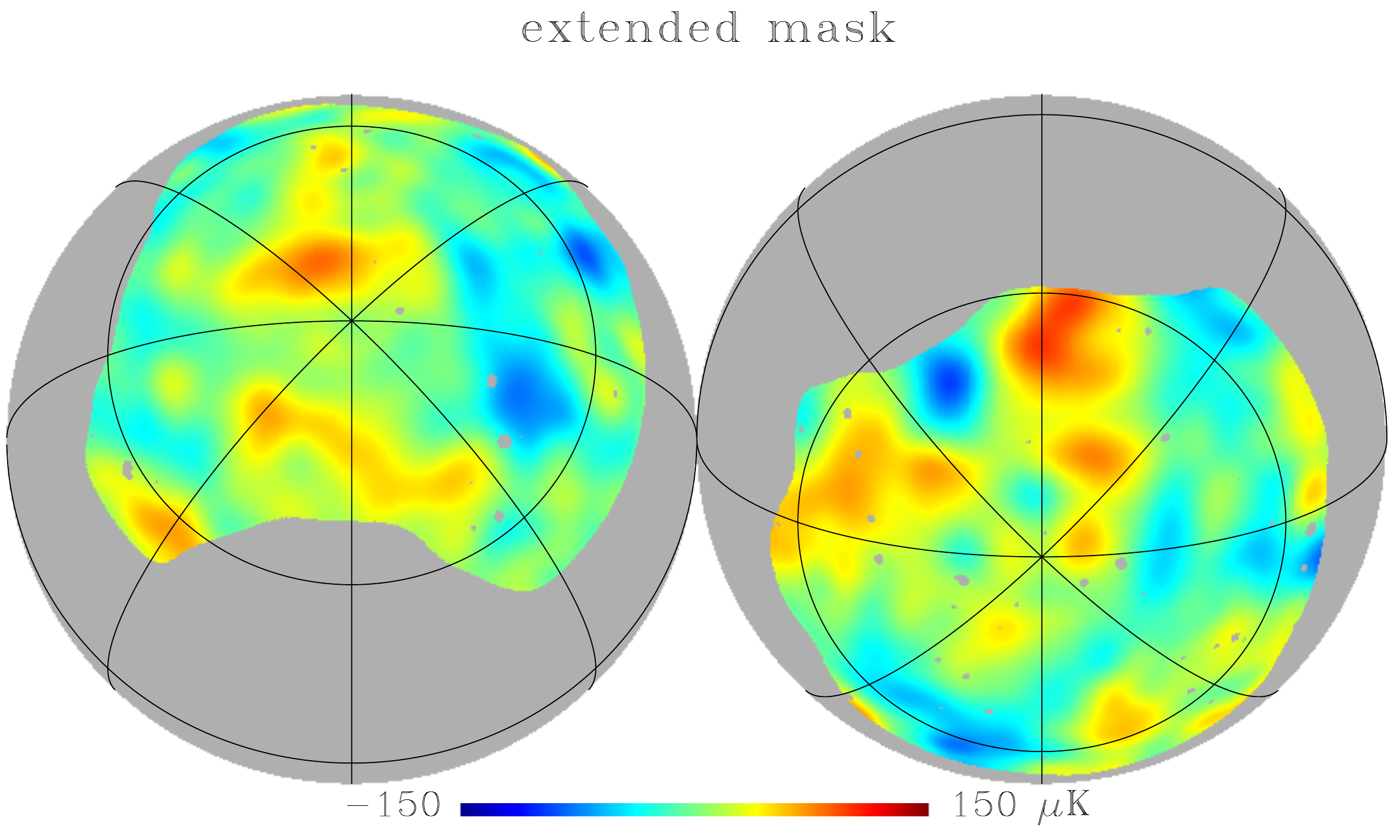} \\
\includegraphics[width=90mm]{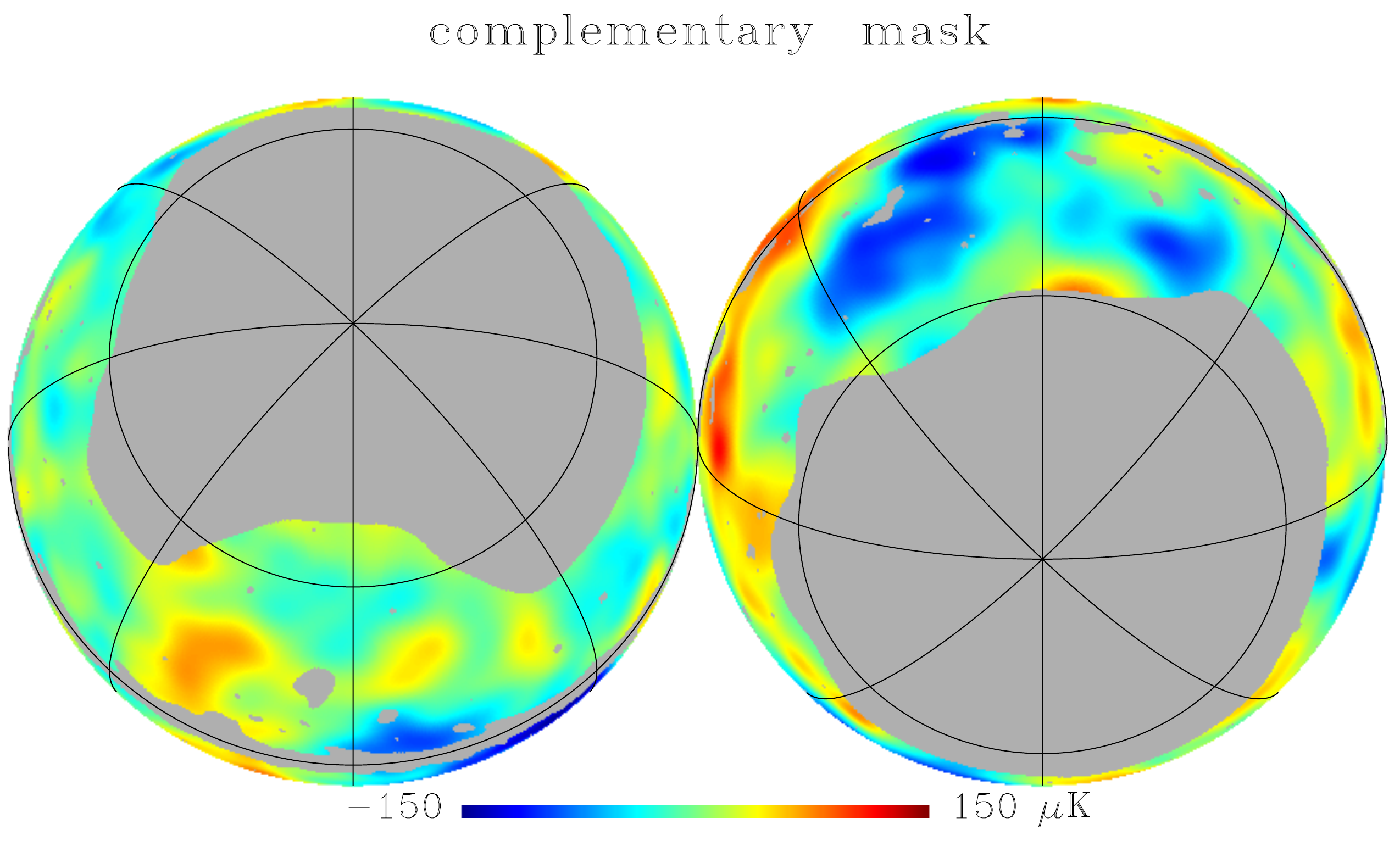} \\
\end{tabular}
\caption{\small The {\sc Planck} CMB temperature map provided by the \texttt{Commander} algorithm, with three different maskings. From upper to lower panel: standard mask (case $a$), extended mask (case $f$), complementary mask (case $j$). The corresponding observed sky fractions are $93.6 \%$, $39.4 \%$ and $54.2 \%$ respectively. Color code units: $\mu K$.}
\label{fig:masks}
\end{figure}

In order to highlight the nature of these choices, fig.~\ref{fig:masks} displays the visible sky for the standard \texttt{Commander} mask in $a$, the visible sky for its blind extension in $f$ that resulted in the most significant determination of $\Delta$ in \cite{Gruppuso:2015xqa}, and the complementary sky in $j$.

We have used the {\sc Planck} temperature CMB map provided by the \texttt{Commander} algorithm, which enters the temperature sector of the low-$\ell$ {\sc Planck} likelihood~\cite{Aghanim:2015xee}.
This was smoothed at $440'$ and was downsampled at {\tt
HEALPix}\footnote{\url{http://healpix.sourceforge.net/}} \cite{gorski} resolution $N_\mathrm{side}=16$. A Gaussian white noise realization with $\sigma = 2\,\mu\mathrm{K}$
was added to that map, spawning a noise covariance matrix $\calN$ with elements $\calN_{ij} \equiv \sigma^2 \delta_{ij}$.

We proceeded along two different lines. We first performed Bayesian parameter estimation, both in the standard $\Lambda$CDM model and in its one--parameter extension, which we denote $\Lambda$CDM$\Delta$. This also allowed to obtain Bayesian estimates of a few quantities that characterize large--scale anomalies:
\begin{itemize}
\item $S_\frac{1}{2}$, defined as~\cite{Spergel:2003cb}
\begin{equation}
S_\frac{1}{2} \ = \ \int_{\pi/3}^{\pi} d \theta \, C (\theta)^2 \sin \theta
\label{s1su2estimator} \ ,
\end{equation}
which is a standard tool to quantify the lack of correlation in the two--point function $C(\theta)$ for temperature CMB anisotropies, where
\begin{equation}
C(\theta) \ \equiv \  \sum_{\ell = 2}^{\ell_{\rm max}} \left( {{2 \ell +1} \over {4 \pi}} \right) P_{\ell}(\cos\theta) \ C_{\ell}^{\rm TT}  \ . \label{CTT}
\end{equation}
Here $\ell_{\rm max}$ (=29 in our analysis) is the highest multipole considered, the $P_{\ell}$ are Legendre polynomials and the $C_{\ell}^{\rm TT}$ are temperature angular power spectrum coefficients. In practice, $S_\frac{1}{2}$ quantifies the distance between $C(\theta)$ and the null function, within the range $60^{\circ} < \theta < 180^{\circ}$, and can be rewritten as~\cite{Copi:2011pe}
\beq
S_\frac{1}{2} \ = \ \sum_{\ell,\ell'} \ C_\ell^{\rm TT}\, {\cal I}_{\ell \ell'}\, C_{\ell'}^{\rm TT} \ ,
\eeq
with
\beq
{\cal I}_{\ell \ell'} \ = \ \frac{(2\ell+1)(2\ell'+1)}{(4\,\pi)^2} \ \int_{-1}^{1/2} \ dx \ P_\ell(x)\, P_{\ell'}(x) \ ;
\eeq

\item the \emph{variance}, used in~\cite{Monteserin:2007fv,Cruz:2010ud,Gruppuso:2013xba},
\begin{equation}
V\left(\ell_{max}\right) \ \equiv \ C(0) \ = \ \sum_{\ell = 2}^{\ell_{max}} \left( {{2 \ell +1} \over {4 \pi}} \right) \, C_{\ell}^{\rm TT}  \ , \label{variance}
\end{equation}
which defines the auto-correlation function up to $\ell_{max}$;

\item the \emph{even-odd asymmetry}~\cite{Kim:2010gf,Kim:2010gd,Gruppuso:2010nd,aluri,Ade:2015hxq}, defined as
\begin{equation}
R(\ell_{max}) \ = \ \frac{C^{+}\left({\ell_{max}}\right)}{C^{-}\left({\ell_{max}}\right)} \ ,
\label{r}
\end{equation}
where
\begin{equation}
C^{\pm}\left({\ell_{max}}\right) \ = \ \frac{1}{N_\pm}\ \sum_{\ell=2}^{\ell_{max}} \ \frac{1 \ \pm \ (-1)^\ell}{2}\  \frac{\ell (\ell + 1) }{2\,\pi} \ C_{\ell}^{\rm TT} \ ,
\label{C+-}
\end{equation}
and
\begin{equation}
N_+ \ = \ \left[\frac{\ell_{max}}{2}\right] \ , \qquad N_- \ = \ \left[\frac{\ell_{max}-1}{2}\right] \ ,
\end{equation}
with square brackets denoting integer parts. The quantity $R$ in eq.~\eqref{r} compares the total amounts of power carried by even and odd multipoles, up to $\ell_{max}$.
\end{itemize}

\begin{figure}[ht]
\centering
\begin{tabular}{c}
\includegraphics[width=85mm]{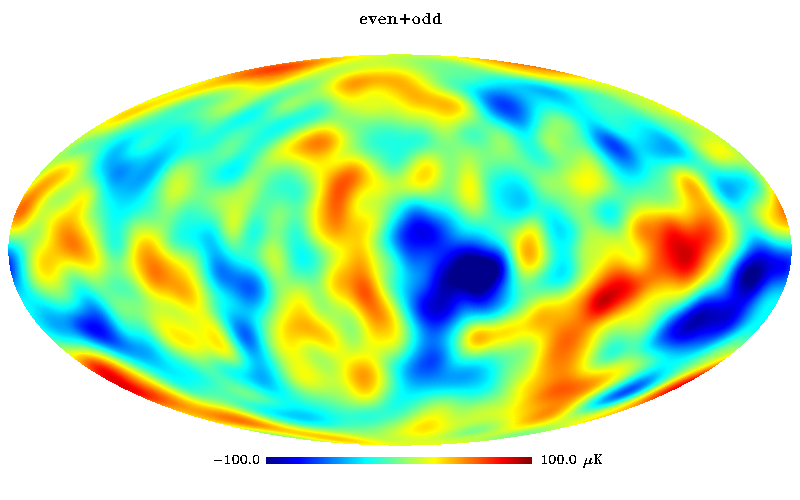}  \\
\includegraphics[width=85mm]{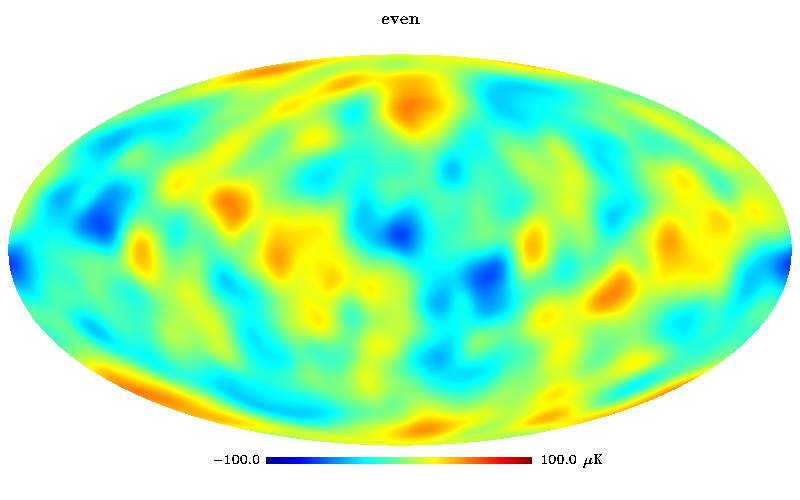} \\
\includegraphics[width=85mm]{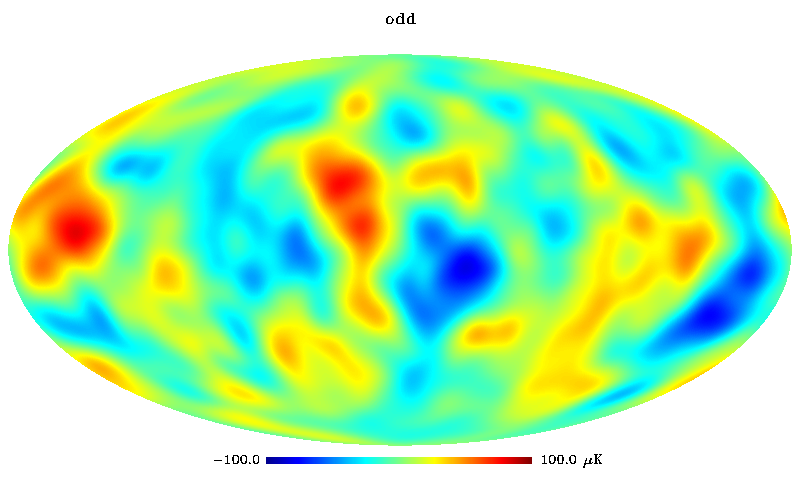} \\
\end{tabular}
\caption{\small The {\sc Planck} CMB temperature map provided by the \texttt{Commander} algorithm, smoothed at $440'$ and reconstructed at $N_{side}=16$.
Upper panel: the full map $\mathbf{m}$; middle panel: its even-parity projection $\mathbf{m_+}$; lower panel: its odd-parity projection $\mathbf{m_-}$. Color code units: $\mu K$. The reader will not fail to spot eye-catching differences between the symmetrized low--$\ell$ map in the middle panel and the others.}
\label{fig:commander}
\end{figure}

In addition to the standard \texttt{Commander} temperature anisotropy map $\mathbf{m}$, we also considered its point--parity even and odd projections $\mathbf{m}_\pm$,
which discretize
\beq
\left(\frac{\delta T}{T}(\hat n)\right)_\pm \ \equiv \ \frac{1}{2} \ \left[ \frac{\delta T}{T}(\hat n) \ \pm \ \frac{\delta T}{T}(- \hat n) \right] \ , \label{maps}
\eeq
where $\hat n$ identifies the direction corresponding to a given pixel.
The resulting full--sky maps are displayed fig.~\ref{fig:commander}.

For given theoretical $C_\ell^{\rm TT}$, the likelihood function of the temperature map is, up to a normalization
\beq
\mathcal{L}\left(\mathbf{m}|C_\ell^{\rm TT}\right) \ \propto \ \frac{1}{|\calC^{\rm TT}|^{1/2}}\ \exp{\left(\,-\,\frac{\mathbf{m}^t \,\left(\calC^{\rm TT}\right)^{-1} \, \mathbf{m}}{2}\right)} \ . \label{like}
\eeq
Here $\calC^{\rm TT} = \calS^{\rm TT}+\calN^{\rm TT}$ is the total temperature covariance matrix, including a signal matrix $\cal S^{\rm TT}$, with elements
\beq
\calS_{ij}^{\rm TT} \ = \ \sum_{\ell = 2}^{4 N_\mathrm{side}} \ \frac{2\ell+1}{4\pi}\,C_\ell^{\rm TT}\,P_\ell (\hat n_i \cdot \hat n_j) \ , \label{sigcov}
\eeq
where $\hat{n}_i$ and $\hat{n}_j$ identify the directions pointing to pixels $i$ and $j$ and the $C_\ell$ for $\ell>29$ are fixed to fiducial values. Given that signal and noise
in the original map are both Gaussian distributed, the even- and odd-parity maps are still Gaussian, and the corresponding likelihood functions are
\beq
\mathcal{L}(\mathbf{m}_\pm|C_\ell^{\rm TT}) \ \propto \ \frac{1}{|\calC_\pm^{\, \rm TT}|^{1/2}}\ \exp{\left(\,-\, \frac{\mathbf{m}_\pm^t\, \left(\calC_\pm^{\rm TT}\right)^{-1} \mathbf{m}_\pm}{2}\right)} \ , \label{like_evodd}
\eeq
where $\calC_\pm^{\,\rm TT} \equiv \calS_\pm^{\,\rm TT} + \calN_\pm^{\,\rm TT}$. Here
\bea
\calS_{\pm,\; ij}^{\,\rm TT} & = & \sum_{\ell = 2}^{4 N_\mathrm{side}} \ \frac{2\ell+1}{4\pi}\,C_\ell^{\rm TT}\, P_\ell (\hat n_i \cdot \hat n_j) \ \frac{1\pm(-1)^\ell}{2} \ , \label{S_evodd}\\[0.3cm]
\calN_{\pm,\; ij}^{\,\rm TT} &= & \frac{1}{2}\, \sigma^2 \left[ \delta_{ij} \pm \delta_{i p_j}\right] \ + \ {\widetilde \sigma}^2 \delta_{ij}\; \label{N_evodd},
\eea
where the first result follows from the behavior under parity of the Legendre polynomials, and where $p_j$ identifies the index labeling the direction opposite to pixel $j$. The reader should notice the addition of a (small) diagonal second noise term:~\footnote{We chose ${\widetilde \sigma}^2=0.01\,\mu K^2$.} it regularizes the covariance matrix, which would be singular otherwise, since for technical reasons we found it convenient to work with the full datasets even in the presence of (anti)symmetrizations. Note also how eq.~\eqref{S_evodd} implies that the temperature likelihood of the parity even (odd) map only depends on the even (odd) theoretical $C_\ell^{\rm TT}$'s.

As explained in~\cite{tegmark}, taking into account the polarization brings along the other non--vanishing correlators $\langle T \, Q \rangle$, $\langle U \, U \rangle$ and $\langle Q \, Q \rangle$.  {This entails some complications, but it is important to explain how the even--odd separation works in this case, also in view of the forecasts of Section~\ref{sec:forecasts}.} The corresponding (anti)symmetrizations lead to
\bea
\langle T^\pm\left( \hat n\right)\,Q^\pm\left( \hat n^\prime\right)\rangle &=& - \ \sum_\ell \frac{(2\ell+1)}{4\,\pi} \ \frac{1\pm(-1)^\ell}{2} \ F_\ell^{10}(z)\ C_\ell^{{\rm TE}} \ , \nonumber \\
\langle Q^\pm\left( \hat n\right)\,Q^\pm\left( \hat n^\prime\right)\rangle &=& \sum_\ell \frac{(2\ell+1)}{4\,\pi}  \biggl[\frac{1\pm(-1)^\ell}{2} F_\ell^{12}(z)\ C_\ell^{{\rm EE}} \nonumber \\ &-& \frac{1\mp(-1)^\ell}{2} F_\ell^{22}(z)\ C_\ell^{{\rm BB}} \biggr] \ , \nonumber \\
\langle U^\pm\left( \hat n\right)\,U^\pm\left( \hat n^\prime\right)\rangle &=& \sum_\ell \frac{(2\ell+1)}{4\,\pi}  \biggl[\frac{1\mp(-1)^\ell}{2} F_\ell^{12}(z)\ C_\ell^{{\rm BB}} \nonumber \\ &-& \frac{1\pm(-1)^\ell}{2} F_\ell^{22}(z)\ C_\ell^{{\rm EE}} \biggr]\ , \label{pol_corr}
\eea
which enter the complete signal covariance matrix, and where
\bea
U^\pm\left( \hat n \right) &=& \frac{1}{2} \left[ U\left( \hat n \right) \ \mp U\left( \,-\, \hat n \right)\right]\ ,  \nonumber \\
Q^\pm\left( \hat n \right) &=& \frac{1}{2} \left[ Q\left( \hat n \right) \ \pm Q\left( \,-\, \hat n \right)\right]\ ,
\eea
together with corresponding combinations for the noise.
These results follow from the parity relations
\beq
Q\left( \hat n\right) \ + \ i\, U\left( \hat n\right) \ \longrightarrow \ Q\left( \, - \, \hat n\right) \ - \ i\, U\left( \,-\,\hat n\right)
\eeq
and from the symmetry properties
\bea
F_\ell^{10}\left(- z \right) &=& \left(-1\right)^\ell \ F_\ell^{10}\left(z \right)\ , \nonumber \\
F_\ell^{12}\left(- z \right) &=& \left(-1\right)^\ell \ F_\ell^{12}\left(z \right) \ , \nonumber \\
F_\ell^{22}\left(- z \right) &=& \left(-1\right)^{\ell+1} \ F_\ell^{22}\left(z \right) \ ,
\eea
with $z = \hat n \cdot \hat n^\prime$, which are listed in the Appendix of~\cite{tegmark}. Notice that the correlation functions in eq.~\eqref{pol_corr} are adapted to frames aligned to the maximal circles joining the points identified by $\hat n$ and $\hat n^\prime$. The actual results in standard frames with one axis aligned with the Galactic plane can then be obtained making use of the rotation matrices in~\cite{tegmark}.

We derived parameter estimates using the Monte Carlo engine \texttt{CosmoMC}  \cite{Lewis:2002ah}, interfaced with a modified version of
the Boltzmann code \texttt{camb}  \cite{Lewis:1999bs}, and with a pixel-based low-$\ell$ ($\ell < 30$) likelihood code implementing eqs.~\eqref{like} and \eqref{like_evodd} or, more precisely, their versions extended to include polarization.
We also included information from the high-$\ell$ ($\ell \ge 30$) temperature power spectrum and from the lensing potential from {\sc Planck}, processed through the publicly available
likelihood released by the {\sc Planck} collaboration \cite{Aghanim:2015xee}.
Notice that, since the preceding analysis of \cite{Gruppuso:2015zia,Gruppuso:2015xqa} resulted in the determination of $\Delta$ in eq.~\eqref{delta}, which only impacts very large scales, {we refrain from} (anti)symmetrizing the high--$\ell$ likelihood in this analysis.

We also performed a complementary frequentist analysis, in order to highlight the tension between data and the predictions of $\Lambda$CDM. To this end, we simulated 10000 CMB-plus-noise maps, extracting their signal contributions from the {\sc Planck} fiducial $\Lambda$CDM angular power spectrum. We analyzed these maps and the observed \texttt{Commander} map with an optimal angular power spectrum estimator, {\it BolPol}~\cite{Gruppuso:2009ab}, for the whole sequence of masks. With the resulting spectra, we built the estimators of eqs.~(\ref{s1su2estimator}) and (\ref{variance}), and the joint probability $P[C^+(\ell_{max}),C^-(\ell_{max})]$, where the $C^\pm$ are defined in eq.~(\ref{C+-}), automatically accounting for cosmic variance and for the increase of sampling variance that accompanies reduced sky fractions.

\section{Results}
\label{results}

We can now turn to a description of our results. We first present the constraints on the cutoff scale $\Delta$ and the standard $\Lambda$CDM parameters and then turn to the three estimators of Section~\ref{dataandsims}. We conclude our discussion with a frequentist assessment of the tension between low--$\ell$ data and $\Lambda$CDM, paying special attention to disentangling the contributions of even and odd multipoles, and with some forecasts for future experiments aimed at high--quality polarization data.

\subsection{Constraints on $\Delta$} \label{sec:constraints}

In this section we provide the posterior distributions for the scale $\Delta$ of eq.~(\ref{cutoff}) and the six $\Lambda$CDM parameters, namely
the angle $\theta_s$  subtended by the sound horizon at recombination, the baryon density $\Omega_b h^2$, the cold dark matter density $\Omega_c h^2$,
the logarithmic amplitude $\log(10^{10} A_s)$, the spectral index $n_s$ and the re-ionization optical depth $\tau$.

{As explained in Section~\ref{dataandsims}, for all masks listed in table \ref{tab:masks}, we derived our constraints also considering the low-$\ell$ likelihood ($\ell \leq 29$) based on the full (even plus odd) spectrum. }In detail, we resorted to the extensions of eqs.~\eqref{like} and \eqref{sigcov} that include polarization and to their counterparts corresponding to eqs.~\eqref{like_evodd}, \eqref{S_evodd}  and \eqref{N_evodd}, which only involve even or odd multipoles.
Our results are collected in table \ref{tab:delta}, where we denote with $\Delta$ the estimates derived from the ``even + odd'' dataset and with $\Delta_{+}$ ($\Delta_{-}$) those derived from its ``even'' (``odd'') subsets. We shall stick to this notation also in the following.

\begin{figure}[ht]
\centering
\includegraphics[width=100mm]{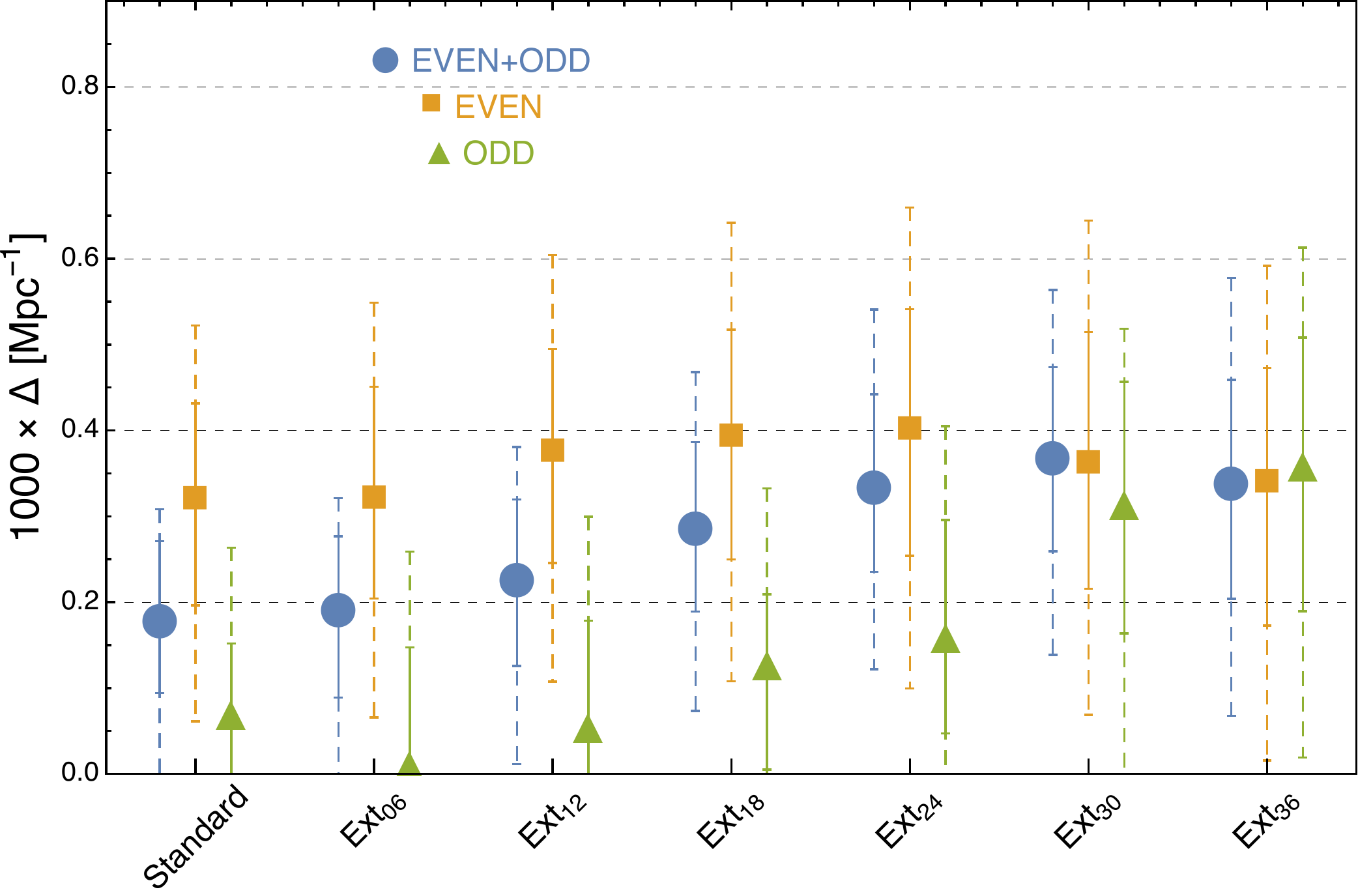}
\caption{\small Constraints on $\Delta$ for the standard and extended masks. Estimates and error bars are shown in blue (``even + odd'') when all the low-$\ell$ temperature information is taken into account and in orange (green) when only the even (odd) multipoles are considered. In all cases, it should be understood that the low-$\ell$ information is complemented by the high-$\ell$ {\sc Planck} TT likelihood. Solid (dashed) error bars stand for
68\% (95\%) Bayesian credible intervals. Notice the slight change of pattern for the ``odd'' determinations around Ext$_{06}$, which will reflect itself on the behavior of the even--odd asymmetry in fig.~\ref{fig:whiskers_D+minusD-}.}
\label{fig:whiskersevenodddelta}
\end{figure}
\begin{figure}[ht]
\centering
\includegraphics[width=100mm]{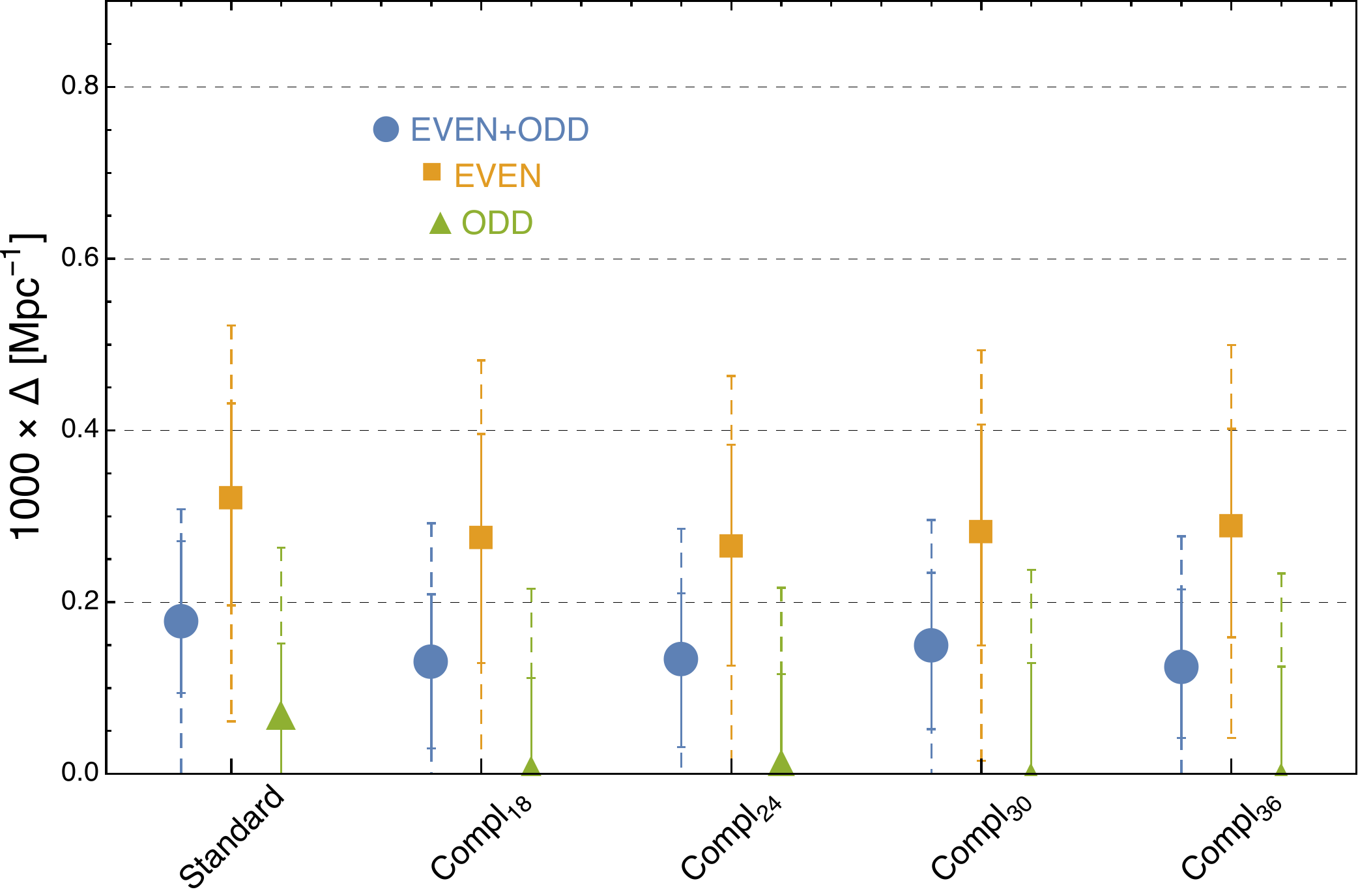}
\caption{\small Same as fig.~\ref{fig:whiskersevenodddelta}, for the standard and complementary masks.}
\label{fig:whiskersevenodddelta_comp}
\end{figure}

\begin{table}
\centering
\caption{\small Constraints on the scale $\Delta$ for the various masks in table~\ref{tab:masks} that were used in our analysis. $\Delta$, $\Delta_+$ and $\Delta_-$ refer to estimates from the ``even+odd'', ``even'', and ``odd'' datasets.  When the estimate of $\Delta$ differs from zero at more than 95\% confidence level, we report it in the form of mean~$\pm$ 68\% uncertainty; otherwise, we only report the 95\% upper limit.}
\vskip 5pt
\label{tab:delta}
\begin{tabular}{cccccc}
\hline
Case & Label & $10^3 \Delta\,[\mathrm{Mpc^{-1}}]$ & $10^3 \Delta_+ \,[\mathrm{Mpc^{-1}}]$ & $10^3 \Delta_- \,[\mathrm{Mpc^{-1}}]$ \\
\hline
$a$ & Standard            & $< 0.31$ & $0.31^{+0.13}_{-0.11}$ & $< 0.26$ \\
$b$ & ${\rm Ext}_{06}$    & $<0.32$ & $0.32^{+0.13}_{-0.12}$ & $< 0.26$\\
$c$ & ${\rm Ext}_{12}$    & $0.22^{+0.10}_{-0.09}$ & $0.36^{+0.13}_{-0.12}$ & $<0.30$ \\
$d$ & ${\rm Ext}_{18}$    & $0.28^{+0.11}_{-0.09}$ & $0.38 \pm 0.13$ & $< 0.33$\\
$e$ & ${\rm Ext}_{24}$    & $0.33 \pm 0.10$ & $0.39^{+0.15}_{-0.14}$  & $< 0.41$ \\
$f$ & ${\rm Ext}_{30}$    & $0.36^{+0.11}_{-0.10}$ & $0.36 ^{+0.16}_{-0.14}$ & $< 0.52$\\
$g$ & ${\rm Ext}_{36}$    & $0.33 \pm 0.13$ & $0.32 \pm 0.15$ & $ 0.34^{+0.17}_{-0.15}$\\
$h$ & ${\rm Compl}_{18}$  & $< 0.29$ & $< 0.48$ & $< 0.22$\\
$i$ & ${\rm Compl}_{24}$  & $<0.29$ & $<0.46$ & $<0.22$\\
$j$ & ${\rm Compl}_{30}$  & $< 0.30$ & $0.27^{+0.13}_{-0.12}$ & $< 0.24$\\
$k$ & ${\rm Compl}_{36}$  & $< 0.28$ & $0.28^{+0.13}_{-0.12}$ & $< 0.23$\\
\hline
\end{tabular}
\end{table}

Fig.~\ref{fig:whiskersevenodddelta} collects the estimates for the parameter $\Delta$ obtained considering the standard and extended masks of cases $a$--$g$ in table~\ref{tab:masks}. We use blue circles when we employ $\mathbf m$ (``even + odd'' multipoles), orange squares when we employ $\mathbf m_{+}$ (``even'' multipoles), and finally green triangles when we employ $\mathbf m_{-}$ (``odd'' multipoles).

Notice that:
\begin{itemize}
\item $\Delta$ shifts toward higher values at higher Galactic latitudes,
attaining a maximum statistical significance in the ${\rm Ext}_{30}$ mask, as already found in our previous work \cite{Gruppuso:2015xqa};
\item the mean value of $\Delta_+$ is always larger than the mean value of $\Delta_-$, with only one exception, in the Ext$_{36}$ mask. Moreover, $\Delta_+$ is statistically larger than zero at more than $ 2 \, \sigma$ in all masks,
while $\Delta_-$ is always statistically compatible with zero within $ 2 \, \sigma$, with one exception in the Ext$_{36}$ mask. All in all, the even multipoles drive the detection of $\Delta$ in \cite{Gruppuso:2015xqa} in a way that is largely independent of the mask;
\item more in detail, $\Delta_-$ is statistically compatible with zero within $ 1 \, \sigma$ in all masks from $a$ to $d$, and then increases monotonically. On the other hand,  $\Delta_{+}$ increases slightly up to the $e$ mask, to then decrease slightly in the $f$ and $g$ masks.
\end{itemize}
\begin{figure}[ht]
\centering
\includegraphics[width=100mm]{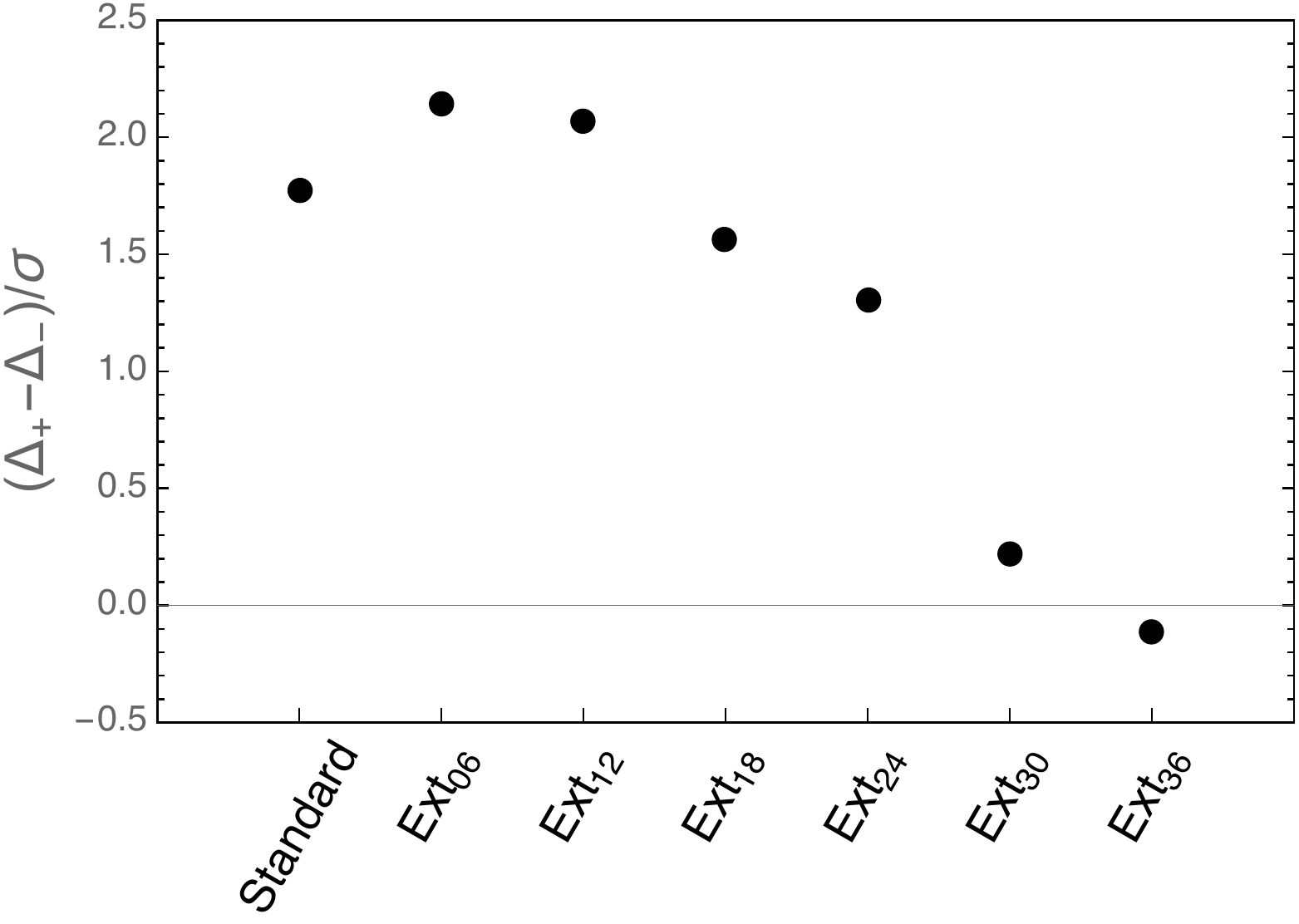}
\caption{\small Even-odd asymmetry traced by the cut-off scale $\Delta$, according to eq.~(\ref{evenoddasymmetry_Delta}).}
\label{fig:whiskers_D+minusD-}
\end{figure}

Fig.~\ref{fig:whiskersevenodddelta_comp} compares the estimates for $\Delta$ obtained in the complementary masks $h$--$k$ of table~\ref{tab:masks}.
The reader should appreciate the striking stability of these results, which conforms to the pattern that already emerged:
\emph{even multipoles contain typically less power than odd ones}.

$\Delta_{+}$ and $\Delta_{-}$ can be used to trace the even-odd asymmetry, which we characterize via
\beq
\frac{\Delta_+ - \Delta_-}{\sqrt{\sigma_+^2 + \sigma_-^2}} \, ,
\label{evenoddasymmetry_Delta}
\eeq
with $\sigma_{+}$ ($\sigma_{-}$) the standard deviation of $\Delta_{+}$ ($\Delta_{-}$).
This quantity, shown in fig.~\ref{fig:whiskers_D+minusD-}, is close to the $2 \, \sigma$ level up to the Ext$_{24}$ mask, and then decreases below the $1 \, \sigma$ level in the Ext$_{30}$ and Ext$_{36}$ masks.

\begin{figure}[ht]
\centering
\includegraphics[width=125mm]{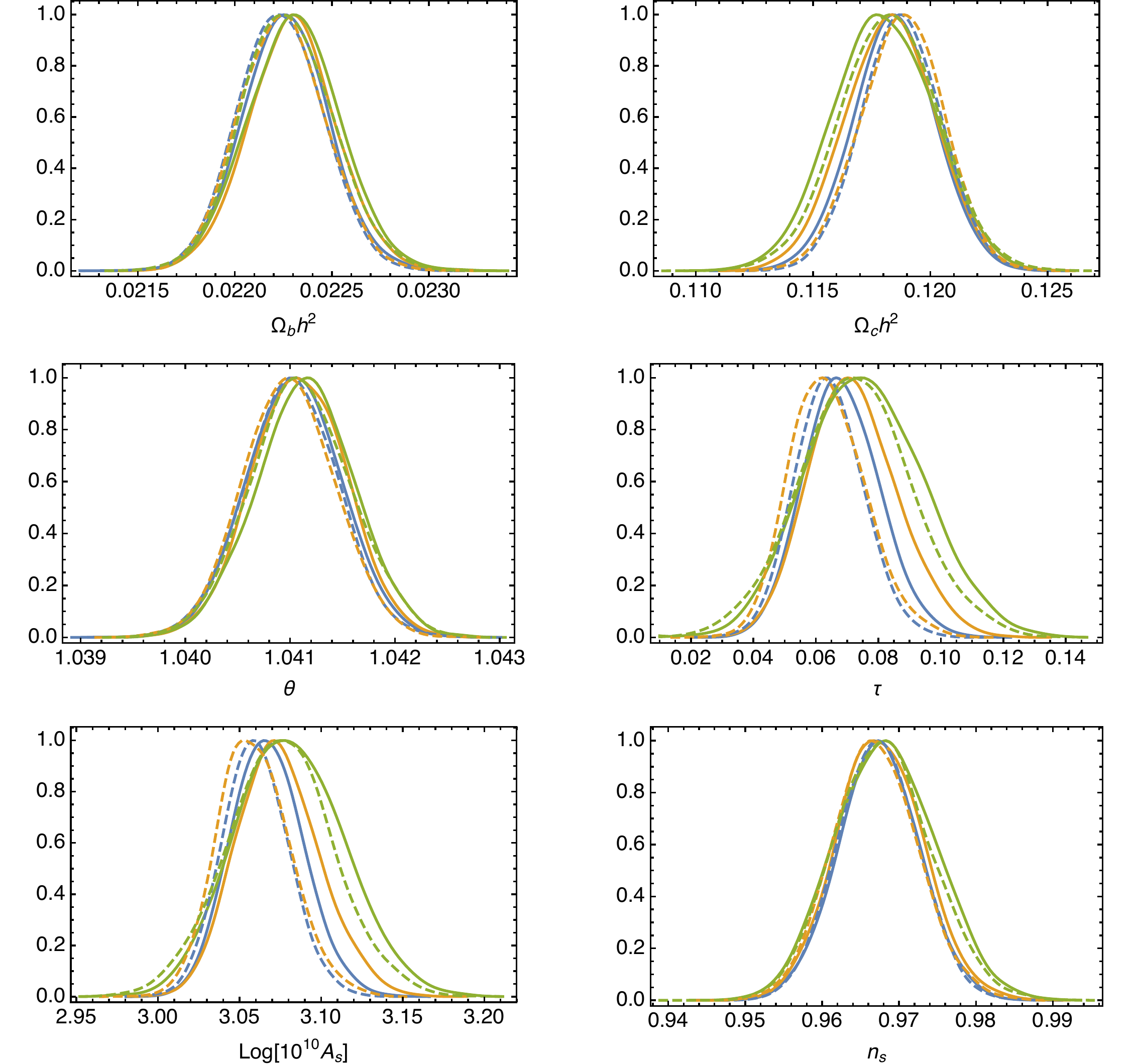}
\caption{\small Posterior distribution functions for the $\Lambda$CDM parameters considering the standard mask.
As before, blue, orange and green curves refer to even+odd, even and odd contribution for the low-$\ell$ likelihood.
Solid curves refer to the $\Lambda {\rm CDM}\Delta$ model, while dashed ones refer to the $\Lambda$CDM model.}
\label{fig:lcdmstandard}
\end{figure}

\begin{figure}[ht]
\centering
\includegraphics[width=125mm]{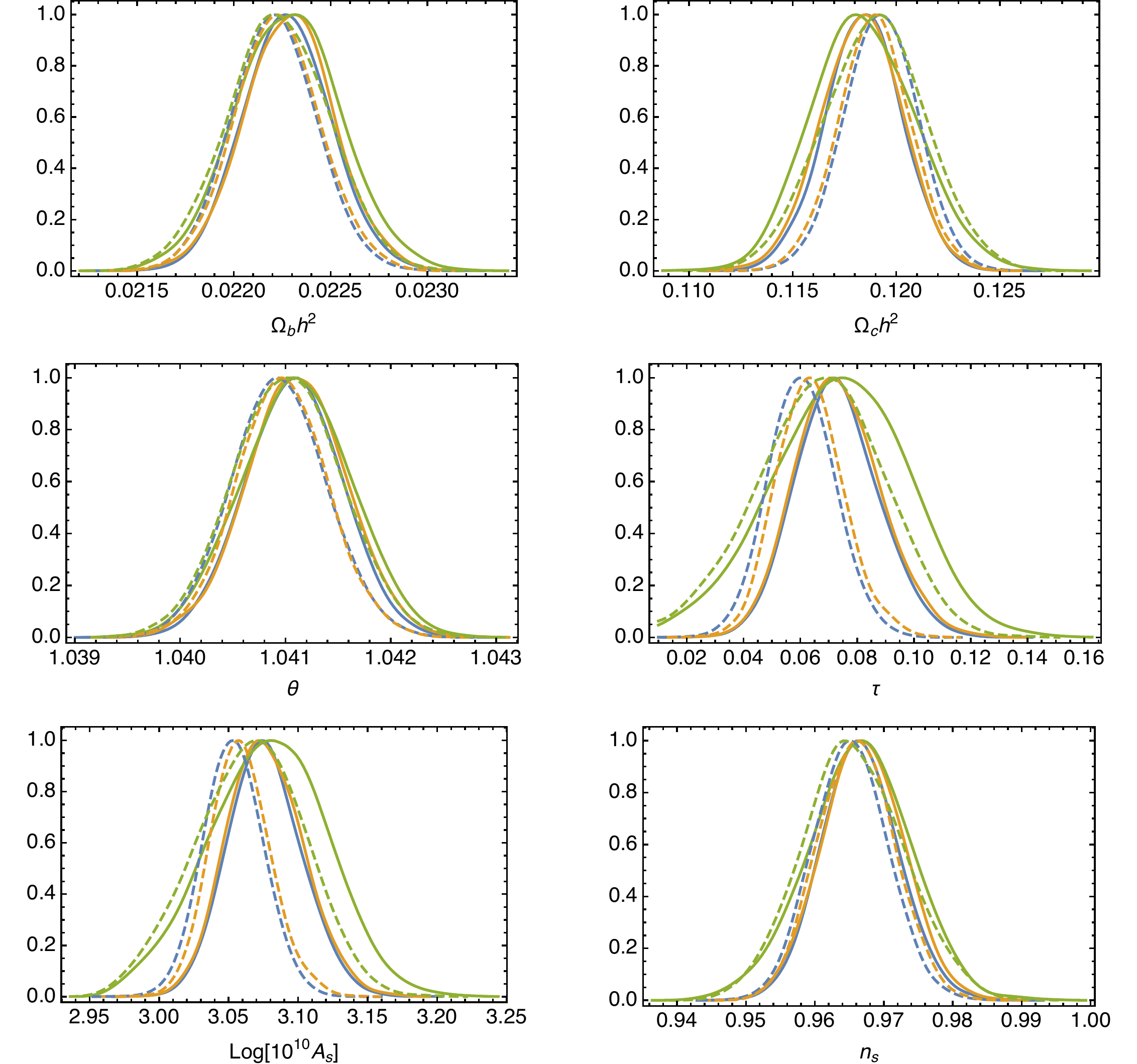}
\caption{\small Posterior distribution functions for the $\Lambda$CDM parameters considering the ${\rm Ext}_{30}$ mask.
Same conventions as in fig.~\ref{fig:lcdmstandard}.}
\label{fig:lcdmext30}
\end{figure}

Figs.~\ref{fig:lcdmstandard}
%, \ref{fig:lcdmext18}
and \ref{fig:lcdmext30} illustrate the behaviour of the six $\Lambda$CDM parameters for three
significant mask choices ($a$
%, $d$
and $f$ of table~\ref{tab:masks}).
Each figure contains 6 posteriors for each parameter with the usual color code: blue, orange and green refer to even+odd, even and odd contributions for the low-$\ell$ likelihood.
Moreover, dashed lines refer to $\Lambda {\rm CDM}$, while solid ones refer to $\Lambda {\rm CDM}\Delta$.
In general, the introduction of $\Delta$, which captures the lack of power, tends to improve the stability of the other parameters.
The continuous lines, which reflect ``even+odd'', ``even'' and ``odd'' contributions, are indeed largely superposed in all cases.
On the other hand, the green dashed lines, which reflect odd multipoles in $\Lambda$CDM, are very close to
the solid ones in figs.~\ref{fig:lcdmstandard}
%and \ref{fig:lcdmext18},
and move only slightly in fig.~\ref{fig:lcdmext30}.
%This is simple to understand: the odd multipoles for the standard and Ext$_{18}$ masks result in a determination of $\Delta$ that is essentially compatible with zero.
Finally, $\Delta$ brings along slight shifts of $n_s$ and $A_s$, consistently with
its role in accounting for the low-$\ell$ lack of power.

%
%
%\begin{figure}[ht]
%\centering
%\includegraphics[width=100mm]{plot/Evevodd_grid1D_ext18mask_paper.pdf}
%%
%\caption{\small Posterior distribution functions for the $\Lambda$CDM parameters considering the ${\rm Ext}_{18}$ mask.
%Same conventions as in fig.~\ref{fig:lcdmstandard}.}
%\label{fig:lcdmext18}
%\end{figure}
%
%

\begin{figure}[ht]
\centering
\includegraphics[width=125mm]{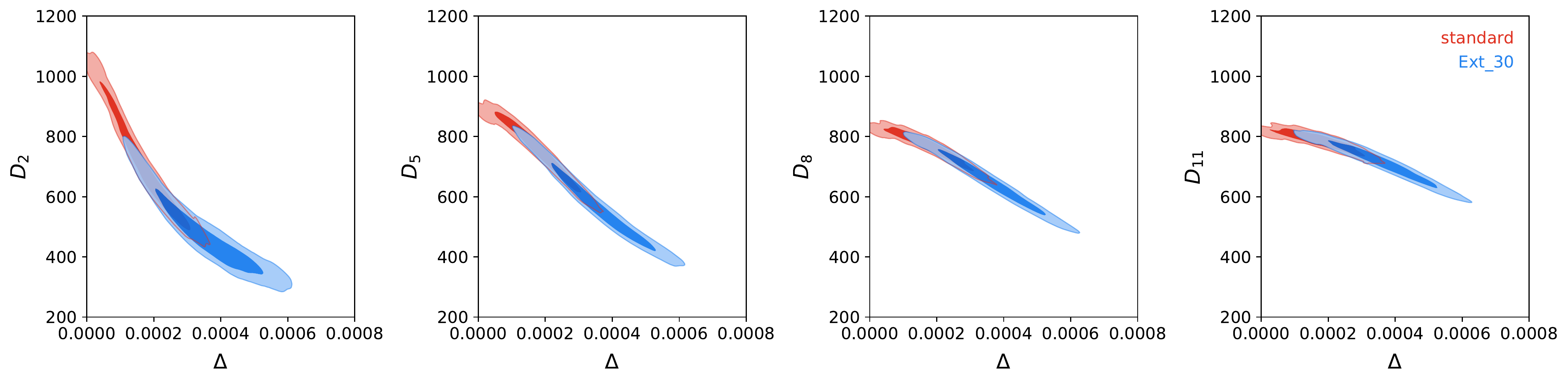}\vspace{0.2cm}
\includegraphics[width=94mm]{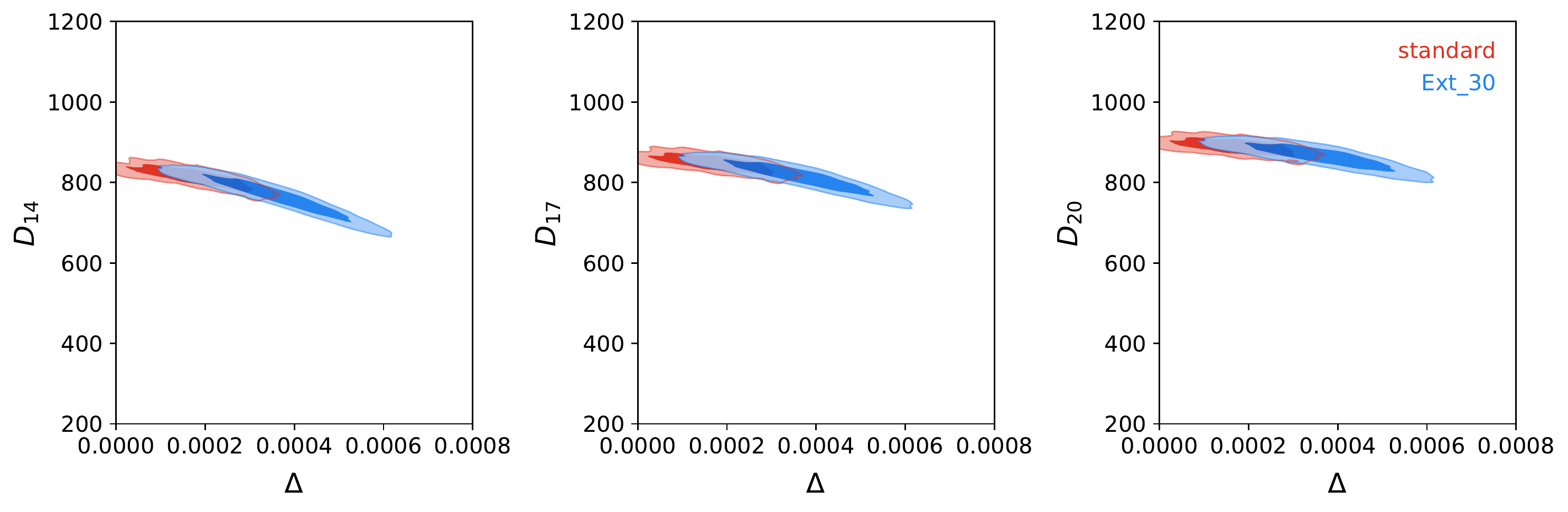}
\caption{\small Two-dimensional contour plots for $\Delta$ {\it vs} $D_{\ell}=\frac{\ell (\ell +1)}{2\, \pi}\, C_{\ell}$, for {$\ell=2,\,5,\,8,\,11,\,14,\,17$ and $20$}. The red contours are for the standard mask,
while the blue ones are for the ${\rm Ext}_{30}$ mask.}
\label{fig:contourDeltaCell}
\end{figure}

As shown in \cite{Gruppuso:2015zia,Gruppuso:2015xqa}, the introduction of the parameter $\Delta$ impacts only the lowest multipoles.
In order to highlight this feature, in fig.~\ref{fig:contourDeltaCell} we have displayed 2D contour plots of $\Delta$ {\it vs}
the band powers $D_{\ell} = \frac{\ell (\ell +1)}{2\, \pi}\, C_{\ell}$, for  { $\ell=2,\,5,\,8,\,11,\,14,\,17$ and $20$}.
There is clearly a high degree of anti-correlation between $\Delta$ and $D_{\ell}$ for the lowest multipoles, which fades out for higher values of $\ell$.
The separation between the two regimes occurs for $\ell \sim 14$ in the standard mask,
and for $\ell \sim 18$ in the ${\rm Ext}_{30}$ mask. Beyond these thresholds, the values of $D_{\ell}$ become very weakly sensitive to $\Delta$.
These results are nicely compatible with what found in fig.~5 of \cite{Gruppuso:2015xqa}.

\begin{figure}[ht]
\centering
\includegraphics[width=125mm]{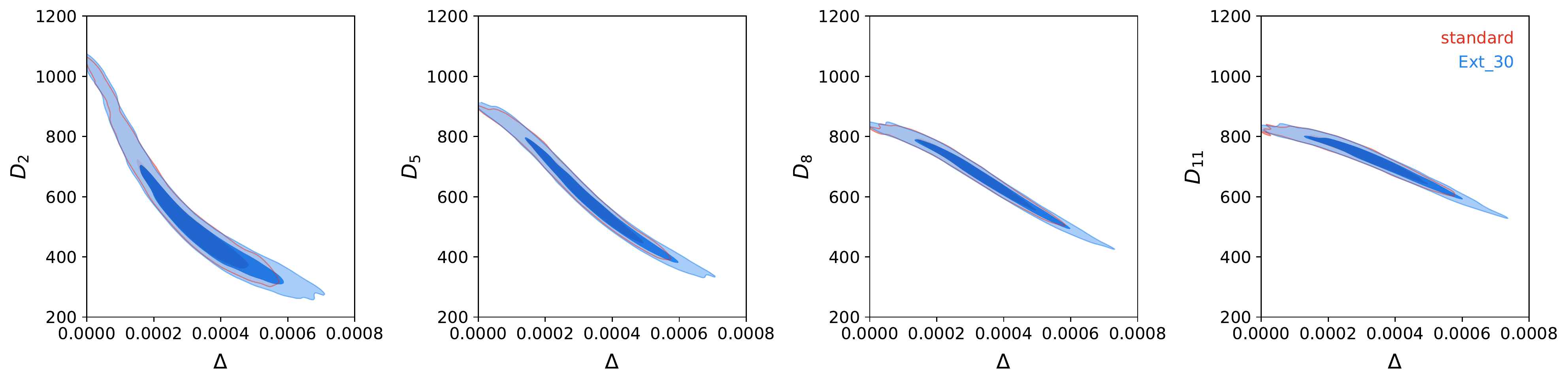}\vspace{0.2cm}
\includegraphics[width=94mm]{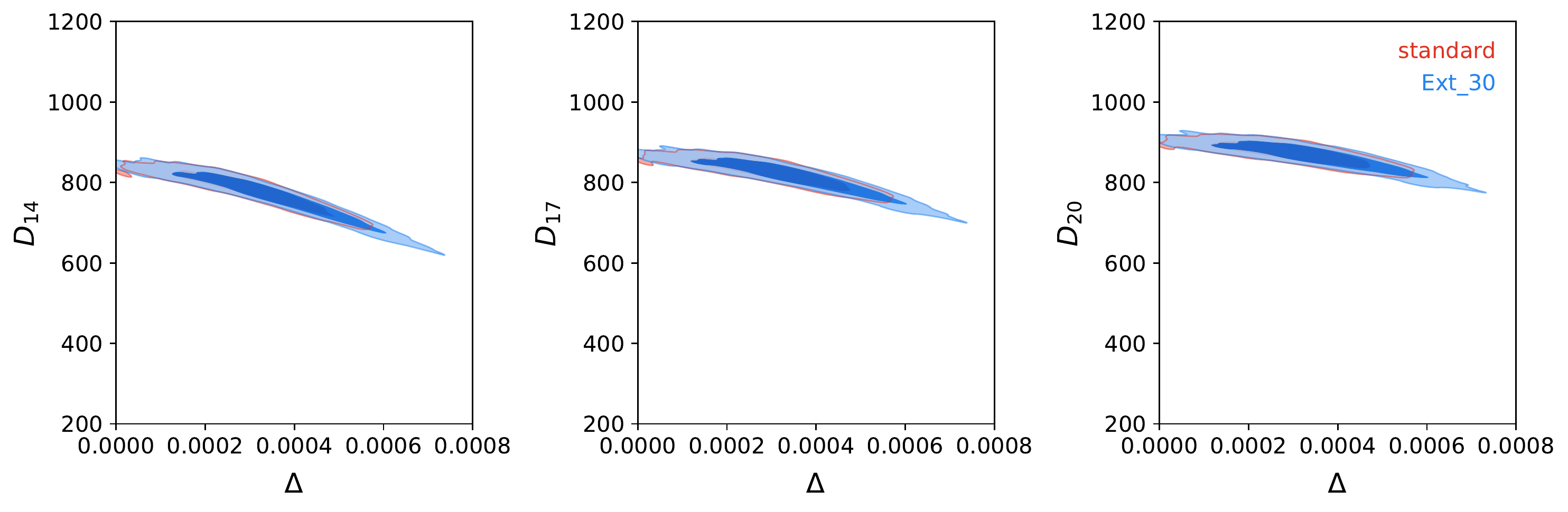}
\caption{\small Two-dimensional contour plots for $\Delta$ {\it vs} $D_{\ell}=\frac{\ell (\ell +1)}{2\, \pi}\, C_{\ell}$, for {$\ell=2,\,5,\,8,\,11,\,14,\,17$ and $20$}, obtained with the ``even'' part of the likelihood function.
The barely visible red contours, which are mostly covered by the larger blue ones, are for the standard mask, while the blue contours are for the ${\rm Ext}_{30}$ mask.}
\label{fig:contourDeltaCelleven}
\end{figure}

\begin{figure}[ht]
\centering
\includegraphics[width=125mm]{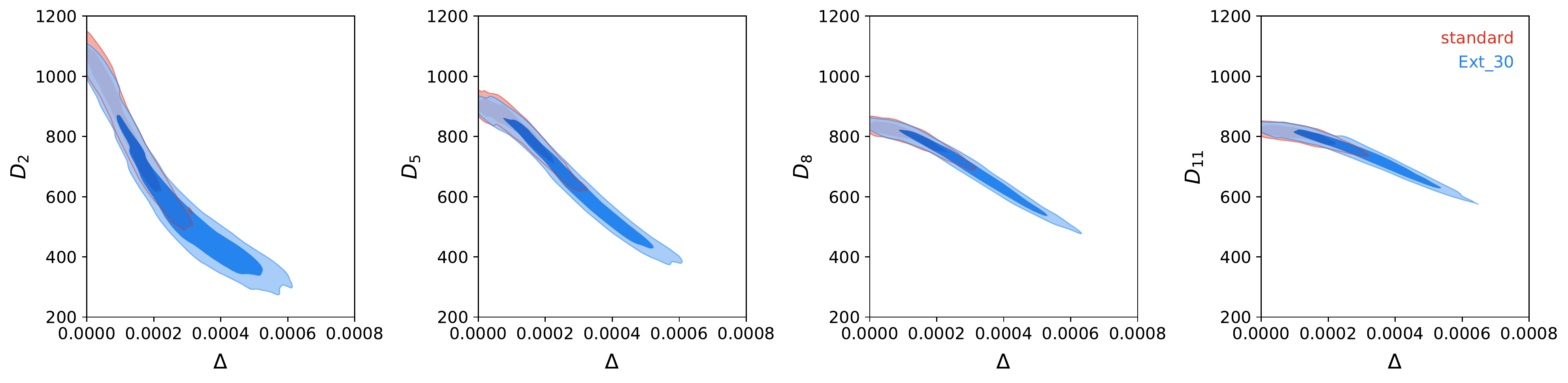}\vspace{0.2cm}
\includegraphics[width=94mm]{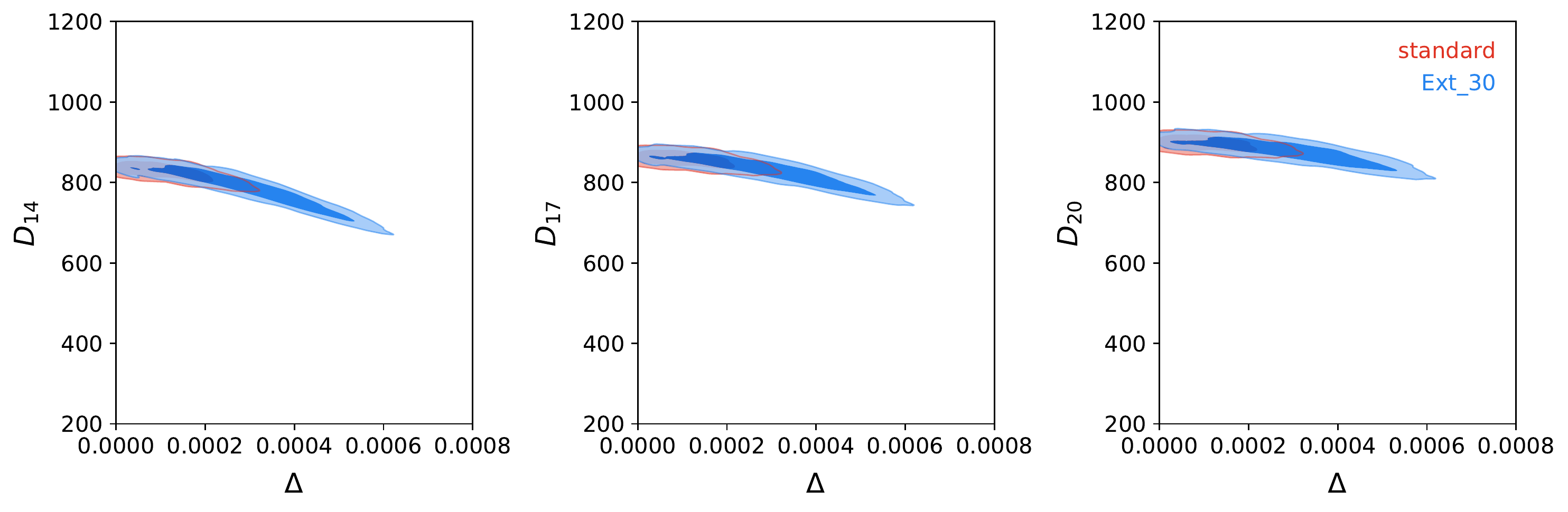}
\caption{\small Two-dimensional contour plots for $\Delta$ {\it vs} $D_{\ell}=\frac{\ell (\ell +1)}{2\, \pi}\, C_{\ell}$, for {$\ell=2,\,5,\,8,\,11,\,14,\,17$ and $20$}, obtained with the ``odd'' part of the likelihood function. The barely visible red contours, which are mostly covered by the larger blue ones, are for the standard mask, while the blue contours are for the ${\rm Ext}_{30}$ mask.}
\label{fig:contourDeltaCellodd}
\end{figure}

Similarly, figs.~\ref{fig:contourDeltaCelleven} and \ref{fig:contourDeltaCellodd} show the 2D contour plots of $\Delta$ {\it vs} $D_{\ell}$
obtained taking into account only the ``even'' or ``odd'' parts of the likelihood function.
Notice that the level of anti-correlation of these cases is similar to what already found in fig.~\ref{fig:contourDeltaCell}.
Moreover, consistently with our previous findings, one can see again that the ``even'' part pushes $\Delta$ towards larger values, making it more significantly different from zero.
On the other hand, the ``odd'' part tends to make $\Delta$ more compatible with zero, although the significance of its detection increases somewhat in wider masks.
\begin{figure}[ht]
\centering
\includegraphics[width=85mm]{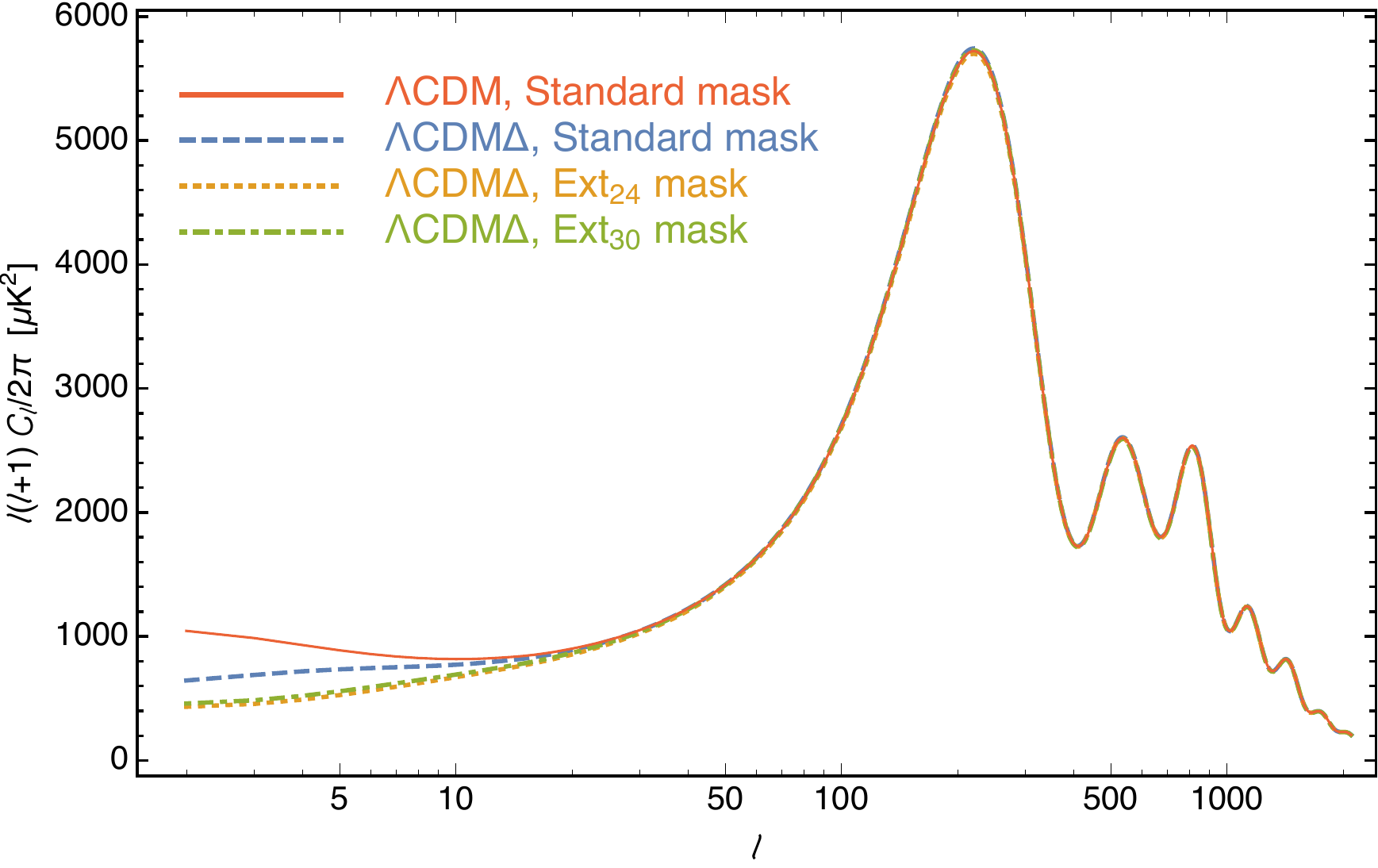}\\
\includegraphics[width=85mm]{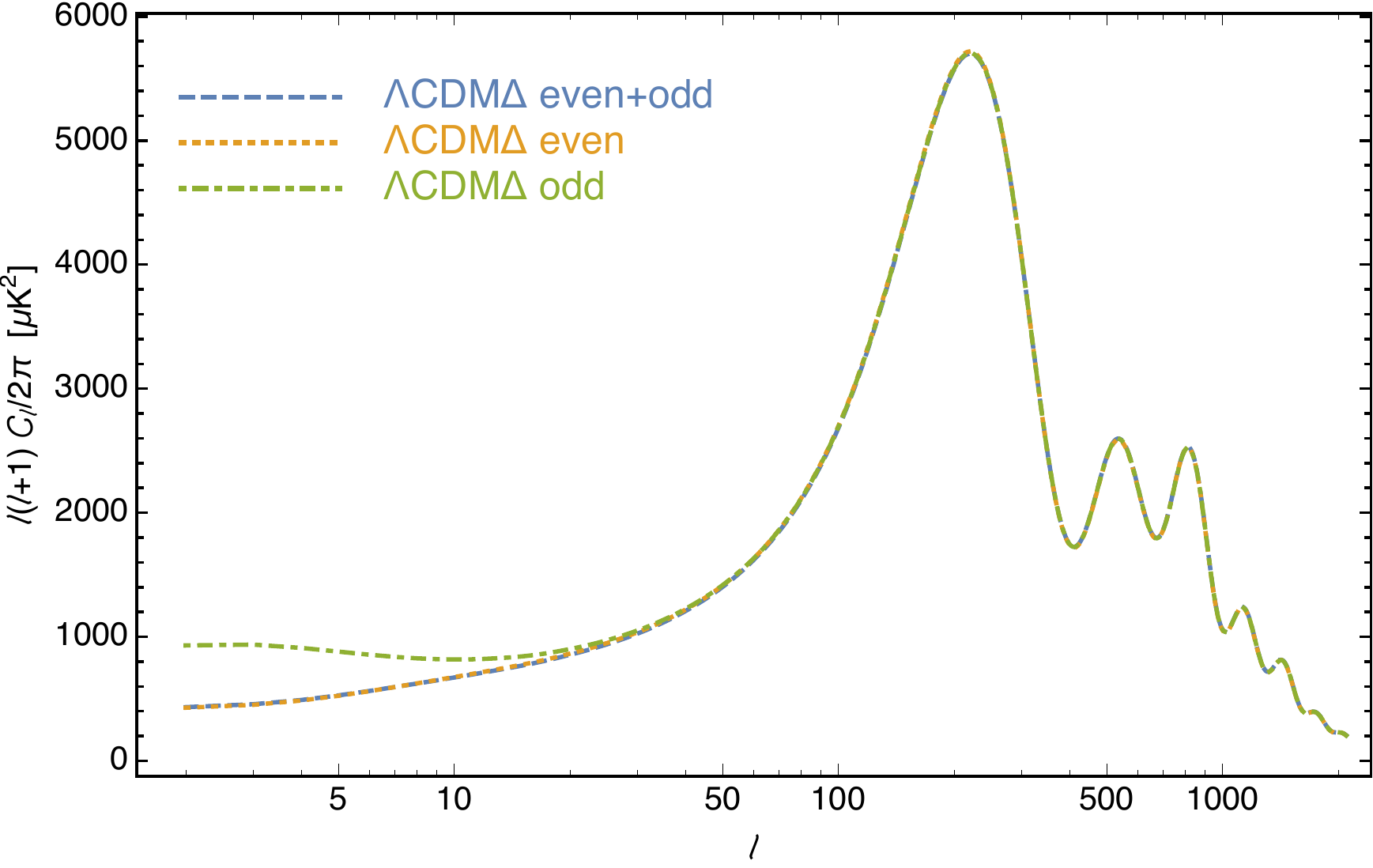}\\
\includegraphics[width=85mm]{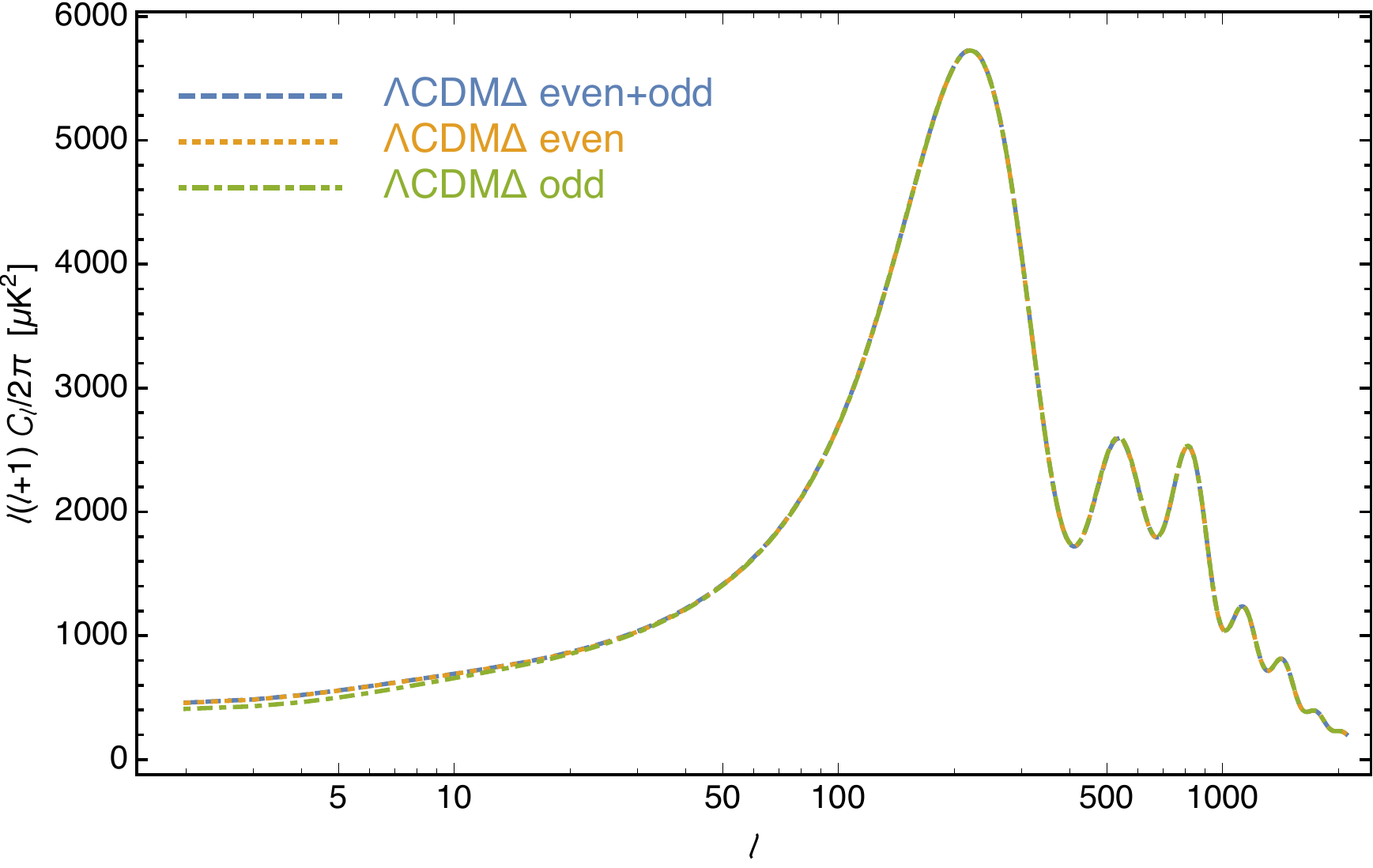}
\caption{\small Fiducial models obtained with even+odd $\ell <30$ datasets corresponding to the Standard mask, to the Ext$_{24}$ mask and to the Ext$_{30}$ mask (upper panel). Comparison of the fiducial models obtained with even+odd, even and odd $\ell <30$ datasets for the Ext$_{24}$ mask (middle panel), and for the Ext$_{30}$ mask (lower panel).}
\label{fig:fiducial}
\end{figure}

The upper panel in fig.~\ref{fig:fiducial} displays the fiducial models obtained with complete $\ell < 30$ datasets corresponding to the Standard mask, to the Ext$_{24}$ mask and to the Ext$_{30}$ mask. The middle panel compares the fiducial models obtained for the Ext$_{24}$ mask using even, odd and even+odd datasets for $\ell <30$. Finally, the lower panel compares the fiducial models obtained for the Ext$_{30}$ mask using even, odd and even+odd datasets for $\ell <30$.
\begin{table}
\centering
\caption{\small Akaike tests for the extended masks in table~\ref{tab:masks} and corresponding probabilities $P$ for the two models that we are considering ($\Lambda {\rm CDM}$ and $\Lambda {\rm CDM} \Delta$). Here $\delta {\rm AIC} ={\rm AIC}_{\Lambda {\rm CDM}} - {\rm AIC}_{\Lambda {\rm CDM}\Delta}$.}
\vskip 5pt
\label{tab:akaike1}
\begin{tabular}{cccrccc}
\hline
Case & Label & dataset & $\delta {\rm AIC}$ & $P(\Lambda {\rm CDM})$ & $P(\Lambda {\rm CDM}\Delta)$ \\
\hline
$a$ & Standard            & full & $ -0.680 $ & $ 0.584 $ & $ 0.416$ \\
$a$ & Standard            & even & $ 1.908 $ & $ 0.278$ & $ 0.722 $ \\
$a$ & Standard            & odd & $ -0.630 $ & $ 0.578 $ & $ 0.422 $ \\
$b$ & ${\rm Ext}_{06}$    & full & $ -1.172 $ & $ 0.642 $ & $ 0.358$ \\
$b$ & ${\rm Ext}_{06}$    & even & $ 3.848 $ & $ 0.127 $ & $ 0.873$ \\
$b$ & ${\rm Ext}_{06}$    & odd & $ -2.522 $ & $ 0.779 $ & $ 0.221$ \\
$c$ & ${\rm Ext}_{12}$    & full & $ 0.540 $ & $ 0.433 $ & $ 0.567$ \\
$c$ & ${\rm Ext}_{12}$    & even & $ 3.752 $ & $ 0.133 $ & $ 0.867$ \\
$c$ & ${\rm Ext}_{12}$    & odd & $ -2.802 $ & $ 0.802 $ & $ 0.198$ \\
$d$ & ${\rm Ext}_{18}$    & full & $ 3.020 $ & $ 0.181 $ & $ 0.819$ \\
$d$ & ${\rm Ext}_{18}$    & even & $ 4.289 $ & $ 0.105 $ & $ 0.895$ \\
$d$ & ${\rm Ext}_{18}$    & odd & $ -2.373 $ & $ 0.766 $ & $ 0.234$ \\
$e$ & ${\rm Ext}_{24}$    & full & $ 2.418 $ & $ 0.230 $ & $ 0.770$ \\
$e$ & ${\rm Ext}_{24}$    & even & $ 3.359 $ & $ 0.157 $ & $ 0.843$ \\
$e$ & ${\rm Ext}_{24}$    & odd & $ -2.199 $ & $ 0.750 $ & $ 0.250$ \\
$f$ & ${\rm Ext}_{30}$    & full & $ 4.750 $ & $ 0.085 $ & $ 0.915$ \\
$f$ & ${\rm Ext}_{30}$    & even & $ 1.205 $ & $ 0.354 $ & $ 0.646$ \\
$f$ & ${\rm Ext}_{30}$    & odd & $ -0.657 $ & $ 0.581 $ & $ 0.419$ \\
$g$ & ${\rm Ext}_{36}$    & full & $ 2.156 $ & $ 0.254 $ & $ 0.746$ \\
$g$ & ${\rm Ext}_{36}$    & even & $ 1.103 $ & $ 0.366 $ & $ 0.634$ \\
$g$ & ${\rm Ext}_{36}$    & odd & $ -0.286 $ & $ 0.536 $ & $  0.464$ \\
\hline
\end{tabular}
\end{table}

\begin{table}
\centering
\caption{\small Akaike tests for the complementary masks in table~\ref{tab:masks} and corresponding probabilities $P$ for the two models that we are considering ($\Lambda {\rm CDM}$ and $\Lambda {\rm CDM} \Delta$). Here $\delta {\rm AIC} ={\rm AIC}_{\Lambda {\rm CDM}} - {\rm AIC}_{\Lambda {\rm CDM}\Delta}$.}
\vskip 5pt
\label{tab:akaike2}
\begin{tabular}{cccrccc}
\hline
Case & Label & dataset & $\delta {\rm AIC}$ & $P(\Lambda CDM)$ & $P(\Lambda {\rm CDM}\Delta)$ \\
\hline
$h$ & ${\rm Compl}_{18}$  & full & $ -1.408 $ & $ 0.669 $ & $ 0.331$ \\
$h$ & ${\rm Compl}_{18}$  & even & $ 0.725 $  & $ 0.410$ & $ 0.590 $\\
$h$ & ${\rm Compl}_{18}$  & odd & $ -1.376 $ & $ 0.666 $ & $ 0.334$ \\
$i$ & ${\rm Compl}_{24}$  & full & $ -0.294 $ & $ 0.537 $ & $ 0.463$ \\
$i$ & ${\rm Compl}_{24}$  & even & $ -0.433 $ & $ 0.554 $ & $ 0.446$ \\
$i$ & ${\rm Compl}_{24}$  & odd & $ -3.203 $ & $ 0.832 $ & $ 0.168$ \\
$j$ & ${\rm Compl}_{30}$  & full & $ -0.290 $ & $ 0.536 $ & $ 0.464$ \\
$j$ & ${\rm Compl}_{30}$  & even & $ 0.540 $ & $ 0.433 $ & $ 0.567$ \\
$j$ & ${\rm Compl}_{30}$  & odd & $ -2.930 $ & $ 0.812 $ & $ 0.188$ \\
$k$ & ${\rm Compl}_{36}$  & full & $ -2.436 $ & $ 0.772 $ & $ 0.228$ \\
$k$ & ${\rm Compl}_{36}$  & even & $ 2.462 $ & $ 0.226 $ & $ 0.774$ \\
$k$ & ${\rm Compl}_{36}$  & odd & $ -2.497 $ & $ 0.777 $ & $ 0.223$ \\
\hline
\end{tabular}
\end{table}

Tables~\ref{tab:akaike1} and \ref{tab:akaike2} contain the results of the Akaike tests, obtained computing
\beq
{\rm  AIC} \ = \ 2\, k \ - \ 2 \, \log {\cal L}_{max}  \ ,
\eeq
where $k$ is the number of parameters used in the fit to obtain the maximum likelihood ${\cal L}_{max}$,
and the corresponding probabilities $P$ for the two models at stake, $\Lambda {\rm CDM}$ and $\Lambda {\rm CDM}\Delta$. Consistently with our other results here and in \cite{Gruppuso:2015zia,Gruppuso:2015xqa}, notice the preference, within $\Lambda {\rm CDM}\Delta$, for the ${\rm Ext}_{30}$ mask in the full case and for the ${\rm Ext}_{18}$ mask in the even case. When only odd multipoles are considered, $\Lambda {\rm CDM}$ is preferred in almost all cases.
\begin{table}
\centering
\caption{\small Detection levels of $\Delta$ for the extended masks of Table~\ref{tab:masks}. There are a few differences with respect to~\cite{Gruppuso:2015xqa} since here we rely, for the low-$\ell$ polarization, on the joint {\sc PlanckLFI}+WMAP dataset described in \cite{Lattanzi:2016dzq}. Some determinations improved slightly, along the lines of what we shall discuss in Section~\ref{sec:forecasts}, due to the larger sky fraction thus available in polarization.}
\vskip 5pt
\label{tab:detection_extended}
\begin{tabular}{ccccc}
\hline
Case & Label & dataset & Detection Level (\%) & Detection Level ($\sigma$)\\
\hline
\hline
$a$ & Standard            & full & $ 93.26$ & 1.83\\
$a$ & Standard            & even & $ 98.59$  & 2.46 \\
$a$ & Standard            & odd & $ 52.52$  & 0.72 \\
$b$ & ${\rm Ext}_{06}$    & full & $ 92.30$  & 1.77 \\
$b$ & ${\rm Ext}_{06}$    & even & $ 98.65$  & 2.47\\
$b$ & ${\rm Ext}_{06}$    & odd & $ 41.03$ & 0.54 \\
$c$ & ${\rm Ext}_{12}$    & full & $ 96.41$ & 2.10 \\
$c$ & ${\rm Ext}_{12}$    & even & $ 99.39$  & 2.74\\
$c$ & ${\rm Ext}_{12}$    & odd & $ 18.93$ & 0.24\\
$d$ & ${\rm Ext}_{18}$    & full & $ 99.15$ & 2.63\\
$d$ & ${\rm Ext}_{18}$    & even & $ 99.23$ & 2.67\\
$d$ & ${\rm Ext}_{18}$    & odd & $ 69.80$ & 1.03\\
$e$ & ${\rm Ext}_{24}$    & full & $ 99.32$ & 2.71\\
$e$ & ${\rm Ext}_{24}$    & even & $ 99.05$ & 2.59\\
$e$ & ${\rm Ext}_{24}$    & odd & $ 81.57$ & 1.33\\
$f$ & ${\rm Ext}_{30}$    & full & $ 99.84$ & 3.16\\
$f$ & ${\rm Ext}_{30}$    & even & $ 98.47$ & 2.43\\
$f$ & ${\rm Ext}_{30}$    & odd & $ 94.37$ & 1.91\\
$g$ & ${\rm Ext}_{36}$    & full & $ 98.60$ & 2.46\\
$g$ & ${\rm Ext}_{36}$    & even & $ 96.27$ & 2.08\\
$g$ & ${\rm Ext}_{36}$    & odd & $ 96.60$ & 2.12\\
\hline
\end{tabular}
\end{table}
\begin{table}
\centering
\caption{\small Detection levels of $\Delta$ for the complementary masks of Table~\ref{tab:masks}.}
\vskip 5pt
\label{tab:detection_complementary}
\begin{tabular}{ccccc}
\hline
Case & Label & dataset & Detection Level (\%) & Detection Level ($\sigma$) \\
\hline
\hline
$h$ & ${\rm Compl}_{18}$  & full & $ 80.29$ & 1.29 \\
$h$ & ${\rm Compl}_{18}$  & even & $ 94.70 $ & 1.94\\
$h$ & ${\rm Compl}_{18}$  & odd & $ 9.36$ & 0.12\\
$i$ & ${\rm Compl}_{24}$  & full & $ 78.54$ & 1.24\\
$i$ & ${\rm Compl}_{24}$  & even & $ 94.43$ & 1.91\\
$i$ & ${\rm Compl}_{24}$  & odd & $ 24.70$ & 0.32\\
$j$ & ${\rm Compl}_{30}$  & full & $ 86.96$ & 1.51\\
$j$ & ${\rm Compl}_{30}$  & even & $ 96.28$ & 2.08\\
$j$ & ${\rm Compl}_{30}$  & odd & $ 0.00$ & 0\\
$k$ & ${\rm Compl}_{36}$  & full & $ 82.46$ & 1.36\\
$k$ & ${\rm Compl}_{36}$  & even & $ 97.41$ & 2.23\\
$k$ & ${\rm Compl}_{36}$  & odd & $ 0.00$ & 0\\
\hline
\end{tabular}
\end{table}
\begin{figure}[ht]
\centering
\includegraphics[width=90mm]{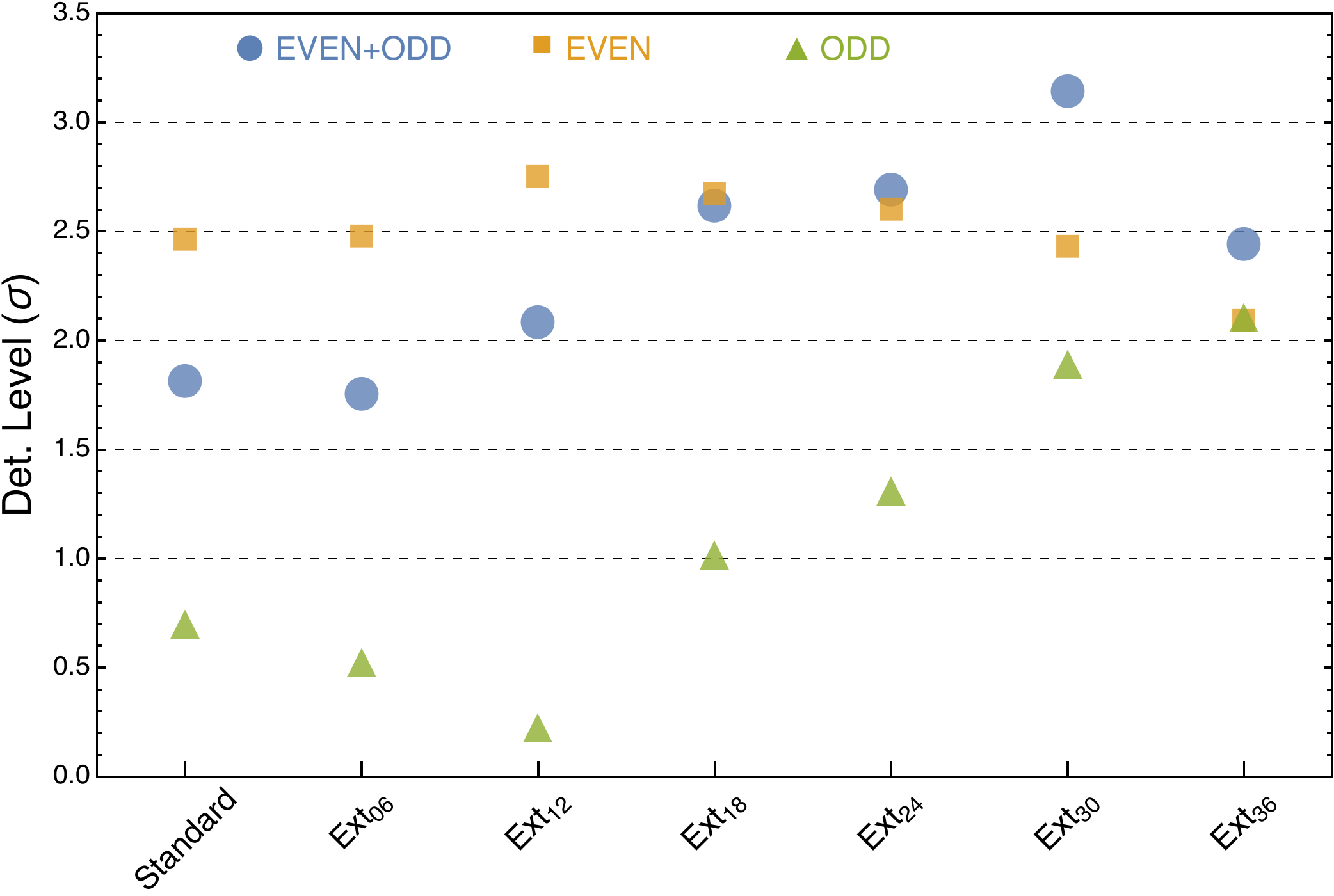}
\caption{\small Detection levels of $\Delta$ in units of $\sigma$ for the \emph{full} dataset (blue circles), \emph{even} dataset (orange squares) and \emph{odd} dataset (green triangles), for the various masks of Table.~\ref{tab:detection_extended}.}
\label{fig:detection_delta_sigma}
\end{figure}

Tables~\ref{tab:detection_extended} and \ref{tab:detection_complementary} collect the detection levels of $\Delta$ for all the extended and complementary masks that we considered, whose properties are summarized in table~\ref{tab:masks}. Fig.~\ref{fig:detection_delta_sigma} displays the results in the last column of table~\ref{tab:detection_extended}.
\begin{figure}[ht]
\centering
\begin{tabular}{ccc}
\includegraphics[width=38mm]{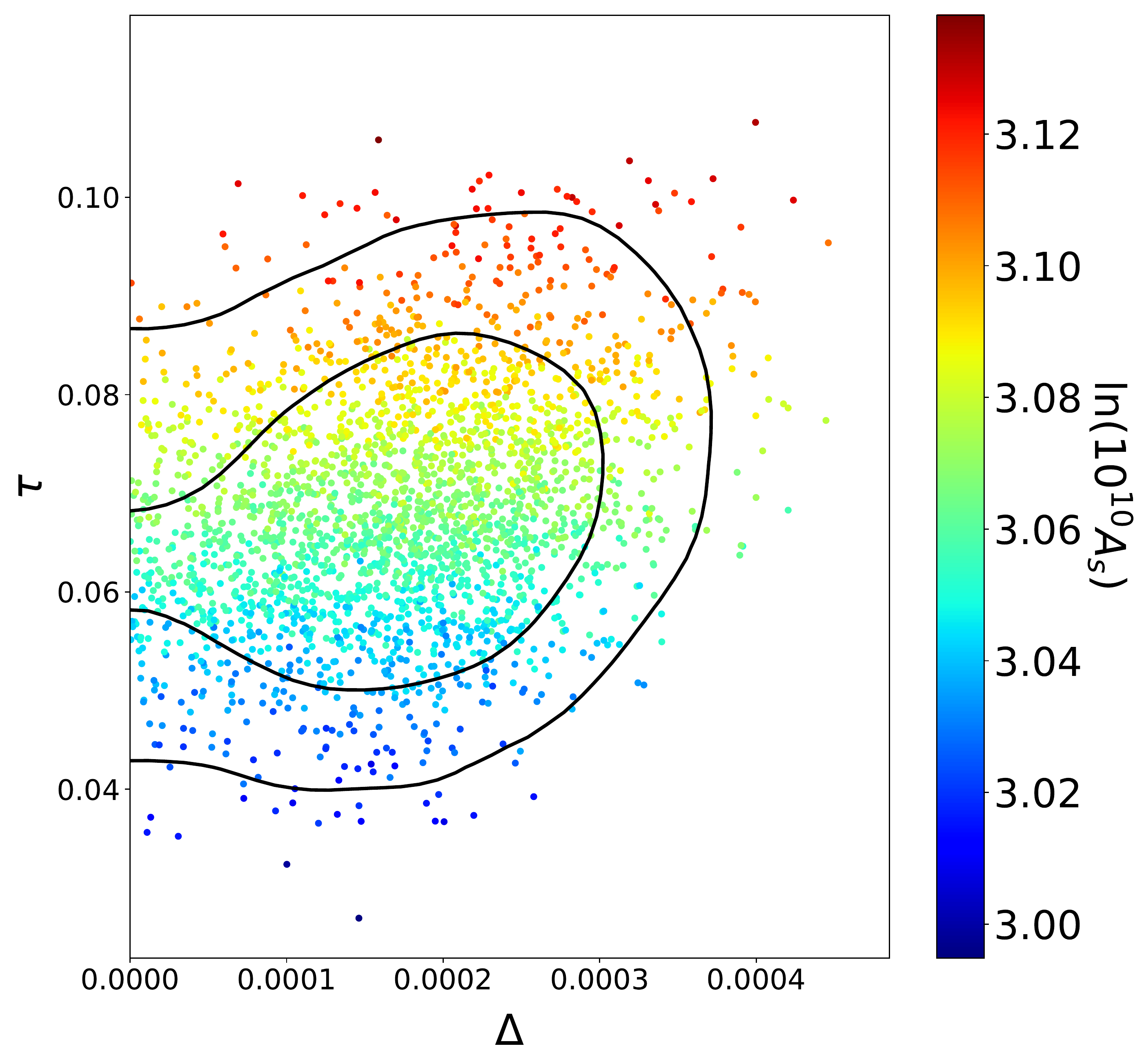} &
\includegraphics[width=38mm]{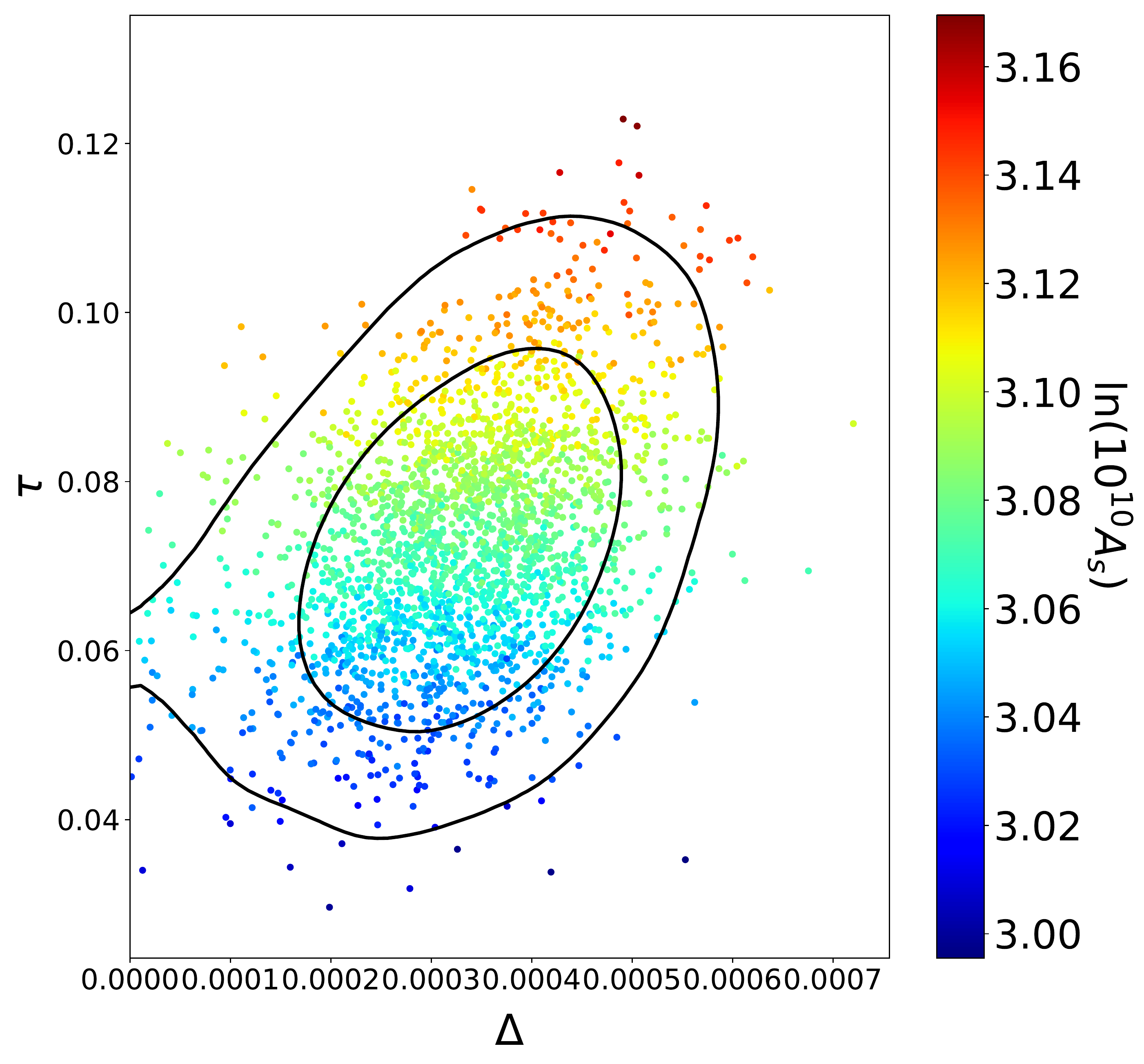} &
\includegraphics[width=38mm]{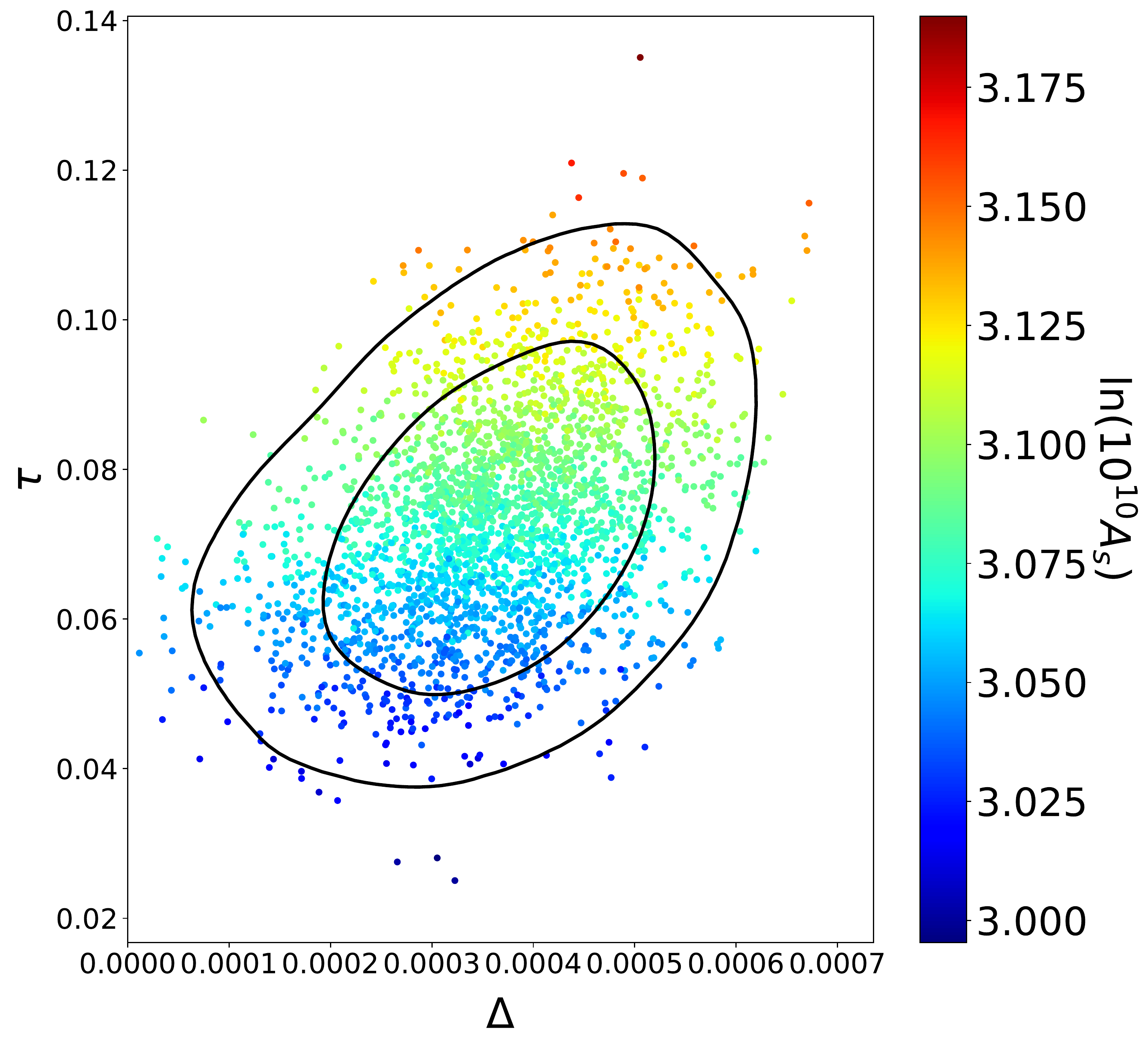}
\end{tabular}
\caption{\small Correlations between $\tau$, $\Delta$ and $A_s$ in the standard mask (left panel), in the ${\rm Ext}_{24}$ mask (middle panel) and in the ${\rm Ext}_{30}$ mask (right panel).}
\label{fig:correlation_effects}
\end{figure}

One may wonder about the relevance of accounting for low--$\ell$ polarization, especially in view of its low signal--to--noise level in both WMAP and {\sc Planck} data. In order to address this question, we repeated our analysis relying only on temperature data and assuming a Gaussian prior for $\tau = 0.070 \pm 0.015$. Notice, in fact, that $\tau$, $A_s$ and $\Delta$ all impact on large--scale power. In particular, the posteriors for $A_s$ and $\tau$ tend to shift when $\Delta$ is sampled, in ways that are more sensitive to temperature for the former and to polarization for the latter. It is thus preferable to use low--$\ell$ polarization and let the data adjust parameters accordingly. In the presence of a $\tau$ prior, instead, the data are bound to accept it, and consequently both $\Delta$ and $A_s$ shift a bit. The $(A_s,\tau,\Delta)$ correlation pattern is manifest in the three panels of fig.~\ref{fig:correlation_effects}. With a $\tau$ prior the detection level of $\Delta$ in the ${\rm Ext}_{24}$ mask would become slightly higher than in ${\rm Ext}_{30}$ mask, which was instead favoured by the complete analysis in~\cite{Gruppuso:2015xqa}, and is favoured again here, where we resort to the joint {\sc PlanckLFI}+WMAP dataset in polarization, following~\cite{Lattanzi:2016dzq}. A similar argument would apply if one left out the lensing likelihood, which impacts on $A_s$, and hence on $\tau$ and $\Delta$. For this reason, we preferred to feed both low--$\ell$ polarization and lensing information in our analysis.

At any rate, had we only relied on priors, our results would have been affected only slightly, and our conclusions would have been essentially the same~\footnote{The {\sc Planck} collaboration published~\cite{Aghanim:2016yuo} the estimate $\tau = 0.055 \pm 0.009$, which is significantly smaller than what the {\sc Planck} (and WMAP9) publicly available polarization likelihood, which we use, would imply. In order to check that this lower value of $\tau$ does not affect our conclusions, we also performed an analysis using only the temperature portion of the likelihood of eqs.~\eqref{like} and \eqref{sigcov}, along with a $\tau$ prior derived from the above estimate. This choice has again no appreciable effects on our conclusions.}. It is however remarkable how low--$\ell$ polarization has noticeable effects on the detection of $\Delta$, even with the present--day low signal--to--noise ratio. The constraining power of the higher--quality polarization data awaited from the next generation of experiments~\cite{LiteBIRD,LSPE,Essinger-Hileman:2014pja}, in general and on $\Delta$ in particular, remains to be ascertained, but it is potentially appealing, as we shall see shortly.

\subsection{Bayesian constraints on $S_\frac{1}{2}$, $V$ and $R$} \label{sec:constraintsestimators}

For each mask in table~\ref{tab:masks} we have built the posterior distribution functions of the estimators $S_\frac{1}{2}$, $V$ and $R$ defined in Section~\ref{dataandsims},
considering ``even+odd'', ``even'' and ``odd'' contributions.

\begin{figure}[ht]
\centering
\begin{tabular}{cc}
\includegraphics[width=58mm]{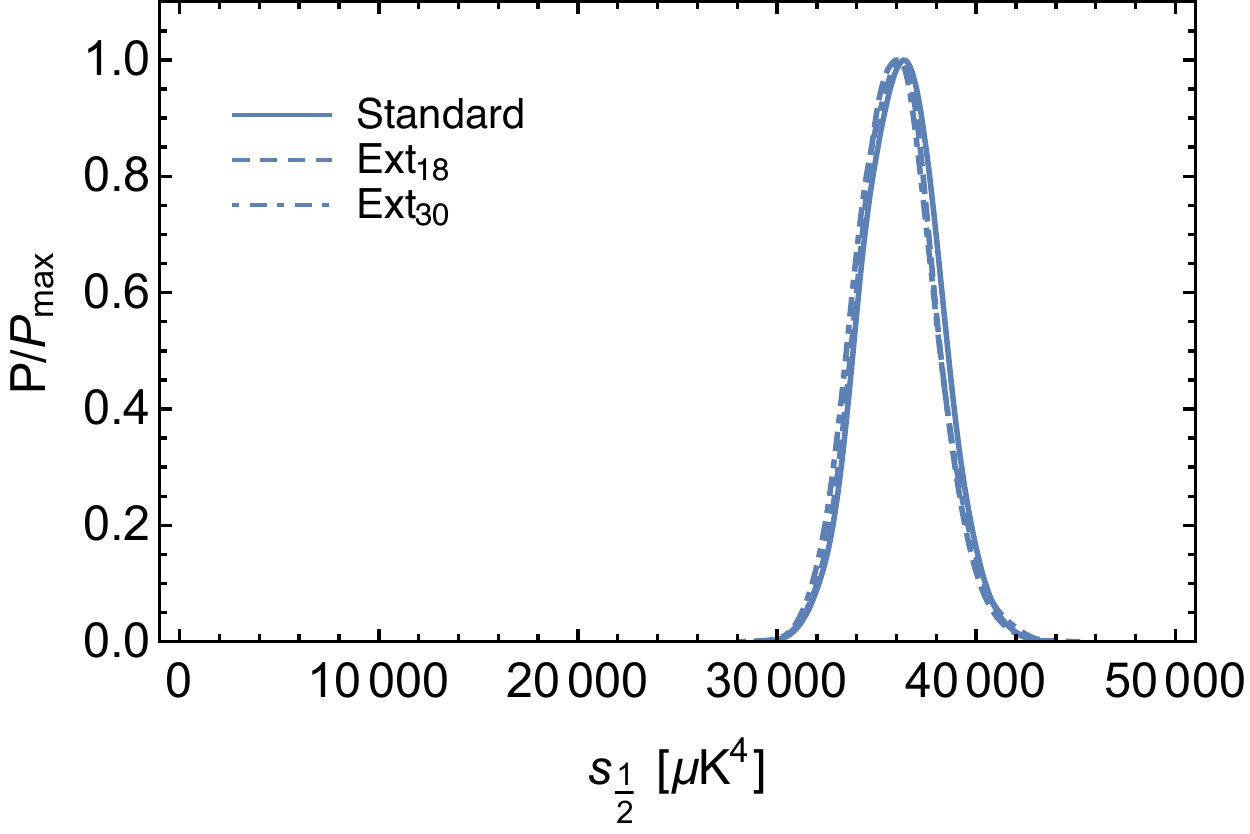}  &
\includegraphics[width=58mm]{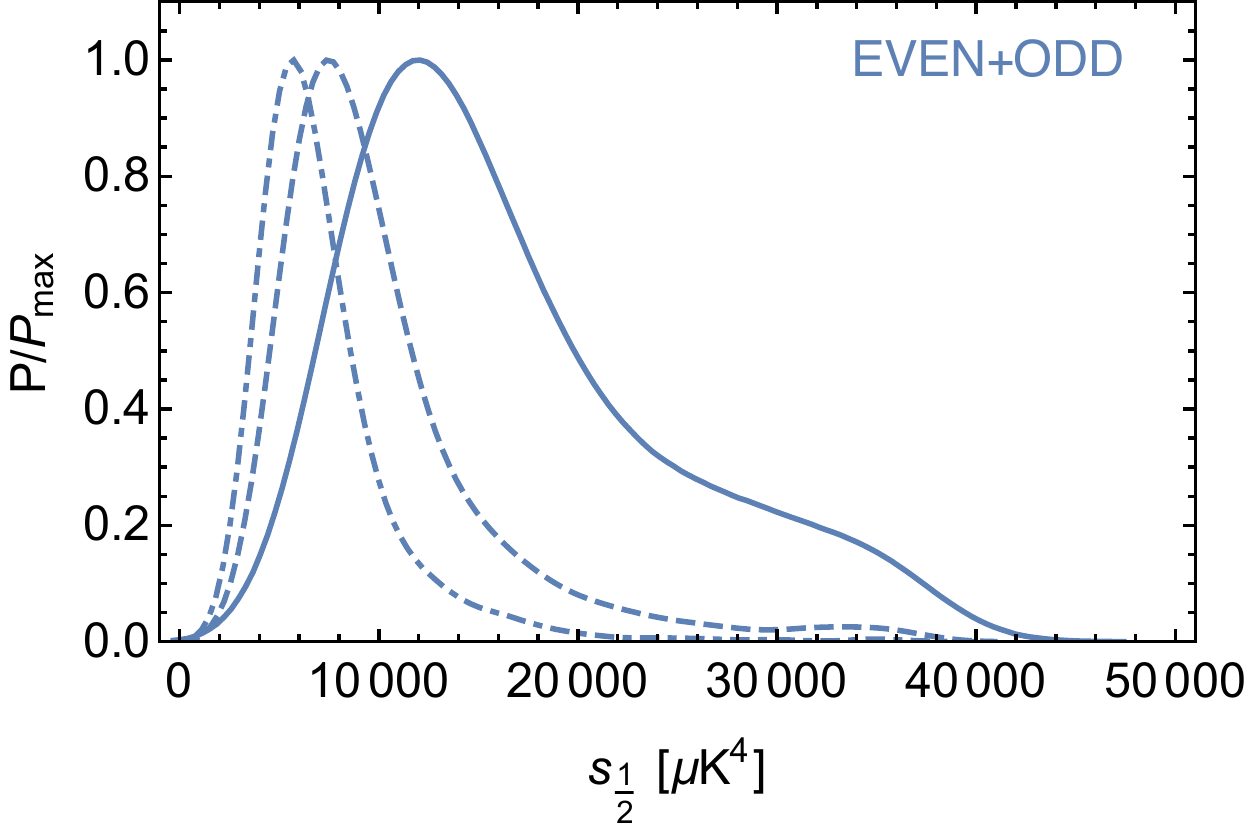} \\
\includegraphics[width=58mm]{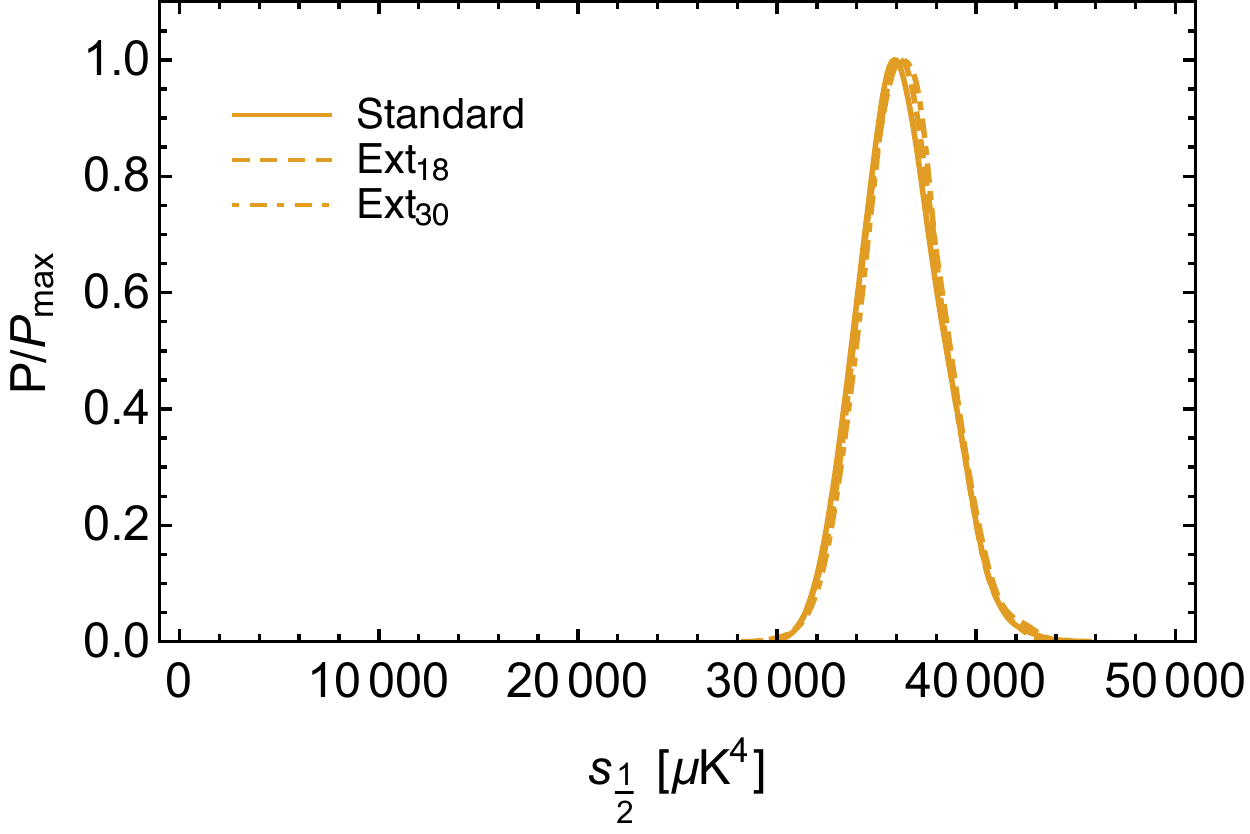} &
\includegraphics[width=58mm]{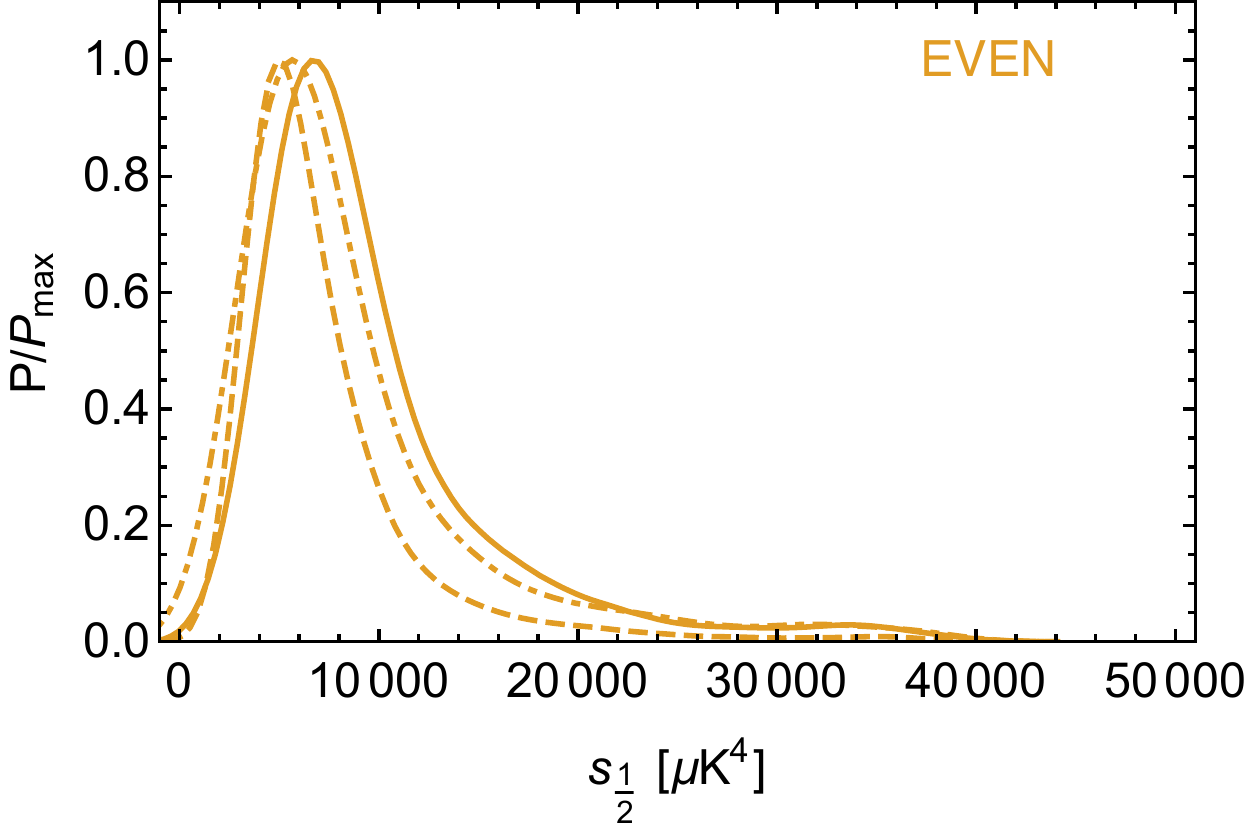} \\
\includegraphics[width=58mm]{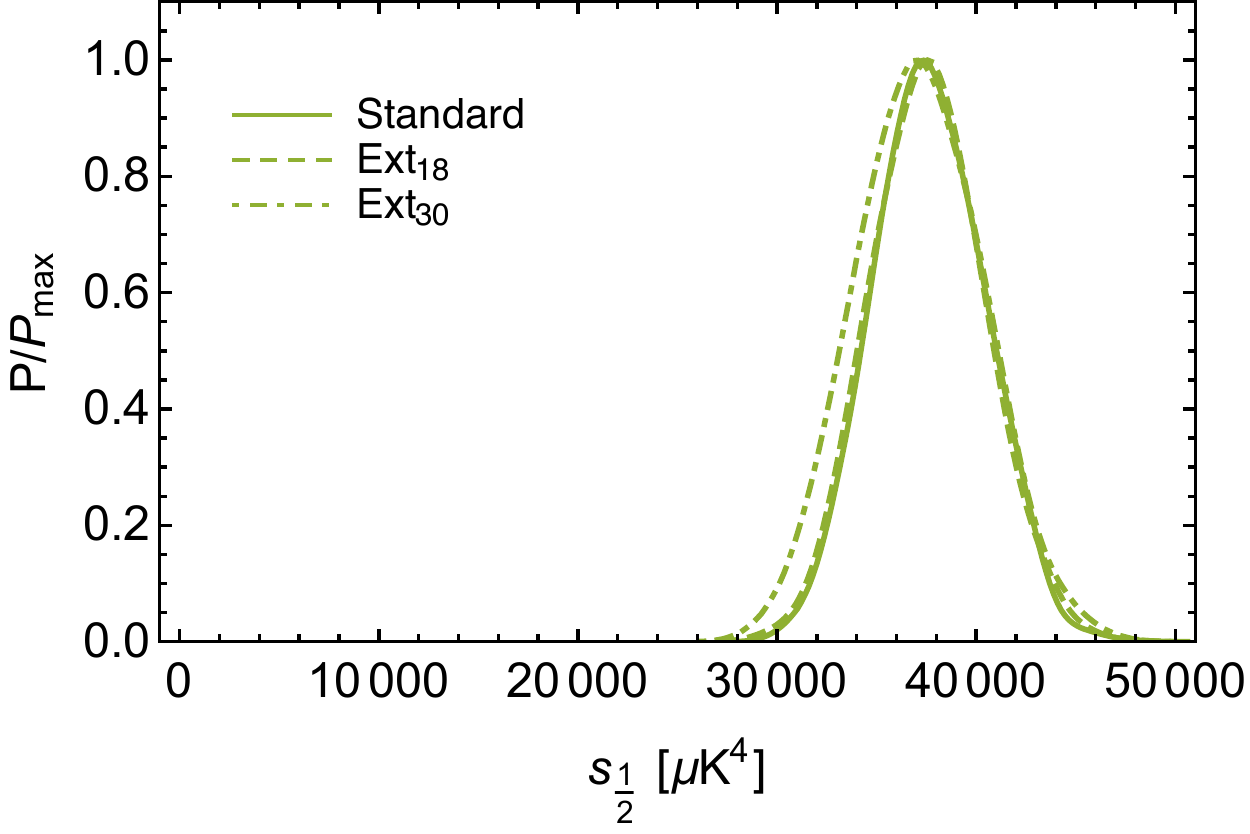} &
\includegraphics[width=58mm]{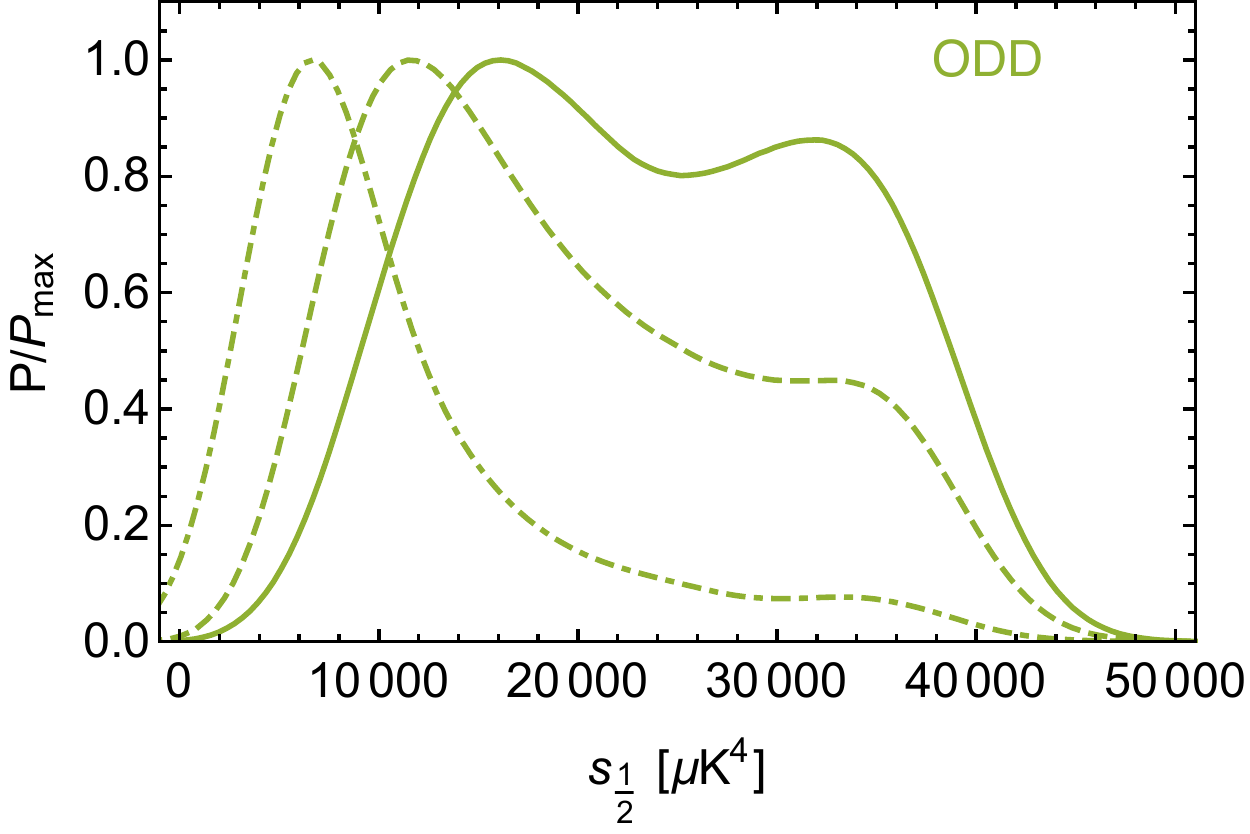}
\end{tabular}
%
%\caption{\small Posterior distribution functions of $S_\frac{1}{2}$. The upper panels refer to $\Lambda$CDM and the lower ones refer to $\Lambda {\rm CDM}\Delta$.
%The left panels are for ``even+odd'', the middle ones are for ``even'' and the right ones are for ``odd''. The three types of curves (continuous, dashed and dashed--dotted) are for the Standard mask, the Ext$_{18}$ mask and the Ext$_{30}$ mask.}
\caption{\small Posterior distribution functions of $S_\frac{1}{2}$. {The left panels refer to $\Lambda$CDM and the right ones refer to $\Lambda {\rm CDM}\Delta$.
The upper panels are for ``even+odd'', the middle ones are for ``even'' and the lower ones are for ``odd''. The three types of curves (continuous, dashed and dashed--dotted) are for the Standard mask, the Ext$_{18}$ mask and the Ext$_{30}$ mask.}}
\label{fig:s1over2}
\end{figure}

\begin{figure}[ht]
\centering
\begin{tabular}{cc}
\includegraphics[width=58mm]{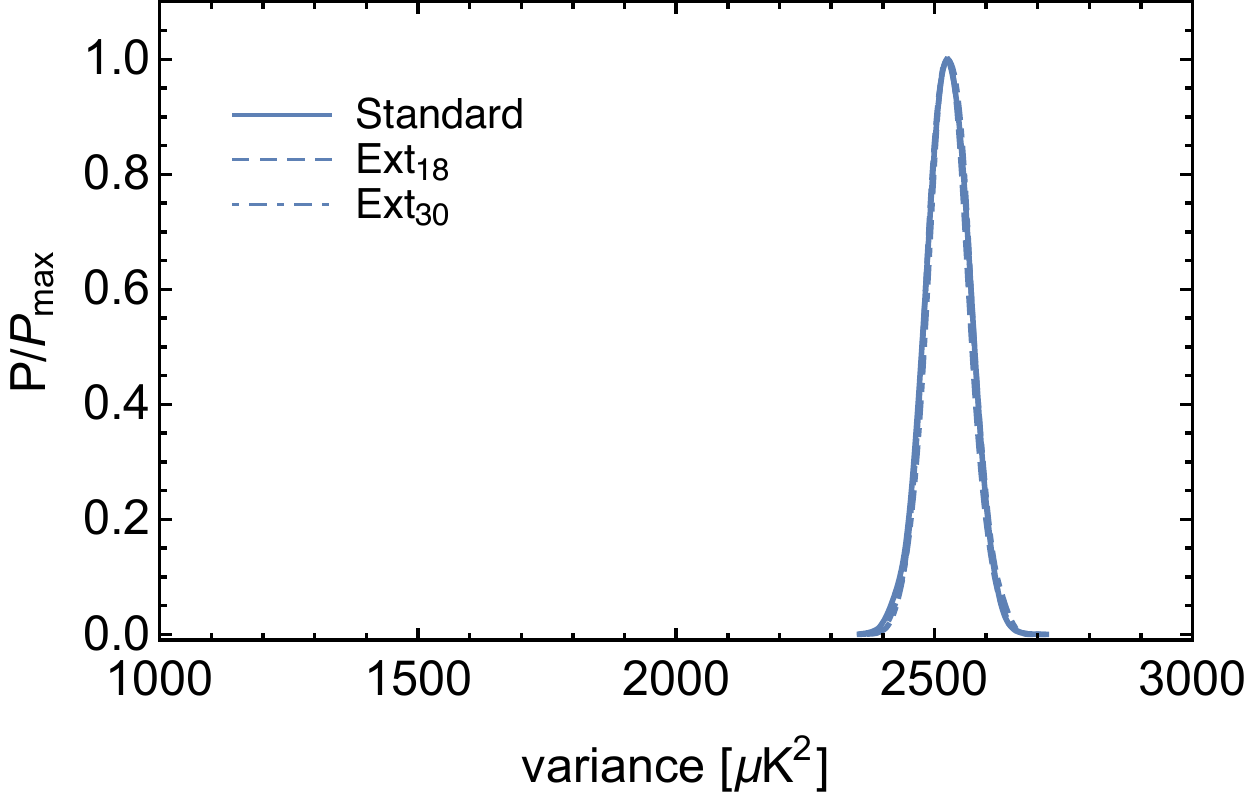}  &
\includegraphics[width=58mm]{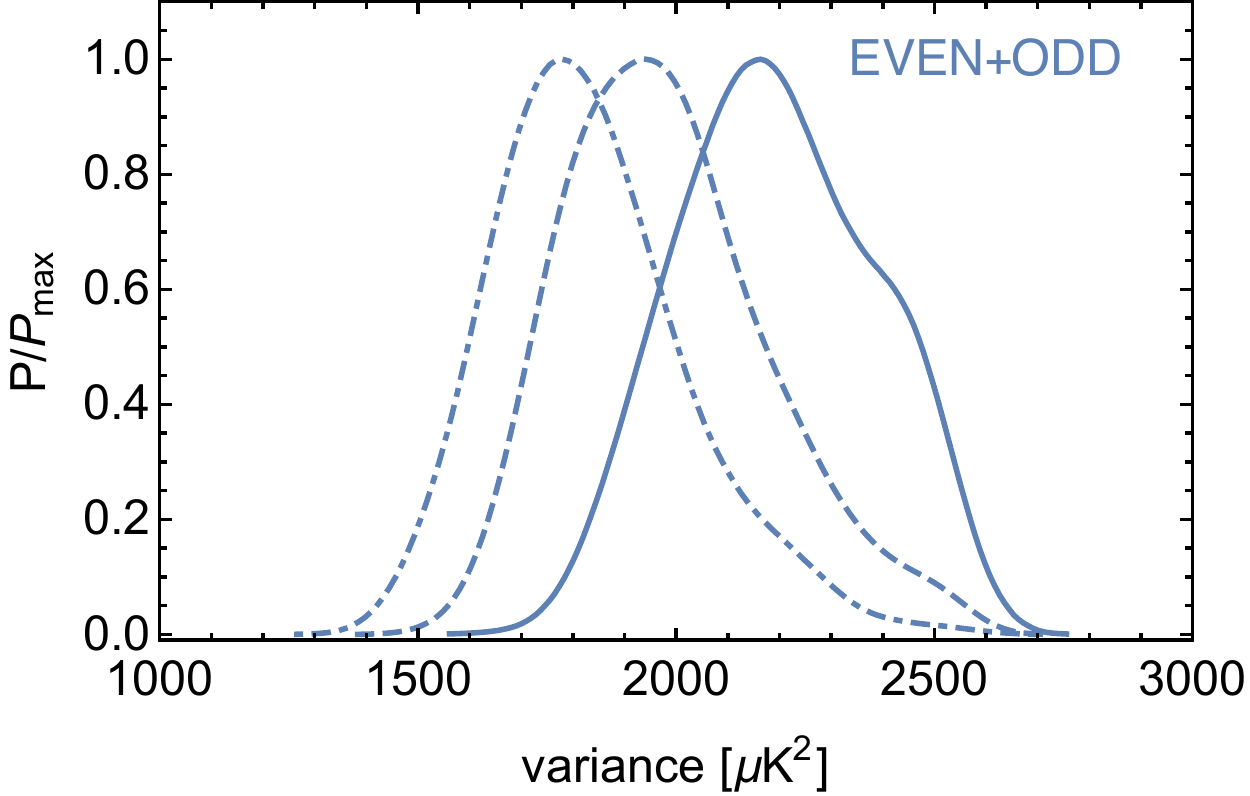}  \\
\includegraphics[width=58mm]{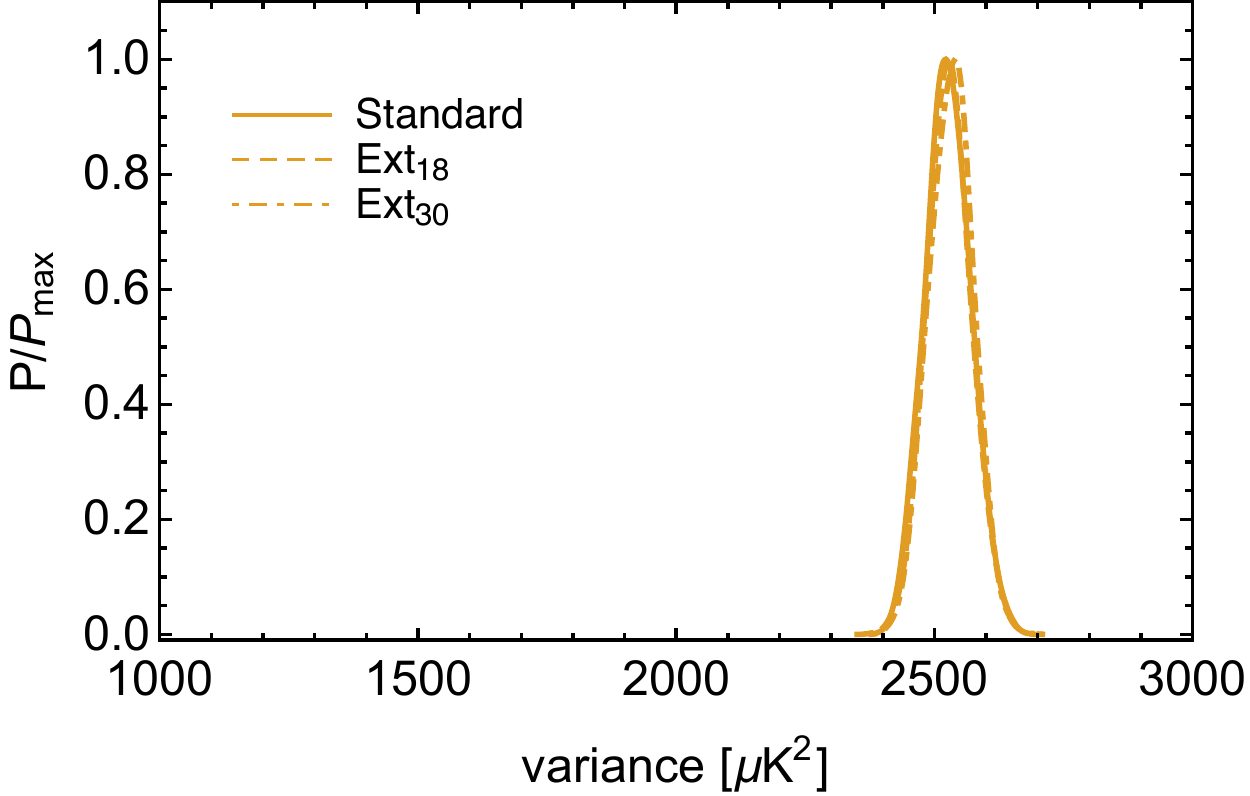} &
\includegraphics[width=58mm]{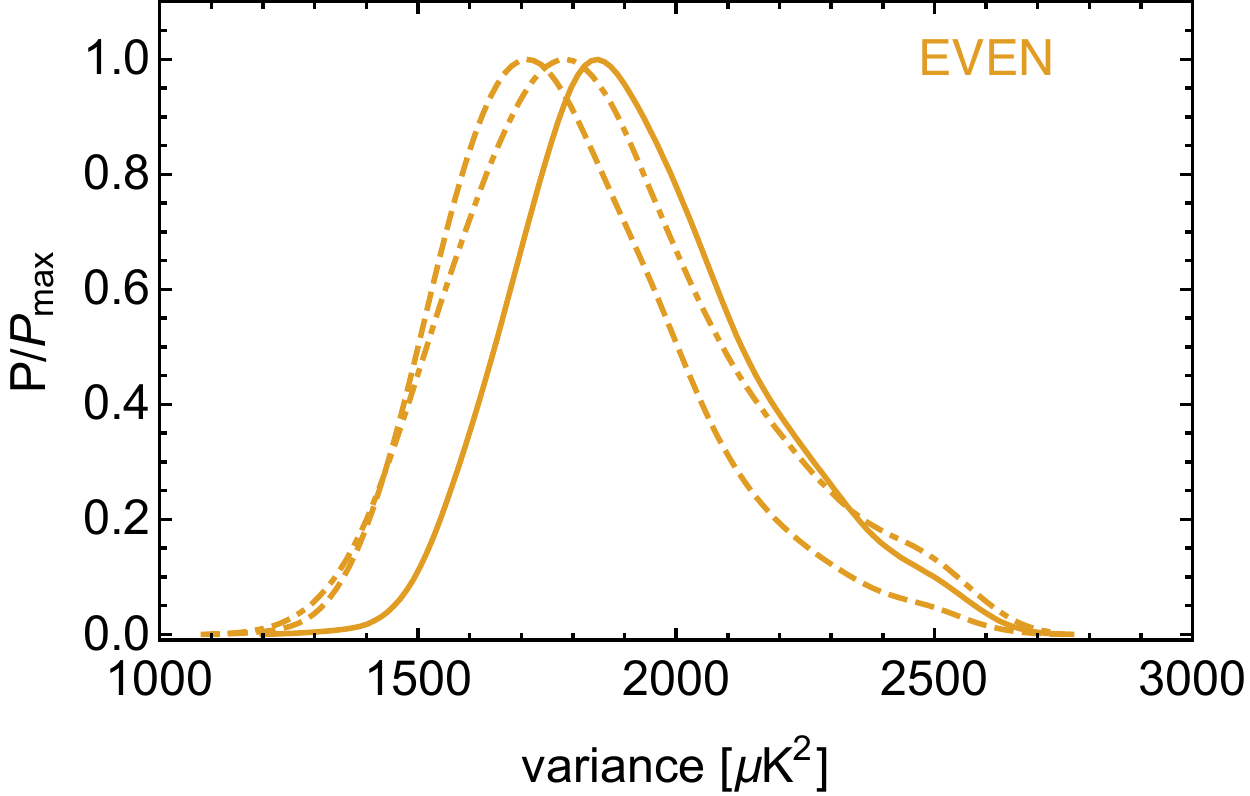} \\
\includegraphics[width=58mm]{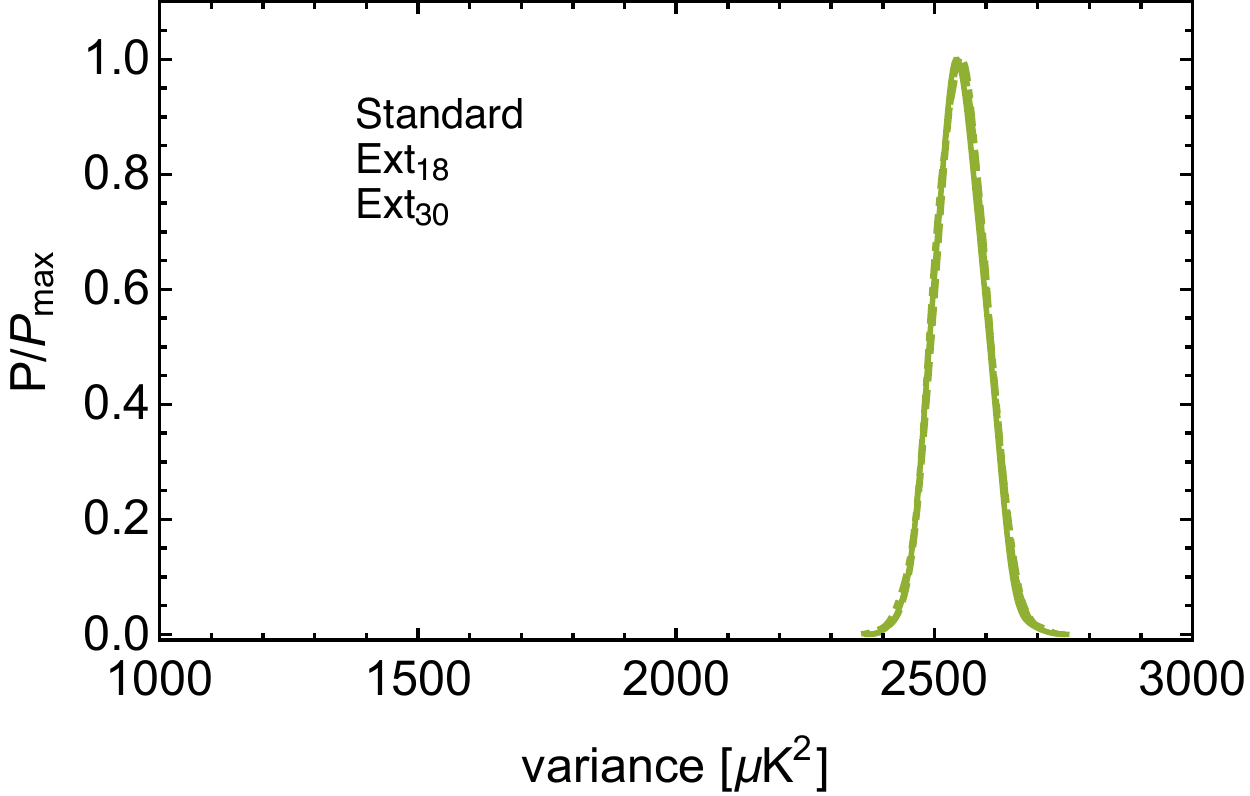} &
\includegraphics[width=58mm]{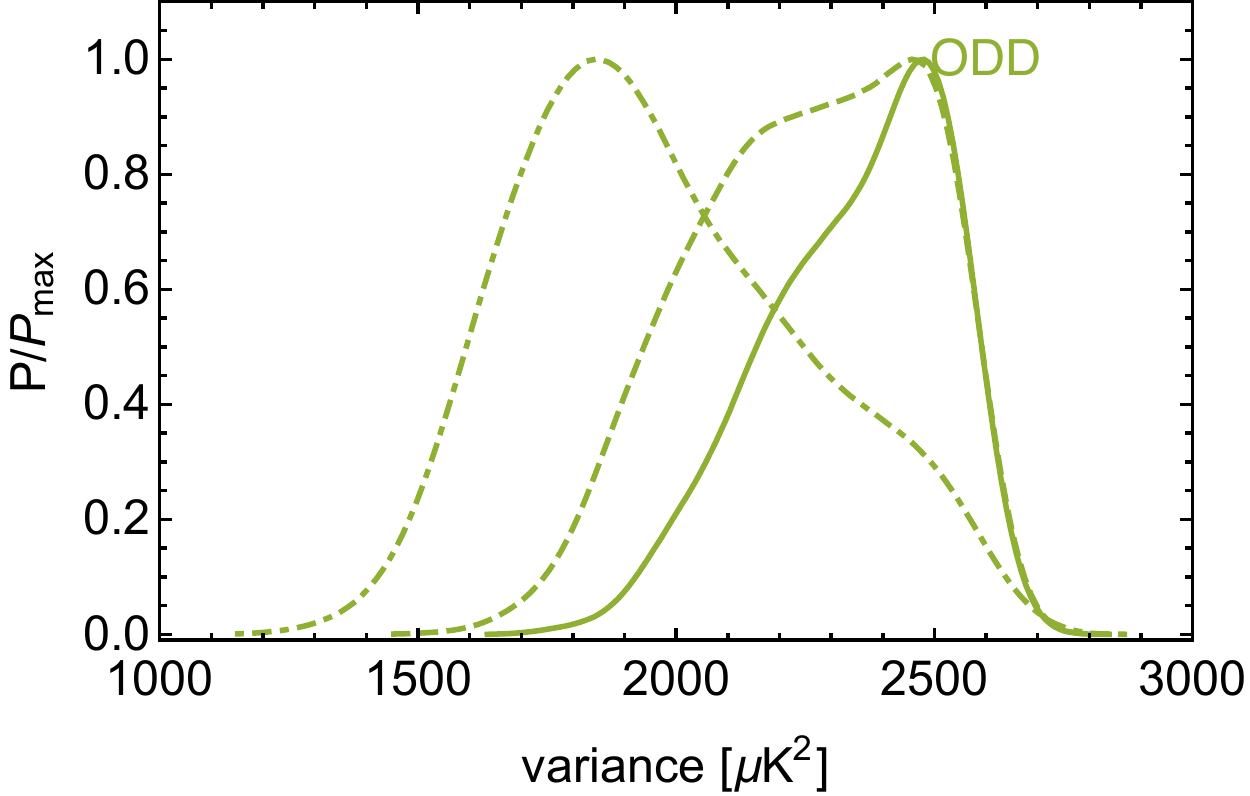}
\end{tabular}
\caption{\small Posterior distribution functions of $V$, with the same conventions as in fig.~\ref{fig:s1over2}.}
\label{fig:V}
\end{figure}

\begin{figure}[ht]
\centering
\begin{tabular}{cc}
\includegraphics[width=58mm]{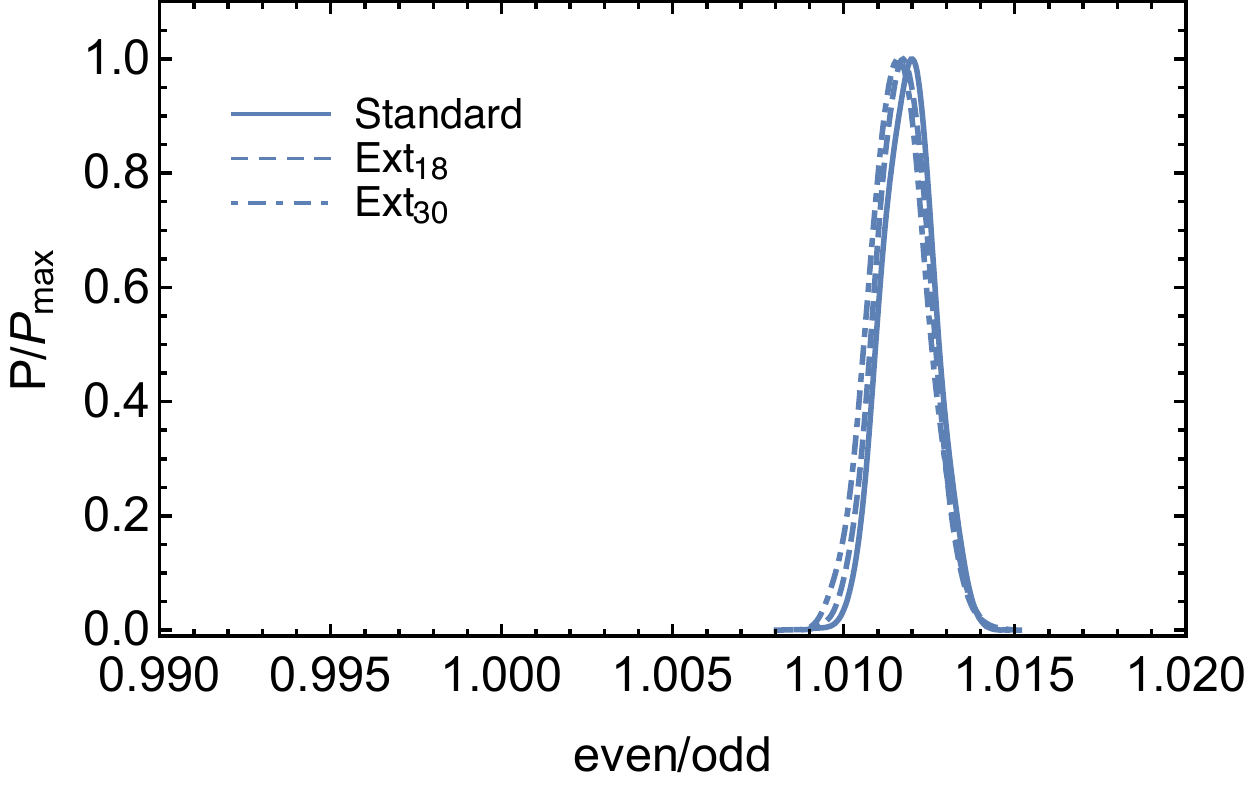}  &
\includegraphics[width=58mm]{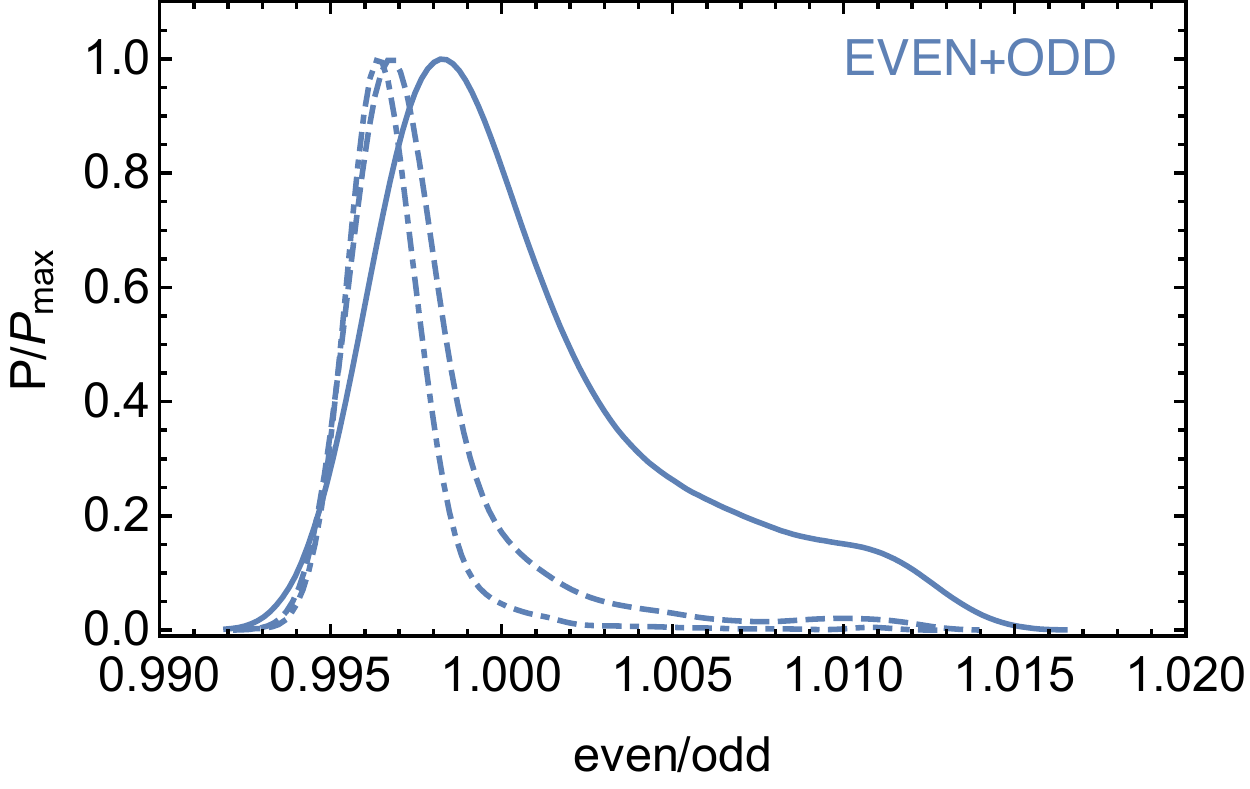}  \\
\includegraphics[width=58mm]{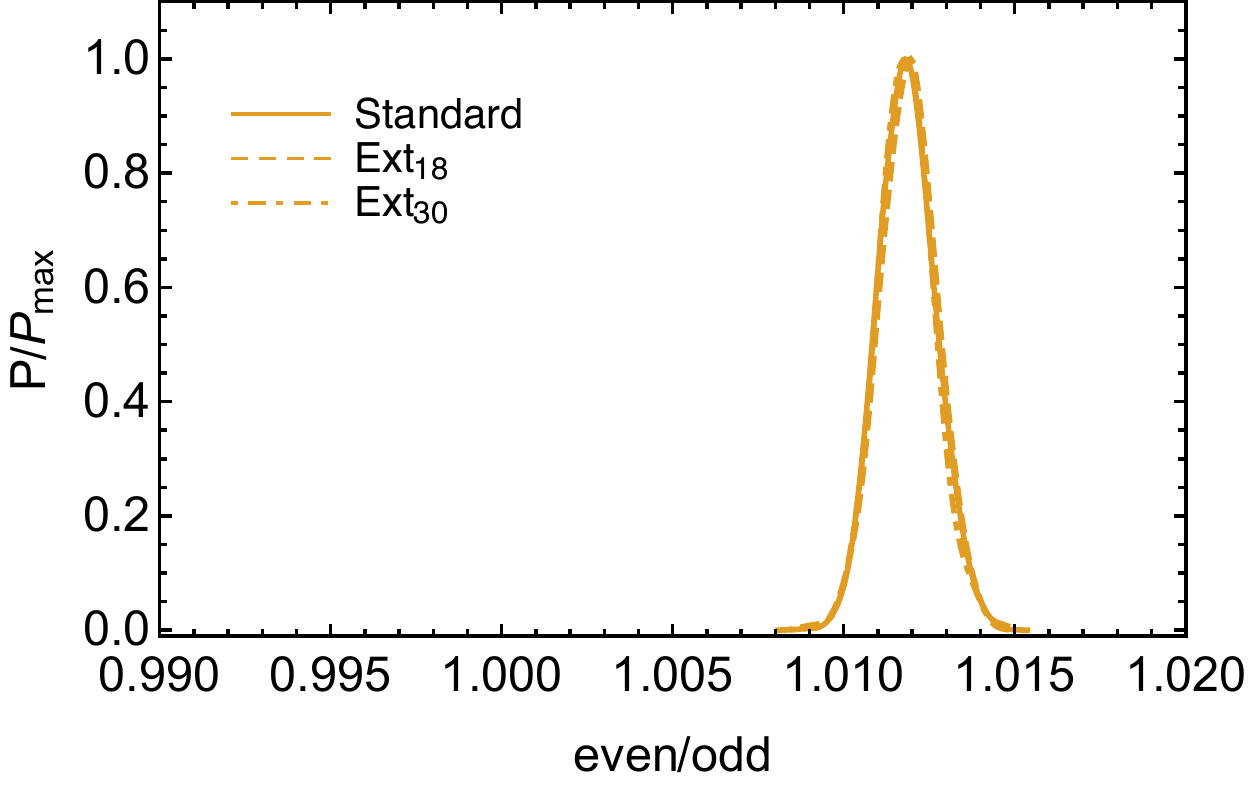} &
\includegraphics[width=58mm]{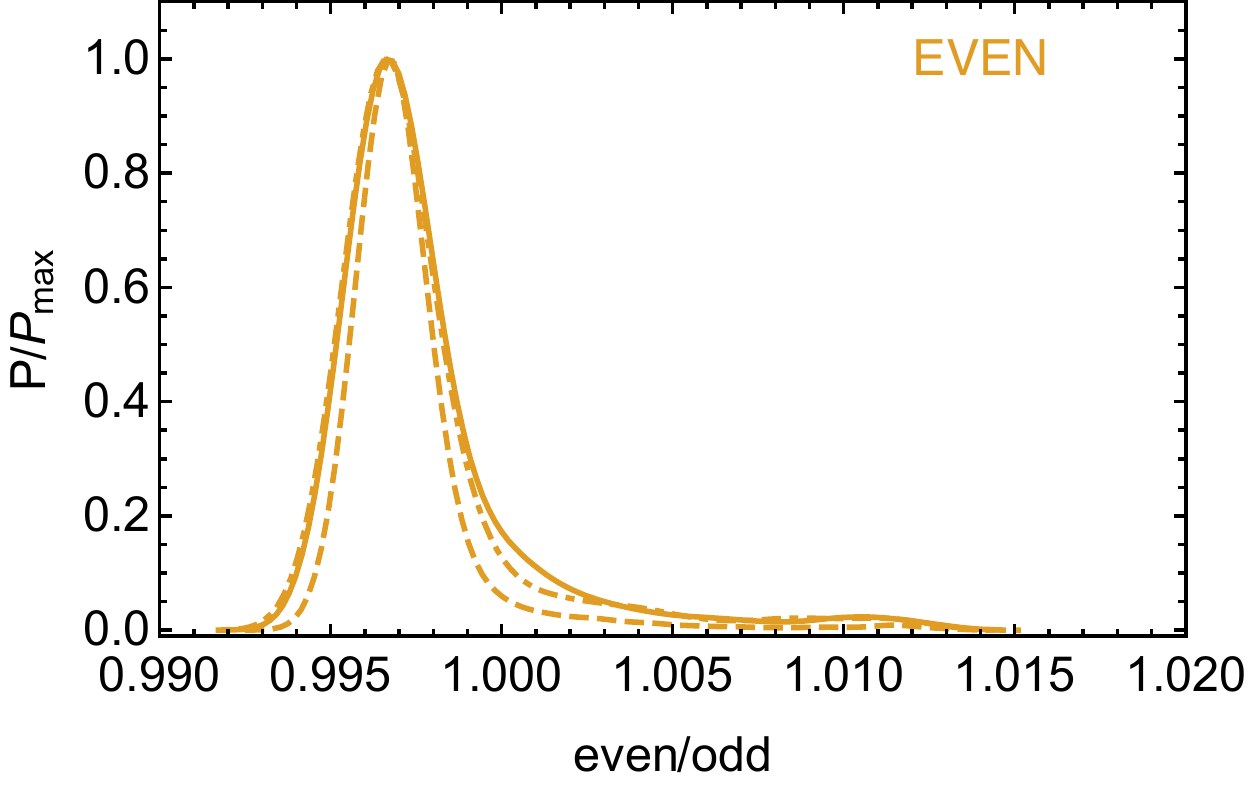} \\
\includegraphics[width=58mm]{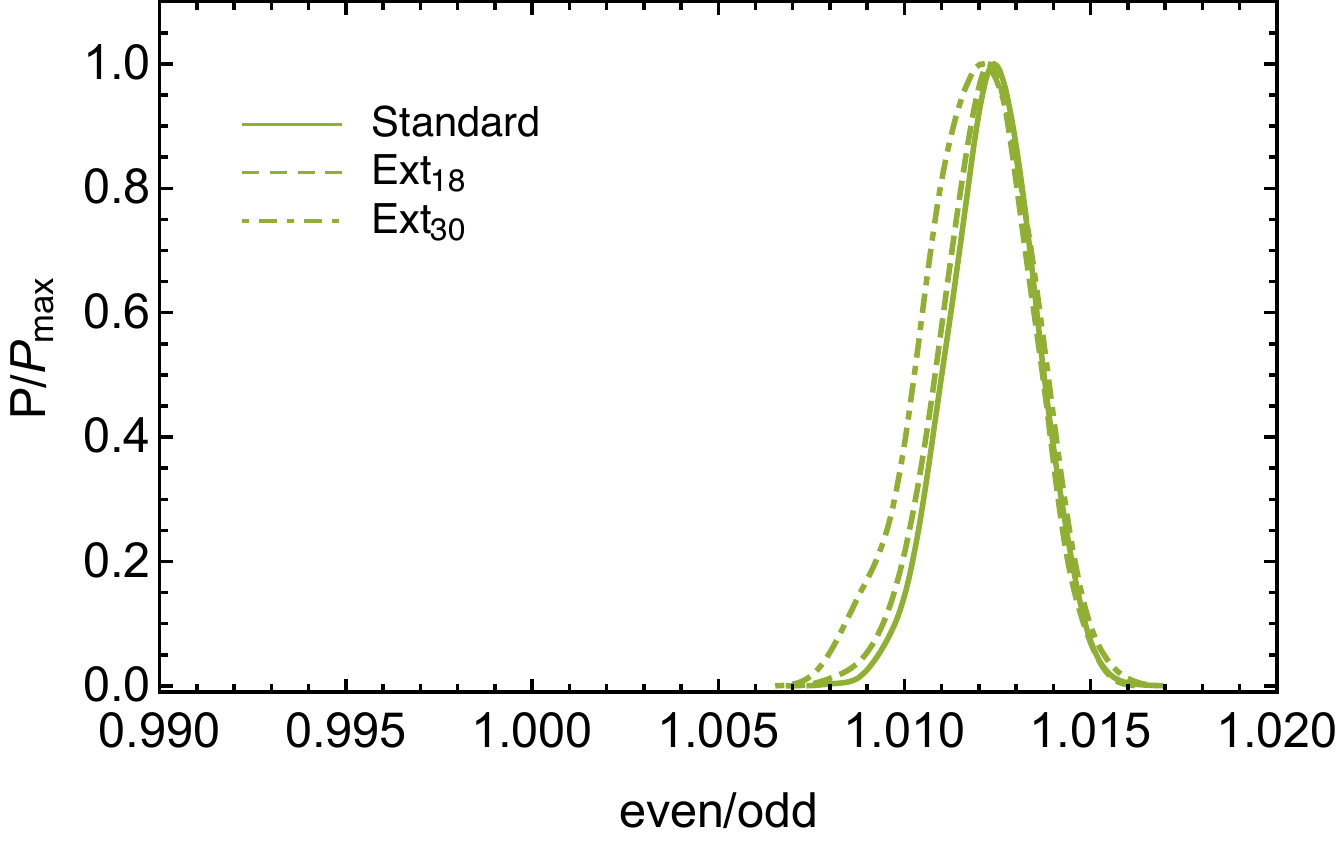} &
\includegraphics[width=58mm]{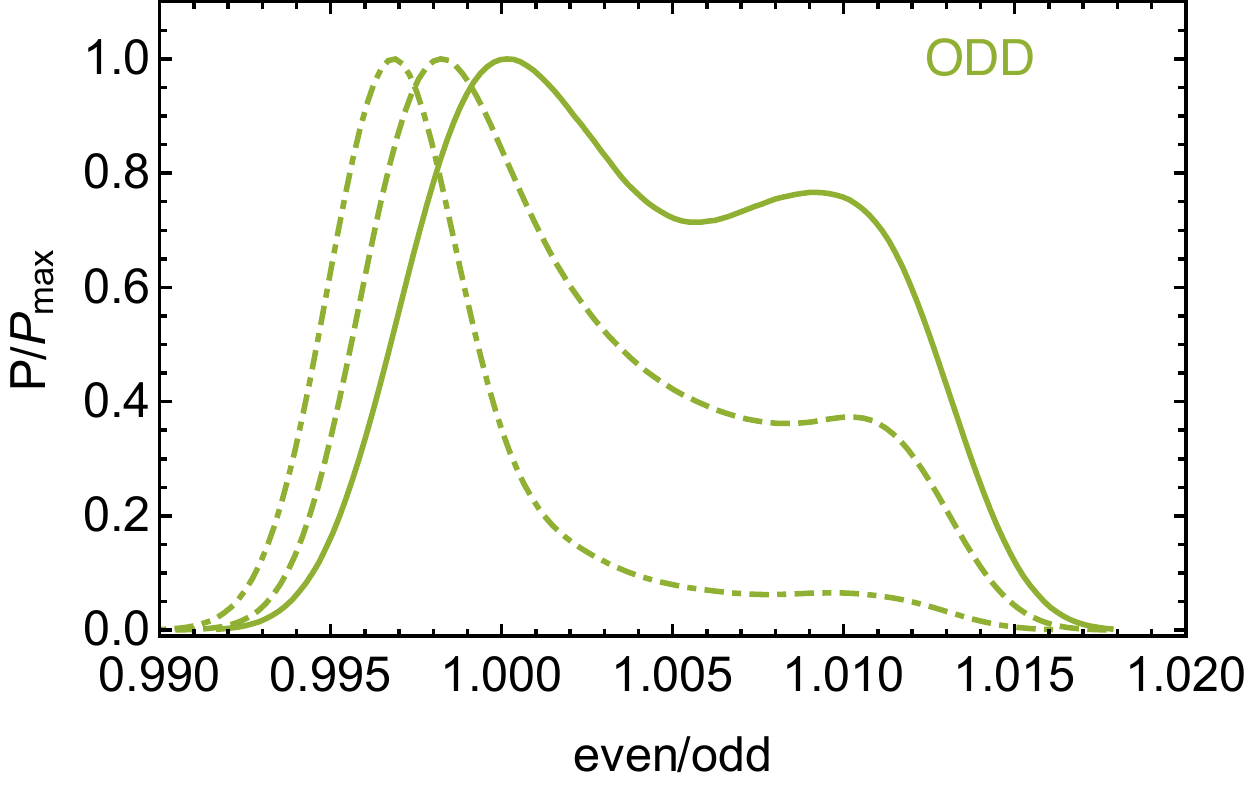}
\end{tabular}
\caption{\small Posterior distribution functions of $R$, with the same conventions as in fig.~\ref{fig:s1over2}.}
\label{fig:R}
\end{figure}

The posteriors of $S_\frac{1}{2}$ are collected in fig.~\ref{fig:s1over2}, where
the first
%row
{column} refers to $\Lambda$CDM while the second refers to $\Lambda {\rm CDM}\Delta$.
In both cases, the
%left
{upper} panels are for ``even+odd'', the middle ones are for ``even'' and the
%right
{lower} ones are for ``odd''.
While the posteriors in the first
%row
{column} are quite stably centered at about $38000 \, \mu$K$^4$,
the others display a richer behaviour, with a second peak around $6000 \, \mu$K$^4$
that becomes readily dominant for wider masks.
Consistently with our preceding findings, the ``odd'' part is more resilient to converging toward the lower peak.
The same pattern presents itself for the other two estimators $V$ and $R$, and the corresponding plots are collected in
figs.~\ref{fig:V} and \ref{fig:R}.

\subsection{Comments on the frequentist approach} \label{sec:constraintsestimatorsfreq}

\begin{figure}[ht]
\centering
\includegraphics[width=85mm]{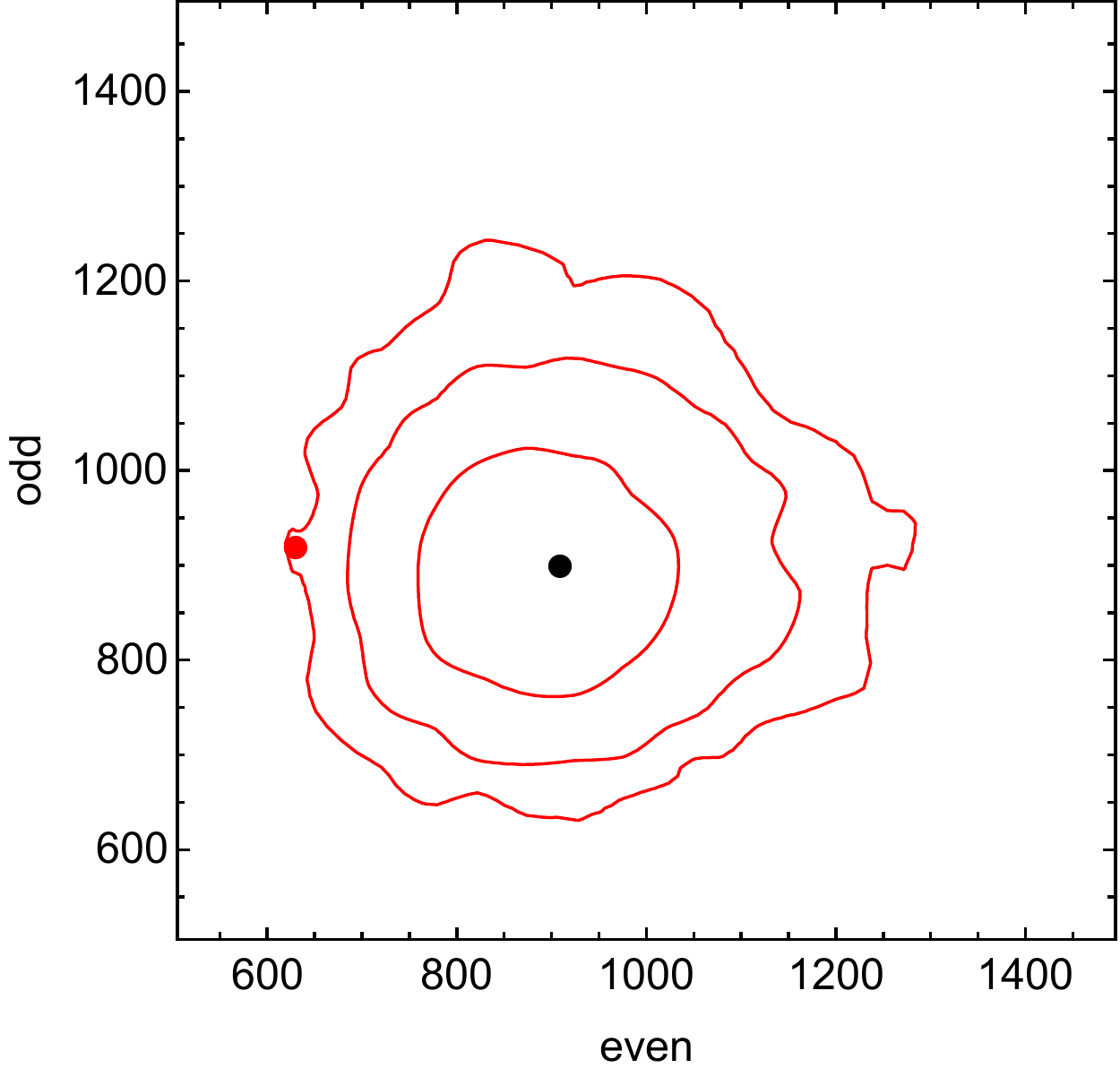}
\caption{\small $C^{+}$ {\it vs} $C^{-}$ in a frequentist approach.
The black dot identifies the expected values of $C^{+}$ and $C^{-}$ in $\Lambda$CDM, while the red one identifies the observed values.
The contours elicit the 1, 2 and 3~$\sigma$ levels expected in $\Lambda$CDM, in units of $\mu$K$^2$.}
\label{fig:c+c-freq}
\end{figure}

In principle, one can also build the preceding estimators via a frequentist approach, in order to test the $\Lambda$CDM model.
In this case, however, wider masks and the corresponding increased sampling variance bring along negative values for some $C_{\ell}^{\,\rm TT}$'s.
This is particularly unpleasant for the ratio $R$ of eq.~\eqref{r}, and all in all this approach is less suitable for our purposes.
Consequently, as an example we show in fig.~\ref{fig:c+c-freq} the 2D contour plot of $C^{+}$ {\it vs} $C^{-}$ for the mask Ext$_{18}$.
The black dot identifies the expected values for $C^{+}$ and $C^{-}$ in a fiducial $\Lambda$CDM model, while the red one
identifies the observed values. Moreover, the contours identify the levels at 1, 2 and 3~$\sigma$ for the empirical distribution expected in $\Lambda$CDM.
The resulting information is along the lines of what we already said: the ``even'' contribution lies well below $\Lambda$CDM,
while the ``odd'' one is largely compatible with it.

\subsection{Forecasts for future experiments} \label{sec:forecasts}

Several projects for all-sky future experiments are currently under development, as the JAXA-led LiteBIRD~\cite{LiteBIRD}, or are being proposed, as the European CORE~\cite{Delabrouille:2017rct}. Their observations should provide cosmic--variance--limited measurements of the CMB $E$-mode polarization field, at least in the low-$\ell$ region. In order to explore the prospects that they offer for better determinations of $\Delta$, or for ruling it out altogether, we model their expected polarization maps adding only a diagonal regularization noise contribution of $0.01~\mu\mathrm{K}^2$ in both $Q$ and $U$ to a pixel windowed, $N_\mathrm{side} = 16$ CMB field. The latter is a realization of a fiducial cosmological model\footnote{In particular, we take as our fiducial the best-fit model obtained in the Ext$_{30}$ mask.} with $\Delta = 0.37\times 10^{-3}\,\mathrm{Mpc}^{-1}$. This treatment assumes ideal component separation and negligible systematic effects at the level of $E$-modes, a goal that lies within reach for the next generation space--borne experimental efforts~\cite{LiteBIRD,Delabrouille:2017rct,Remazeilles:2017szm,Natoli:2017sqz}.

\begin{figure}[ht]
\centering
\includegraphics[width=90mm]{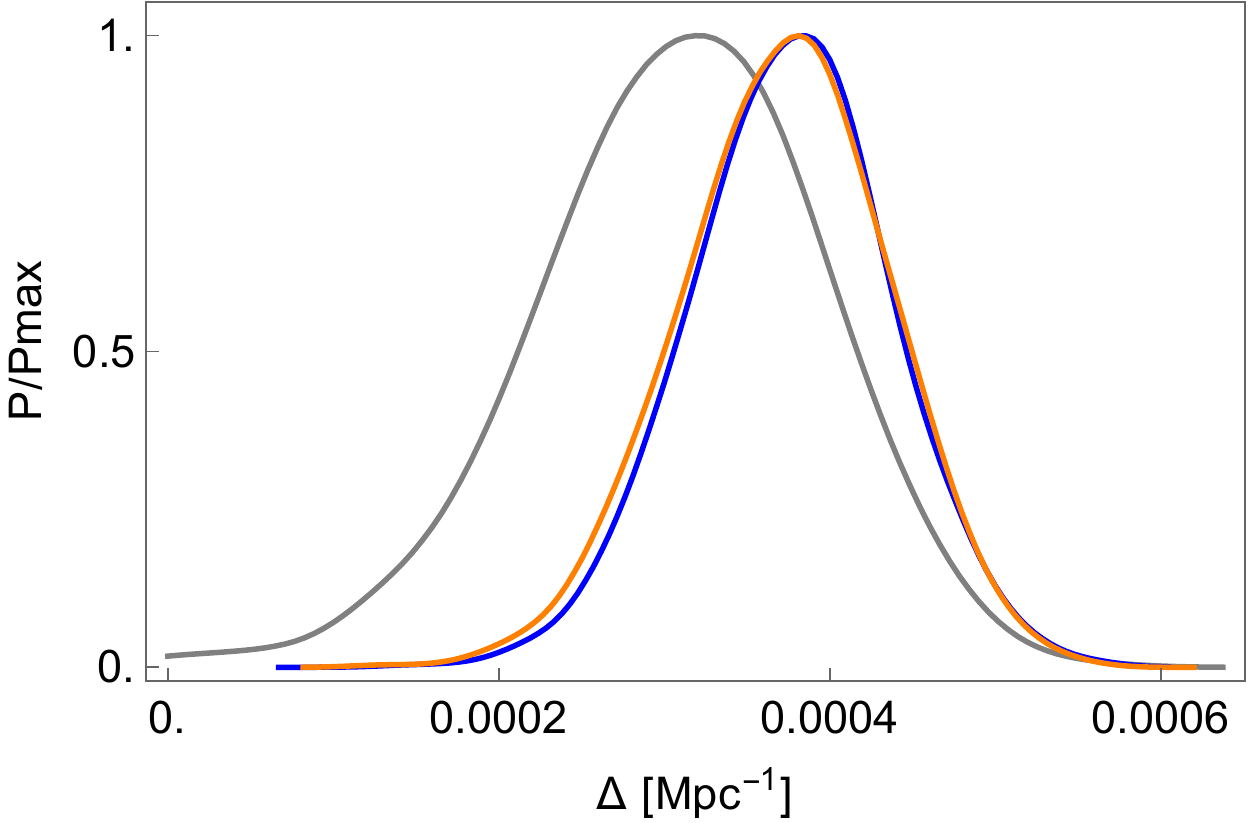}
\caption{\small Expected estimates for $\Delta$ from future polarization--oriented experiments. The curves share the same underlying CMB realization, generated from a fiducial cosmological model with $\Delta = 0.37\times 10^{-3}$. With {\sc Planck}-like noise,
standard mask in temperature and union mask in polarization, the detection level for $\Delta$ could grow up to about 3.5 $\sigma$ (grey curve).
With cosmic--variance--limited temperature and polarization data at large scales and the Ext$_{30}$ masking, the detection level for $\Delta$ could rise up to about 6 $\sigma$ (orange curve). Finally, with the same cosmic--variance--limited data, the standard mask in temperature and no mask in polarization, the detection level could
increase even slightly beyond 6 $\sigma$ (blue curve).}
\label{fig:forecast}
\end{figure}

Our results are presented in fig.~\ref{fig:forecast}, where we collect some forecasts for the detection levels of $\Delta$ within the setup just discussed.
All results are obtained from simulated low--$\ell$ datasets that share the same underlying CMB realization, but are associated to different noise levels and different maskings. These simulated low-$\ell$ datasets are combined with the actual high-$\ell$ and lensing {\sc Planck} data.
To begin with, the gray curve refers to a dataset with the same noise properties as real data, but using the standard and union masks in temperature and polarization, which could result in a detection level of $\Delta$ of about 3.5~$\sigma$. On the other hand an ideal experiment with the same masking, but measuring the large--scale polarization down to the cosmic variance limit could raise the detection level for $\Delta$ up to about $6\,\sigma$ (orange curve). Finally, the same cosmic-variance limited data, analyzed through the standard {\sc Planck} mask in temperature and with no masking in polarization could result in detection levels even slightly beyond $6 \sigma$ (blue curve).

\section{Final remarks}
\label{sec:final_remarks}

We have presented new evidence that some statistical properties of the CMB anisotropy depend sizeably on the Galactic latitude.
To begin with, we have confirmed that at higher latitudes the observed lack of power is more pronounced than in standard masking.
We have characterised this feature via the  primordial power spectra
\beq
P(k) \ \sim \ \frac{k^3}{\left[k^2 \ + \ \Delta^2 \right]^{2\,-\,\frac{n_s}{2}}} \ , \label{cutoff2}
\eeq
which approach $P(k) \ \sim k^{n_s -1}$ for large values of $k$, and where the scale $\Delta$ accounts for the low-$\ell$ lack of power.
Its determination translates, as  in \cite{Gruppuso:2015xqa}, in departures from $\Lambda$CDM
that depend on Galactic masking, and become sizeable below $\ell \sim 7$ for the standard mask and below $\ell \sim 15$ for the extended mask Ext$_{30}$ of table~\ref{tab:masks}.
Moreover, we have discovered that \emph{the latitude dependence is largely due to the odd multipoles}, which contain more power around the Galactic plane than away from it.
On the other hand, the even multipoles appear relatively stable in this respect, and lead to a determination of $\Delta$ that is essentially latitude independent, consistently with its possible cosmological origin. This behavior translates into an even--odd asymmetry that is largely concentrated around the Galactic plane.

One cannot exclude that low--$\ell$ anomalies be a statistical fluke at the 3--$\sigma$ level. The even--odd differences that we are highlighting could also be induced by unidentified systematic contaminations at very large scales that fade out as one moves away from the Galactic plane. However, it seems difficult to envisage systematic effects that could impart an even-odd asymmetry on the CMB sky. For instance, standard diffuse Galactic emissions are typically symmetric with respect to the Galactic plane, and ought to enhance even multipoles with respect to odd ones (see eq.~\eqref{CTT}), a behaviour that is opposite to what we are detecting. In other words, if this signature were due to unaccounted foregrounds, these ought to be distributed \emph{antisymmetrically}, in particular with respect to antipodal points around the Galactic plane.
Residuals of unaccounted asymmetric foregrounds, however, are not the only candidates for such an even-odd pattern, which might also originate from systematic effects of instrumental origin, or from a tricky combination of both. The fact that WMAP and {\sc Planck} observe consistent large--angle CMB temperature patterns, however, restricts the possibilities for anomalies of purely instrumental origin to features shared by the two experiments. This still leaves room, say, for calibration issues that insist on the CMB dipole.

These investigations were inspired by pre-inflationary scenarios that occur in String Theory, where the inflaton decelerates to slow-roll after bouncing against a steep exponential potential.
\emph{The transition to slow-roll introduces a universal $k^3$ -- cut at low frequencies in primordial power spectra, which is captured by eq.~(\ref{cutoff2}).} The cut is generally accompanied
by a narrow peak and a few oscillations, whose detailed features are model dependent. In \cite{Gruppuso:2015xqa} we found at most a scant evidence for a peak located at the transition:
these local effects, if they are there, appear currently beyond reach. Still, the oscillations would be a natural origin of the even--odd asymmetry.
At any rate, relics of the approach to slow--roll would open up an enticing window on the earliest stages of inflation, when sizable primordial inhomogeneities were possibly around. Or, perhaps less enticingly, the power depression might be revealing local features of the potential that were experienced by the inflaton at a later epoch.
\begin{figure}[ht]
\centering
\begin{tabular}{ccc}
\includegraphics[width=35mm]{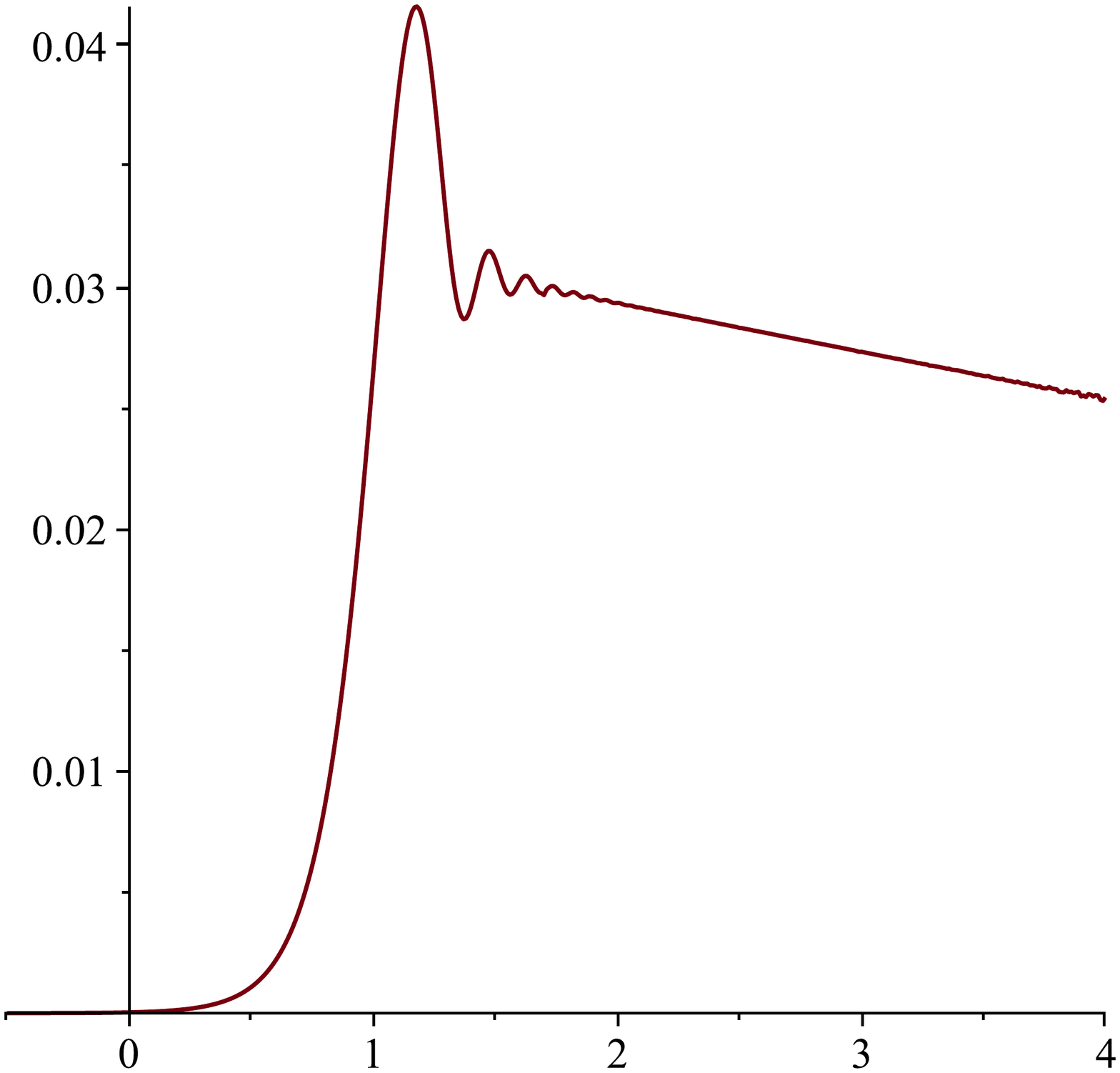}  &
\includegraphics[width=35mm]{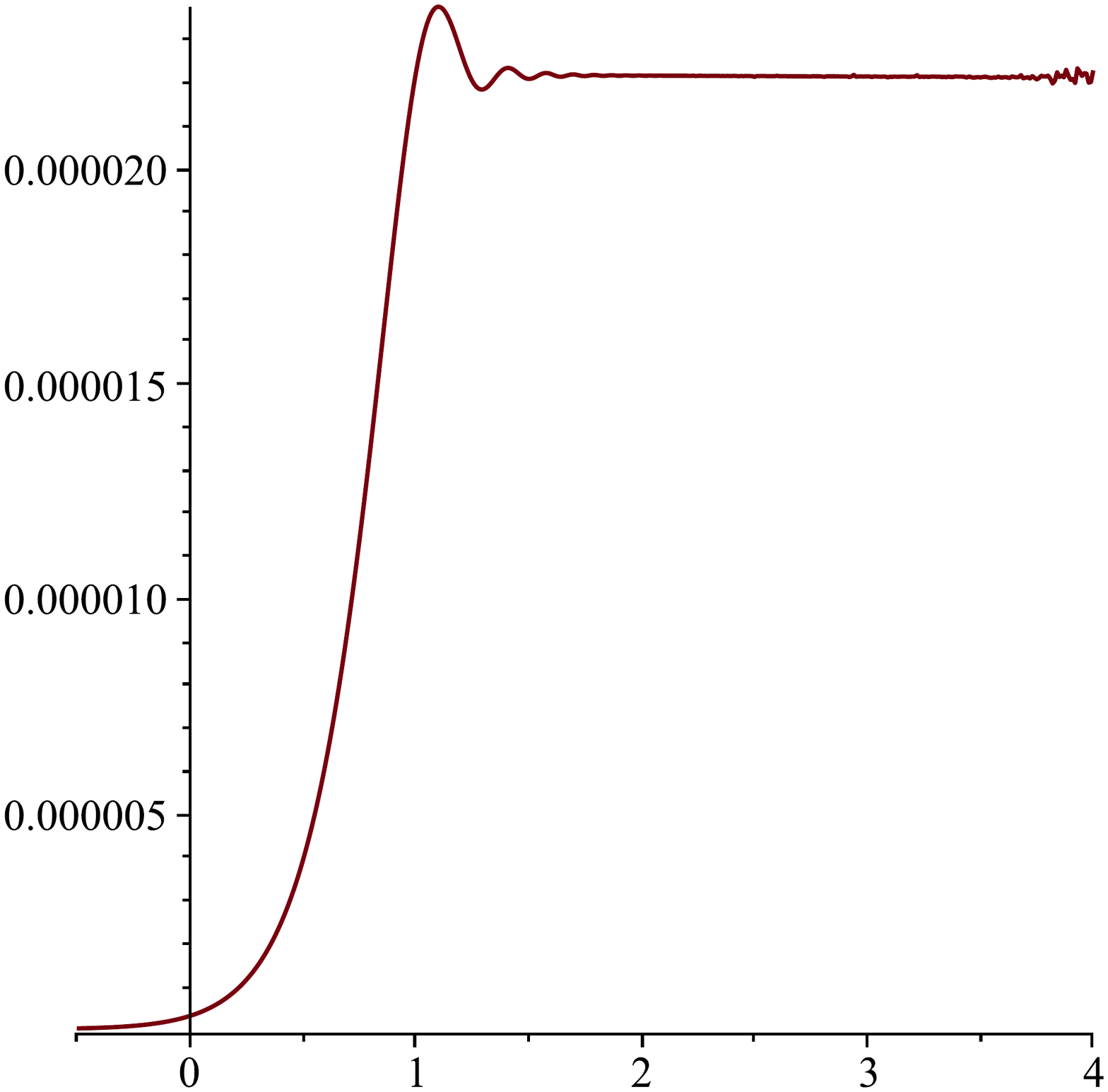} &
\includegraphics[width=35mm]{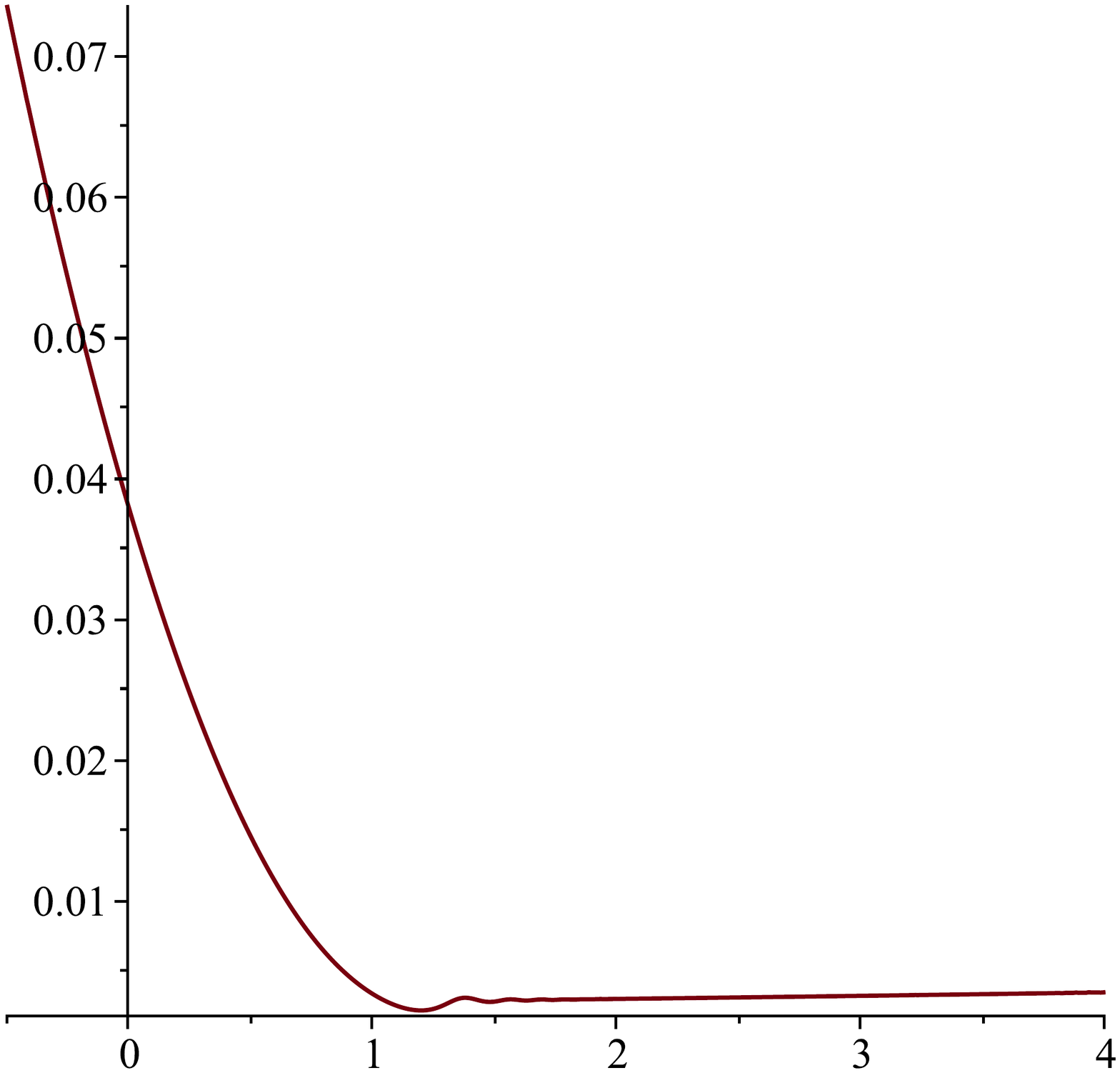}
\end{tabular}
\caption{\small Primordial power spectrum of scalar perturbations (left), of tensor perturbations (center), and corresponding behavior of the tensor--to--scalar ratio $r$ (right) for a Starobinsky model with a ``critical'' exponential wall, as in~\cite{climb}.
The vertical units are arbitrary, but are the same for tensor and scalar signals. The horizontal unit is $\log(k/k_0)$, with $k_0$ corresponding to the scale $\Delta$ of eq~\eqref{cutoff2}.
In this example $r$ would increase by a factor 5 -- 7 for $\ell \lesssim 20$.}
\label{fig:tensortoscalar}
\end{figure}

Pre-inflationary scenarios involving a decelerating inflaton, and in particular the string--inspired ones of \cite{climb}--\cite{erice15}, would have another signature that is potentially interesting: \emph{larger values of the tensor-to-scalar ratio $r$ should present themselves in the low-$\ell$ region} ($\ell \lesssim 20$, according to the typical values of $\Delta$ identified in~\cite{Gruppuso:2015xqa} or here), \emph{where however the amplitudes of both scalar and tensor perturbations would be somewhat reduced}. For instance, in~\cite{depression} we showed that when an inflaton approaches slow--roll in an exponential potential, $r$ is larger than its limiting value, and we found numerically {a similar behavior in other models}. These include a canonically normalized inflaton field in Starobinsky potentials~\cite{inflation}, with also an exponential wall contribution that is  ``critical'' in the sense of \cite{climb}. An example where the inflaton experiences a mild bounce on a ``critical'' exponential~\cite{depression} is displayed in fig.~\ref{fig:tensortoscalar}. Even ignoring the peak in the scalar spectrum, which is not modelled by eq.~\eqref{cutoff2}, here $r$ can grow by a factor 5 within a decade that, with a short inflation, could correspond to the first few multipoles, while accounting somehow for a peak structure the growth would enhance to a factor of about 7. Moreover, larger values of $r$ at low $\ell$ are favoured if the deceleration occurs in steeper regions of the potential. All in all, an enhancement of $r$ by about one order of magnitude in the low--multipole region $\ell \lesssim 20$ appears conceivable for this type of dynamics: it would allow tensor-to--scalar ratios $r \sim 3 \times 10^{-2}$ at large scales even in the Starobinsky--like scenarios that are favoured by {\sc Planck} data. Alternatively, an enhancement mechanism of this type could help one place tighter constraints on inflationary scenarios than what was achieved so far by {\sc Planck}.

The enhancement of $r$ is potentially interesting, together with the corresponding high tilt at low $\ell$, as in fig.~\ref{fig:tensortoscalar}. Regardless of any specific scenario, B--mode detection is a long--awaited result, because it would determine the inflationary scale, but above and beyond all this, because it would also provide some clear, if indirect, evidence for Quantum Gravity. Its effects on interactions are extremely feeble at accessible energies, but its zero--point fluctuations, classicalized by inflation according to~\cite{cm}, ought to be imprinted on the CMB sky. All in all, a properly enhanced $r$ for values of $k \sim \Delta \sim 0.35 \times 10^{-3} {\rm Mpc}^{-1}$ might bring B--mode detection within the reach of planned sub--orbital probes aiming at the CMB re--ionization bump, including the Italian LSPE~\cite{LSPE} and the CLASS~\cite{Essinger-Hileman:2014pja} telescope array, and of the aforementioned satellites LiteBIRD~\cite{LiteBIRD} and  CORE~\cite{Delabrouille:2017rct}. In this respect, we have also shown how the high--quality low--$\ell$ polarization data expected from these future surveys may help to improve the determination of $\Delta$ up to the 6 $\sigma$ level.

Independently of its dynamical origin, a primordial power spectrum modified along the lines of eq.~(\ref{cutoff2}) would also have important consequences for very large galaxy clusters (see, for instance, \cite{finelli}), whose distribution ought to drop at distance scales corresponding to $\Delta$. If further studies of systematics of astrophysical or instrumental origin will lend support to our findings, the results of \cite{Gruppuso:2015zia} and \cite{Gruppuso:2015xqa}, obtained largely from high--latitude Galactic regions, will deserve more attention.%, while the whole {\sc Planck} analysis might be susceptible of improvements.

\section{Conclusions}
\label{sec:conclusions}

Let us conclude with a brief summary of the main points touched upon in the preceding sections, which rest largely on a Bayesian approach:
\begin{itemize}
\item the \emph{even} low--$\ell$ CMB multipoles are largely suppressed with respect to their predicted values in $\Lambda$CDM, in a way that \emph{does not depend sizably on the choice of Galactic masking}, and thus on the portion of sky under scrutiny. This is confirmed by an analysis that rests on several estimators defined in Section~\ref{sec:constraintsestimators}, and reflects itself in detection levels of $\Delta$ via even multipoles alone that are essentially independent of Galactic masking;

\item the \emph{odd} low--$\ell$ CMB multipoles are not suppressed, up to a slight decrease at high Galactic latitudes;

\item the combined behavior of even and odd low--$\ell$ CMB multipoles reflects itself in determinations of the cutoff scale $\Delta$ of eq.~\eqref{cutoff2} that improve at higher latitudes, reaching the 3--$\sigma$ level in a blind $+30^\circ$ extension of the standard {\sc Planck} mask (the slight improvement with respect to~\cite{Gruppuso:2015zia,Gruppuso:2015xqa} is due to a different treatment of low--$\ell$ polarization data, obtained following~\cite{Lattanzi:2016dzq});

\item
String Theory and Supergravity can associate to these results enticing scenarios: $\Delta$ \emph{might be a relic of a deceleration that resulted in the onset of the inflationary phase}, some glimpses of which could be accessible to us with a small--enough number of $e$--folds. If this were the case, the low--frequency cut of both scalar and tensor spectra would be accompanied, in the same low--$\ell$ region, by a sharp \emph{increase of the tensor--to--scalar ratio $r$, with a high tilt}, by almost an order of magnitude. Finally, if $\Delta$ captures a real deviation from an almost scale invariant primordial power spectrum, it should be impinging on structure formation at the largest scales~\cite{finelli}, which ought to drop accordingly;

\item the available low--$\ell$ polarization data have played a role in our determination of the cutoff scale $\Delta$. The next generation of space--borne experiments aimed at high--quality polarization data has the potential to raise it up to the 6--$\sigma$ level (or perhaps to rule it out altogether). A positive result could further our understanding of the inflationary paradigm.
\end{itemize}
\paragraph{Acknowledgments}

We acknowledge the use of computing facilities at NERSC (USA), of the HEALPix package \cite{gorski}, and of the {\sc Planck} Legacy Archive (PLA).
This research was supported by ASI through the Grant 2016-24-H.0 (COSMOS) and through the ASI/INAF Agreement I/072/09/0 for the Planck LFI Activity of Phase E2, and by INFN (I.S. GSS- Pi, FlaG, InDark).
NK was supported in part by the JSPS KAKENHI, Grant Number 26400253 and in part by Scuola Normale, while AS was supported in part by Scuola Normale.
We would like to thank APC-Paris VII, Scuola Normale Superiore, the University of Ferrara and IASF--Bologna for their kind hospitality.
%
%\end{acknowledgments}
%
%
%

%
\end{document}